\documentclass[12pt,reqno]{amsart}
\usepackage{tgpagella, verbatim, float}
\usepackage{amsmath,amssymb,amsthm}
\usepackage{graphicx,epsfig,epstopdf}
\usepackage{url,mathtools}
\usepackage{colortbl,framed}
\usepackage{enumitem}
\usepackage{natbib}
\usepackage{hyperref}
\hypersetup{colorlinks=true,urlcolor=blue,linkcolor=blue,citecolor=blue}
\usepackage{caption}
\usepackage{subcaption}
\usepackage{float}
\usepackage{booktabs}
\usepackage{threeparttable}
\usepackage{subcaption}

\renewcommand{\qed}{\hfill{\tiny \ensuremath{\blacksquare} }}%

\newcommand{\R}{\mathbb{R}}

\newcommand{\mD}{\mathcal{D}}
\newcommand{\mW}{\mathcal{W}}
\newcommand{\mF}{\mathcal{F}}

\newcommand{\mX}{\mathcal{X}}

\newcommand{\vS}{\boldsymbol{S}}

\renewcommand{\qed}{\hfill {\tiny {\ensuremath{\blacksquare}}}}

\newtheorem{theorem}{Theorem}[section]
\newtheorem{corollary}{Corollary}[section]
\newtheorem{lemma}{Lemma}[section]

\newtheorem{assumption}{Assumption}[section]

\theoremstyle{definition}

\newtheorem{algorithm}{Algorithm}
\newtheorem{step}{Step}
\newtheorem{remark}{Comment}[section]
\numberwithin{remark}{section}

\numberwithin{equation}{section}
\numberwithin{theorem}{section}

\newcommand{\eps}{\varepsilon}

\usepackage{url}

\usepackage{pagecolor,lipsum}% http://ctan.org/pkg/{pagecolor,lipsum}

\newcommand{\Ep}{{\mathrm{E}}}

\renewcommand{\Pr}{{\mathrm{P}}}

\renewcommand{\hat}{\widehat}
\newcommand{\En}{{\mathbb{E}_n}}

\renewcommand{\Pr}{{\mathrm{P}}}
\newcommand{\RR}{\mathbb{R}}

\renewcommand{\hat}{\widehat}
\renewcommand{\leq}{\leqslant}

\newcommand{\diag}{{\rm diag}}

\renewcommand{\qed}{\hfill{\tiny \ensuremath{\blacksquare} }}%
\newcommand{\mR}{\mathcal{R}}
\newcommand{\mY}{\mathcal{Y}}

\newcommand{\dd}{\mathrm{d}}

\renewcommand{\qed}{\hfill {\tiny {\ensuremath{\blacksquare}}}}

\theoremstyle{definition}

\numberwithin{remark}{section}
\numberwithin{equation}{section}
\numberwithin{theorem}{section}

\usepackage{url}

\usepackage{pagecolor,lipsum}% http://ctan.org/pkg/{pagecolor,lipsum}

\renewcommand{\Pr}{{\mathrm{P}}}

\renewcommand{\hat}{\widehat}

\renewcommand{\Pr}{{\mathrm{P}}}
\renewcommand{\dd}{{\mathrm{d}}}

\renewcommand{\hat}{\widehat}
\renewcommand{\leq}{\leqslant}

\usepackage{wallpaper}
\usepackage{graphicx}

\usepackage{setspace}
\usepackage{lipsum,appendix}
\usepackage{placeins}



\usepackage{etoolbox}
\patchcmd{\section}{\scshape}{\bfseries}{}{}
\makeatletter
\renewcommand{\@secnumfont}{\bfseries}
\makeatother

\usepackage{inputenc}
\usepackage{soul}
\usepackage{lineno}
\newcommand{\green}[1]{{\color[rgb]{0,0.501961,0} #1}} % for highlighting changes in green

\begin{document}

\normalsize
\onehalfspacing

\title[BDR]{Bivariate distribution regression; theory, estimation and an application to intergenerational mobility} 
\author[Chernozhukov, Fern\'andez-Val, Meier, van Vuuren \and Vella]{Victor Chernozhukov \and Iv\'an Fern\'andez-Val \and Jonas Meier \and Aico van Vuuren \and Francis Vella$^\dag$}\thanks{$^\dag$ Chernozhukov: Department of Economics and Center for Statistics and Data Science, MIT; Fern\'andez-Val: Department of Economics, Boston University; Meier: Swiss National Bank; van Vuuren: Faculty of Economics and Business, University of Groningen; Vella: Department of Economics, Georgetown University. Aico van Vuuren received financial support from Vetenskapsrådet, registration number 2020-02423.} 

%\author{Iv\'an Fern\'andez-Val \ \quad  Jonas Meier \ \quad  Aico van Vuuren \ \quad  Francis Vella}

% Boston University, Department of Economics,  ivanf@bu.edu.
% University of Amsterdam, Amsterdam School of Economics, j.c.meier@uva.nl.
% Georgetown University, Department of Economics, fgv@georgetown.edu.
% University of Groningen, Faculty of Economics and Business, a.p.van.vuuren@rug.nl.

\date{\today}
\maketitle

% no page number on title page, reset page numbering
\thispagestyle{empty} 
%\renewcommand{\thepage}{\arabic{page}} \setcounter{page}{0} 

%\vspace{0.1cm}

\sloppy

%\begin{center}\textbf{Abstract}\end{center}
\noindent 
%Although distribution regression (DR) was introduced in the 1970's as a tool for estimating conditional distributions, only recently has it been more actively employed in empirical economic investigations. This recent popularity reflects the more rigorous consideration of the functionals which can be obtained via DR. 
We employ distribution regression (DR) to estimate the joint distribution of two outcome variables conditional on chosen covariates. While Bivariate Distribution Regression (BDR) is useful in a variety of settings,
%where the focus is on the joint distribution of two outcomes conditional on covariates
it is particularly valuable when some dependence between the outcomes persists after accounting for the impact of the covariates.  
Our analysis relies on a result from \cite{Chernozhukov2018distribution} which shows that any conditional joint distribution has a local Gaussian representation. We describe how BDR can be implemented and present some associated functionals of interest. As modeling the unexplained dependence is a key feature of BDR, we focus on functionals related to this dependence. We decompose the difference between the joint distributions for different groups into composition, marginal and sorting effects. %The composition effect describes the difference due to covariate values and the marginal captures that from differences in the respective marginal distributions. The sorting effect reflects the impact of the unexplained dependence. 
We provide a similar decomposition for the transition matrices which describe how location in the distribution in one of the outcomes is associated with location in the other. Our theoretical contributions are the derivation of the properties of these estimated functionals and appropriate procedures for inference. 
Our empirical illustration focuses on intergenerational mobility.  Using the Panel Survey of Income Dynamics data, we model the joint distribution of parents' and children's earnings. 
%We assess the role of the remaining local correlation reflecting the relationship between the unobservables influencing each of the outcomes. %The latter indicates how strongly earnings correlate at a specific location in the distribution once controlled for covariates. 
By comparing the observed distribution with constructed counterfactuals, we isolate the impact of observable and unobservable factors on the observed joint distribution. We also evaluate the forces responsible for the difference between the transition matrices of sons' and daughters'.
%Employing administrative data from Switzerland, we find that the distribution of father's and son's earnings considerably differ from the one father's and daughter's. In particular, the magnitude and the shape of the local correlation vary substantially.

\vspace{0.2cm}

\vspace{0.2cm}

\vfill
\begin{singlespacing}
\begin{scriptsize}
%\noindent\textbf{Acknowledgements: } We would like to thank seminar participants at the University of Amsterdam. 
%Finally, we are thankful to have been granted access to the WiSIeR data, a collection of datasets provided by the Swiss Federal Social Insurance Office. 
\end{scriptsize}
\end{singlespacing}

\newpage
\onehalfspacing

% no TOC
%\tableofcontents 
%\newpage

\onehalfspacing

\section{Introduction} \label{s:intro}
Although distribution regression (DR) was introduced in the 1970s by \cite{Williams1971}, %for ordinal outcomes 
and  revisited by \cite{Foresi1995} in the 1990s, it has only relatively recently been more actively employed in empirical economic investigations.  For example, inference for DR with a continuous outcome variable was only developed in \cite{Chernozhukov2013inference} and for discrete, mixed discrete or continuous outcomes in \cite{Chernozhukovs2020}.  These papers have resulted in an examination of the economically interesting functionals which can be obtained from an estimated distribution of an economic variable conditional on a vector of exogenous variables. This has resulted in the use of DR in a range of different models and settings. These include, for example, models with endogeneity \citep{Chernozhukov2020nonseparable}, sample selection \citep{Chernozhukov2018distribution, FVV2024}, dynamic panels \citep{fernandezval2023dynamic}, and most recently difference-in-difference estimation \citep{fernandezval2024}.

While these extensions cover a wide class of models they,  with the exception of 
%\cite{fernandezval2024},
some studies we note below, are restricted to models with univariate outcomes. However, for many empirical questions it is necessary to examine how the joint distribution of two outcomes varies in response to changes in the conditioning variables. For example, in investigating household labor supply responses to changes in tax rates, one should quantify how the hours worked by each of the spouses respond and, perhaps more importantly, how changes in their individual labor supplies are correlated, thereby affecting overall household labor supply. Alternatively, in examining the impact of changes in the minimum wage it would be useful to investigate how both wages and hours of work individually and jointly respond. These examples are just illustrative of the large number of economic questions in which it is important to examine more than one outcome and, perhaps more importantly, how the joint distribution of the outcomes is affected.
%We employ DR to investigate settings in which the outcome of interest is the joint distribution of two outcome variables conditional on a set of covariates. Moreover, we
%describe how bivariate distribution regression (BDR) can be implemented and potentialluy interesting functionals which can be obtained via BDR estimates.
%We also describe the construction of counterfactual distributions. 
%Existing work has considered some of the issues we consider here. For example, \cite{Chernozhukov2018distribution} consider bivariate sample selection and  \cite{fernandezval2023} propose a bivariate DR in the presence of multivariate sample selection. 

The estimation of models with multiple outcomes by DR has been considered in \cite{Meier2020} and \cite{Wang2022}. 
\cite{Meier2020} extends univariate DR to the multivariate setting by using a multivariate indicator function at every location of the distribution.  \cite{Wang2022} propose a fast factorization method that also employs univariate DR. However, this transformation to a univariate approach is potentially restrictive as it requires a single outcome variable to be 
constructed from the multiple outcomes of interest. As DR then employs this constructed outcome, it does not enable each of the individual outcomes to be modeled separately. Moreover, it also bypasses the modeling of the correlation between the unobservables driving each of the outcome variables.% Thus, it is a very restricted approach to modeling the joint distribution of bivariate outcomes.
In contrast, BDR enables each of the individual outcomes to be explicitly modeled and facilitates an analysis of the local dependence structure. Furthermore, the choice of the link function in \cite{Meier2020} and \cite{Wang2022} might provide a poor approximation to the underlying distribution. In contrast, a result from \cite{Chernozhukov2018distribution}, guarantees that the joint distribution is locally Gaussian and BDR is able to approximate it.
%and \cite{fvvp2024} consider the conditional joint distribution of two censored outcomes.\green{???? which publication ????} 

We employ DR to investigate settings in which the outcome of interest is the joint distribution of two outcome variables conditional on a set of covariates. Moreover, we describe how bivariate distribution regression (BDR) can be implemented, as well as the potentially interesting functionals that can be derived from BDR estimates. We separately analyze the roles of the marginal distributions and the dependence structure. 
%Relying on a result from \cite{fernandezval2023} we estimate local parameters and examine their impact on their joint distribution.
%\green{????? which publication ????} This is useful for examining the relationship between the  unobservables across outcomes while allowing it to  differ across the  distributions for each outcome.
%While BDR is essentially a device to model the joint distribution of two outcome variables conditional on a set of covariates, 
%However, the separable specification of the marginal distributions and the dependence is not feasible with the previous models. 
As each is modeled separately in BDR, a high degree of flexibility is incorporated between the role of both observable and unobservable factors. 
%BDR also models the dependence between the unobservables across the two outcomes. 
%While BDR is an attractive approach to modelling a joint distribution 
%Moreover, this allows an assessment of the  respective contributions to inequality. Finally, our proposed method enables us to evaluate how much local dependence remains unexplained, even after controlling for observables. This is informative to address the role played by unobservables, i.e., skill or networks.
Alternative approaches to modeling conditional joint distributions exist. These include the use of copulas \citep[i.e.,][]{Klein2022} and non-parametric estimators \citep[i.e.,][]{Bouzebda2019}. Due to its semi-parametric properties, BDR can be seen as a mixture between these approaches. In contrast to copula models, BDR allows for greater flexibility in incorporating the impact of the covariates and requires relatively weaker parametric assumptions. Non-parametric approaches are more likely to suffer from the curse of dimensionality. A detailed overview of modeling choices for multivariate distributions can be found in \cite{Meier2020}.

Bivariate or multivariate distribution regression has been employed elsewhere. For example, \cite{fernandezval2023} model the joint distribution of spouses' wages in the presence of selection rules, \cite{fernandezval2024} employ it in implementing difference-in-difference procedures with bivariate outcomes, while \cite{fernandezval2022cpscouples} employ it to describe the joint distribution of spouses' annual earnings. The first two papers provide a theoretical contribution related to the literature on which they are focused and the third is purely a descriptive analysis. This paper provides important original theoretical contributions related to inference, the validity of the bootstrap employed, and the decompositions of the joint distribution and transition matrices.

%In this paper, we introduce and analyze various aspects of BDR that make it a useful tool for empirical work, particularly when the focus is on objects that pertain to the bivariate distribution. These objects include bracket CDFs and a bivariate Gini coefficient. We also examine how the conditional rank-rank relationship proposed by \cite{chernozhukov2025} is affected by the treatment of the unobservables.In addition to defining these functionals we provide the associated estimation and inference methods. 

Our empirical illustration is related to intergenerational mobility. This is a research area with potentially important policy implications on a range of topics such as tax policy and inequality reduction. It has received significant attention from policymakers and has generated a substantial academic literature. %including both theoretical and empirical studies. 
In a seminal paper, \cite{Chetty2014} showed that schools, neighborhoods, and family stability are influential determinants of mobility. Others have shown that race \citep{Chetty2020}, immigration status \citep{Abramitzky2021}, human capital \citep{Adermon2021}, or wealth \citep{Adermon2018} are correlated with various mobility measures (see \cite{Mogstad2021} for a recent overview). This suggests that one should include many covariates when modeling intergenerational mobility noting that the appropriate manner to do so is not straightforward given that the relationship between the covariates and earnings appears to vary over the earnings' distributions.
%For example, the mentioned studies typically present estimates of relative mobility across subsamples of the data or promote a linear regression design where the child's outcomes (expressed as ranks in income, human capital, or wealth) are explained by parents' analog and (sometimes) covariates. 
%By estimating the conditional joint distribution of earnings we can explore the role of covariates at different points of the distribution. 
Previous studies typically present estimates of relative mobility across subsamples of the data or employ a linear regression approach in which the child's outcomes (rank of income, human capital, or wealth) are regressed on the  parents' analog and (sometimes) covariates. 
%We extend these existing approaches.
By employing BDR to model the joint distribution of fathers' and children's income 
%We model the latter using a flexible bivariate distribution regression (BDR) approach to serve this purpose. 
we provide an additional methodological approach for evaluating intergenerational mobility. First, we allow for the covariates to have varying effects at different points of the earnings' distributions. Second, we can allow for different covariates to have an impact on each of the marginal distributions and the dependence structure. Third, as the joint distribution captures the entire dependence structure between the respective income measures, standard mobility measures, such as the rank-rank relationship and the conditional expected rank, can be directly derived from the BDR estimates. Fourth, we can explore how these mobility measures are influenced by the dependence in the unobservables and how this dependence may vary across the income distribution. Finally, BDR estimates can be employed to produce counterfactual bivariate distributions corresponding to scenarios capturing alternative values of the local correlation. 

The remainder of this paper is organized as follows. Section \ref{s:methodological_contribution} formally presents our methodological contribution.  In Section \ref{s:functionals} we present a number of functionals which are potentially of interest in empirical work. Section \ref{s:estimation} outlines the associated estimation procedure. Section \ref{sec:theory} presents the asymptotic theory.  Section \ref{s:empirical} contains our empirical application. Section \ref{s:conclusion} concludes.

\section{Bivariate Distribution Regression}\label{s:methodological_contribution}

Let $F_{Y,W \mid X}$ be the joint distribution of $(Y,W)$ conditional on $X$. We consider the bivariate distribution regression (BDR) model:
\begin{equation}\label{eq:drmodel}
  F_{Y,W \mid X}(y,w \mid x) = \Phi_2(x'\mu_y, x'\nu_w; g(x'\delta_{yw})),   
\end{equation}
where $\Phi_{2}(\cdot, \cdot; \rho)$ is the distribution of the standard bivariate normal with parameter $\rho$, and $u \mapsto g(u)$ is a known link function with range $[-1,1]$ such as the Fisher transformation $g(u) = \tanh(u)$.  In \eqref{eq:drmodel} the marginal distributions of $Y$ and $W$ conditional on $X$ follow distribution regression models:
$$
F_{Y \mid X}(y \mid x) = \Phi(x'\mu_y), \quad F_{W\mid X}(w \mid x) = \Phi(x'\nu_w),
$$
where $\Phi$ is the distribution of the standard normal. The function $(y,w,x) \mapsto g(x'\delta_{yw})$ measures the local dependence or sorting between $Y$ and $W$ at $(Y,W,X) =  (y,w,x)$. This sorting can vary with respect to observed covariates $X$, and along the distribution as indexed by $(y,w)$. The simplest case is $g(x'\delta_{yw}) = g(\delta_{yw})$, where the sorting only varies with respect to unobservables.

The BDR model  can be motivated by the local Gaussian representation (LGR) of 
\cite{Chernozhukov2018distribution}, which establishes that for any conditional joint distribution,
\begin{equation}
F_{Y,W \mid X}(y,w \mid x) = \Phi_2(\mu(y \mid x), \nu(w \mid x); \rho(y,w \mid x)), \label{eq:LGR1}
\end{equation}
for some functions $\mu$, $\nu$ and $\rho$. The LGR is the right-hand-side of the previous equation and is unique, that is there is a one-to-one mapping between a conditional joint distribution and its LGR. In the LGR the marginal distributions of $Y$ and $W$ conditional on $X$ are represented by nonparametric distribution regression models, that is
$$
F_{Y \mid X}(y \mid x) = \Phi(\mu(y \mid x)), \quad F_{W \mid X}(w \mid x) = \Phi(\nu(w \mid x)).
$$
 The BDR model can be seen as a semiparametric specification for the LGR where the nonparametric functions  $\mu$, $\nu$ and $\rho$ are replaced by (generalized) linear indexes with function-valued  parameters. In particular, the parameters $y \mapsto \mu_y$ and $w \mapsto \nu_w$ measure the effect of the covariates on the marginal distributions of $Y$ and $W$, and $(y,w) \mapsto \rho_{yw}$ measures the effect of the covariance on the local dependence (copula) between $Y$ and $W$.

Let $\bar \mY$ and $\bar \mW$  denote strict subsets of $\mY$ and $\mW$, the supports of $Y$ and $W$, respectively. We impose the following restrictions on the coefficients at the tails to facilitate estimation and inference,
$$
\mu_{y,1} = \mu_{\bar y_y, 1} + (y - \bar y_y)\alpha_{\bar y_y}, \quad \mu_{y,-1} = \mu_{\bar y_y,-1},\quad  y \in \mY \setminus\bar{\mY},
$$
$$
\nu_{w,1} = \nu_{\bar w_w,1} + (w - \bar w_w)\alpha_{\bar w_w}, \quad \nu_{w,-1} = \nu_{\bar w_w,-1},\quad  w \in \mW \setminus\bar{\mW},
$$
and
$$
\delta_{y,w} = \delta_{\bar y_y \bar w_w},\quad  y \in \mY \setminus\bar{\mY}, \quad w \in \mW \setminus\bar{\mW},
$$
where $\bar y_y := \arg \min_{y' \in \bar{\mY}} |y-y'|$,  $\alpha_{\bar y_y} > 0$, $\bar w_w := \arg \min_{w' \in \bar{\mW}} |w-w'|$,  $\alpha_{\bar w_w} > 0$, and the subscript $1$ and $-1$ denote the first element of a vector and its complement. For example, $\mu_{y,1}$ is the first element of $\mu_y$ and $\mu_{y,-1}$ is a vector with the rest of the elements. Here, we follow \cite{chernozhukov2025} and postulate that the random variables $Y$ and $W$ behave in the tails like random variables with distribution $\Phi$, after subtracting the location shifts $x'\mu_{\bar y}$ and $x'\nu_{\bar w}$, and dividing by the scales 
$\alpha_{\bar y_y}$ and $\alpha_{\bar w_w}$, which are different at the upper and lower tails. For the local dependence parameter $\delta_{yw}$, we postulate that it is constant at the tails. We could  allow for additional tail parameters for the intercept of $\delta_{yw}$, but we set them to zero for simplicity.\footnote{Note that, unlike the intercepts of $\mu_y$ and $\nu_w$, the intercept of $\delta_{yw}$ does not need to satisfy any monotonicity restriction at the tails.}

%[IFV: I THINK THIS PARAGRAPH BELOWS TO THE INTRODUCTION] In related work, \cite{Meier2020} and \cite{Wang2022} employ univariate DR to provide estimates of the joint distribution. \cite{Meier2020}, extends univariate DR to the multivariate setting by using a multivariate indicator function at every location of the distribution. In the bivariate case, this implies the estimation of single DRs on a grid of $\mathcal{Y} \times \mathcal{W}$ -- a flexible but computationally costly approach. \cite{Wang2022} propose a fast factorization approach that also employs univariate DRs. These approaches are potentially restrictive as they require a single outcome variable to be discrete and this abstracts from interaction effects between the covariates and the discrete outcome. In contrast, BDR enables the explicit analysis of the local dependence structure. Furthermore, the choice of the link function in \cite{Meier2020} and \cite{Wang2022} might provide a poor approximation to the underlying distribution. With BDR, the LGR in \eqref{eq:LGR1} guarantees that the joint distribution is, locally, Gaussian. 

\section{Functionals of Interest}\label{s:functionals}
%We exploit the rich information contained in the conditional bivariate distribution. 
Rather than examine statistics which can be obtained in the univariate setting, we exploit the rich information contained in the conditional bivariate distribution and focus on functionals which highlight the dependence structure in the data. Accordingly, the local dependence parameter $\delta_{yw}$ features in each of the objects that follow. 

\subsection{Joint Distribution and Decomposition}

\label{ss:marginal} The joint distribution of $Y$ and $W$ can be represented in terms of the BDR model as
$$
F_{Y,W}(y,w) = \int \Phi_2(x'\mu_y,x'\nu_w;g(x'\delta_{yw})) \dd F_X(x),
$$
where $F_X$ is the distribution of $X$.     
Using this representation, we can decompose the difference between the joint distribution of two groups, say $D=1$ and $D=0$. To do so, it is convenient to introduce counterfactual distributions that combine BDR parameters and distributions of $X$ from different groups. Let
\begin{equation}\label{eq:counter-cdf}
    F^{(j,k,l,m)}_{Y,W}(y,w) := \int \Phi_2(x'\mu^j_y,x'\nu^k_w;g(x'\delta^l_{yw})) \dd F^m_X(x),
\end{equation}
where $\mu_y^i$ is the parameter $\mu_y$ in group $i$ and the rest of the terms are defined analogously. In this notation the actual distribution in group $d$ is $F^{(d,d,d,d)}_{Y,W}$.

The difference in the actual distributions between groups $1$ and $0$ can then be decomposed as
\begin{multline}\label{eq:cdf-decomp}
   \underset{\text{Total}}{\underbrace{ F^{(1,1,1,1)}_{Y,W}(y,w) - F^{(0,0,0,0)}_{Y,W}(y,w) }}=  
    \underset{\text{Composition}}{\underbrace{\left[ F^{(1,1,1,1)}_{Y,W}(y,w) - F^{(1,1,1,0)}_{Y,W}(y,w) \right]}} \\
    +  \underset{\text{Sorting}}{\underbrace{\left[ F^{(1,1,1,0)}_{Y,W}(y,w) - F^{(1,1,0,0)}_{Y,W}(y,w)  \right]}} 
    +  \underset{\text{Marginals}}{\underbrace{\left[  F^{(1,1,0,0)}_{Y,W}(y,w) - F^{(0,0,0,0)}_{Y,W}(y,w) \right]}},
\end{multline}
where the first term in brackets is a composition effect, the second term is a dependence or sorting effect, and the third term is the effect of the difference in the marginal distributions. The last term can be further decomposed into 
\begin{equation}
 \underset{\text{Marginal of $W$}}{\underbrace{\left[  F^{(1,1,0,0)}_{Y,W}(y,w) - F^{(1,0,0,0)}_{Y,W}(y,w) \right]}}
+  \underset{\text{Marginal of $Y$}}{\underbrace{\left[ F^{(1,0,0,0)}_{Y,W}(y,w) - F^{(0,0,0,0)}_{Y,W}(y,w)  \right]}},
\end{equation}
where the first term is the marginal effect related to $W$ and the last term is the marginal effect related to $W$. 

\subsection{Transition Matrices and Decomposition}\label{ss:bracket_cdf}
We can also construct transition matrices and decompose differences across those for different groups into composition, sorting and marginal distribution components using counterfactual distributions. Let $\{y_j: 0 \leq j \leq J\}$ and $\{w_k: 0 \leq k \leq K\}$ be grids of values covering the supports of $Y$ and $W$, where we set $y_0=w_0 = - \infty$ and $y_K = w_K = +\infty$. Then, the $(j,k)$ element of the counterfactual transition matrix $T^{(k,d,r,s)}$ is
\begin{equation}
T_{jk}^{(k,d,r,s)} := F^{(k,d,r,s)}_{Y,W}(y_j,w_k) - F^{(k,d,r,s)}_{Y,W}(y_{j-1},w_k)  - F^{(k,d,r,s)}_{Y,W}(y_j,w_{k-1})  + F^{(k,d,r,s)}_{Y,W}(y_{j-1},w_{k-1}). 
\label{eq:tm}
\end{equation}
for $j = 1,\ldots,J$ and $k = 1,\ldots,K$. These matrices provide a parsimonious representation of the joint distribution of $Y$ and $W$.

The difference in the actual transition matrices between groups $1$ and $0$ can then be decomposed as
\begin{multline*}\label{eq:cdf-decomp-tm}
   \underset{\text{Total}}{\underbrace{ T^{(1,1,1,1)} - T^{(0,0,0,0)} }}=  
    \underset{\text{Composition}}{\underbrace{\left[ T^{(1,1,1,1)} - T^{(1,1,1,0)} \right]}} 
    +  \underset{\text{Sorting}}{\underbrace{\left[ T^{(1,1,1,0)} - T^{(1,1,0,0)}  \right]}} \\
    +  \underset{\text{Marginal of $W$}}{\underbrace{\left[  T^{(1,1,0,0)} - T^{(1,0,0,0)} \right]}}   +  \underset{\text{Marginal of $Y$}}{\underbrace{\left[  T^{(1,0,0,0)} - T^{(0,0,0,0)} \right]}},
\end{multline*}
We provide examples of this decomposition in Section \ref{s:empirical}.

\section{Estimation} \label{s:estimation} 

Rather than employing probit, as is done in univariate DR, BDR estimates a sequence of bivariate probits. Let $I^y := 1(Y \leq y)$, $\bar I^y := 1 - I^y$, $J^w := (W \leq w)$ and  $\bar J^w := 1 - J^w$, for $(y,w) \in \mY_n \times \mW_n$, where $\mY_n$ and $\mW_n$  are grids of points in the supports of $Y$ and $W$, respectively. 

Under the BDR model, the joint probability function of $I^y$ and $J^w$ conditional on $X=x$ is
\begin{align}%\label{eq:prob}
f_{I^y,J^w \mid X}(i,j \mid x, \mu_y,\nu_w, \delta_{y,w}) &=
\Phi_{2}(x'\mu_y, x'\nu_w; \rho(x'\delta_{yw})^{ij} \nonumber \\
&\times \Phi_{2}(x'\mu_y, -x'\nu_w; -\rho(x'\delta_{yw})^{i(1-j)} \nonumber \\
&\times \Phi_{2}(-x'\mu_y, x'\nu_w; -\rho(x'\delta_{yw})^{(1-i)j} \nonumber \\
&\times \Phi_{2}(-x'\mu_y, -x'\nu_w; \rho(x'\delta_{yw})^{(1-i)(1-j)} \\
& i,j \in \{0,1\}. \nonumber
\end{align}
Given a random sample of size $n$, $\{(Y_i,W_i,X_i)\}_{=1}^n$, the average conditional log-likelihood at $(y,w)$ is
\begin{align}\label{eq:loglik1}
\ell^{yw}(\mu, \nu,\delta) = \frac{1}{n} \sum_{i=1}^n \log f_{I^c,J^f \mid Z}(I_i^c, J_i^f \mid X_i, \mu,\nu, \delta).
\end{align}
The conditional maximum likelihood estimator (CMLE) of $(\mu_y,\nu_w, \delta_{yw})$ is
\begin{align}\label{eq:cmle}
(\hat \mu_y,\hat \nu_w,\hat \delta_{yw}) \in \arg\max_{\mu,\nu,\delta} \ell^{y,w}(\mu,\nu,\delta).
\end{align}
Several remarks should be noted regarding computation. First, when $\rho(x'\delta_{yw}) = \rho(\delta_{yw})$, then the CMLE is the bivariate probit estimator of $(I^y,J^w)$ on $X$. Second, the CMLE can be computed in two steps. In the first step, we obtain the estimators of $\mu_y$ and $\nu_w$ by probit distribution regression of $I^y$ on $X$ and $J^w$ on $X$, respectively. In the second step, we plug in the estimators of  $\mu_y$ and $\nu_w$ from the first step in the average log-likelihood and maximize with respect to $\delta$, that is
\begin{align}\label{eq:loglik2}
\hat \delta_{yw} \in \arg\max_{\delta} \ell^{y,w}(\hat \mu_y,\hat \nu_w,\delta),
\end{align}
where $\hat \mu_y$ and $\hat \nu_w$ are the distribution regression estimators of $\mu_y$ and $\nu_w$. Finally, to trace the joint and marginal distributions of $Y$ and $W$, we need to obtain the CMLE for a grid of values of $(y,w)$ in the support of $(Y,W)$. For example, we can use the tensor product of two grids for $Y$ and $W$, including sampling percentiles from the $1$ or $2\%$ to the $99$ or $98\%$.

Let $\En$ denote the empirical expectation, that is $\En := n^{-1} \sum_{i=1}$.
\begin{algorithm}[BDR Estimator]\label{alg:bdr} Let  $d_x := \dim X$,  $\RR_n$   denote the set containing the observed values of $R$  and $\bar{\RR}_n = \RR_n \cap \bar{\RR}$, for $R \in \{Y,W\}$.
\begin{enumerate}
\item Estimate $\mu_y$ at  $y \in \bar{\mY}_n$ and $\nu_w$ at $w \in \bar{\mW}_n$ by DR, that is, 
$$
\hat \mu_y \in \arg\max_{\mu  \in \mathbb{R}^{d_x}} \ell^y(\mu) = \En [ \ell_i^{y}(\mu)], \quad \ell_i^{y}(\mu) := I_i^y \log \Phi(X_i'\mu) + \bar I_i^y \log \Phi(-X_i'\mu),
$$
and
$$
\hat \nu_w \in \arg\max_{\nu \in \mathbb{R}^{d_x}} \ell^w(\mu) =  \En[ \ell_i^{w}(\nu)], \quad \ell_i^{w}(\mu) := J_i^w \log \Phi(X_i'\nu) + \bar J_i^w \log \Phi(-X_i'\nu).
$$
\item Estimate $\mu_y$ at $y \in \mY_n \setminus \bar{\mY}_n$ and $\nu_w$ at $w \in \mW_n \setminus \bar{\mW}_n$ by restricted DR, that is, 
$$
\hat \mu_y = (y -\bar y_y)\hat \alpha_{\bar y_y}  + x'\hat \mu_{\bar y_y} \text{ and } \hat \nu_w = (w -\bar w_w)\hat \alpha_{\bar w_w}  + x'\hat \nu_{\bar w_w},$$ where $\bar y_y := \arg \min_{y' \in \bar{\mY}_n} |y-y'|$, $\bar w_w := \arg \min_{w' \in \bar{\mW}_n} |w-w'|$,  
\begin{footnotesize}
    \begin{equation*}
    \begin{split}
\hat \alpha_{\bar y_y} & \in \arg\max_{a \in \mathbb{R}}  \En[  \ell_i^{y_0}(a,\hat \mu_{\bar y_y})], \\  \ell_i^{y_0}(a,\mu) & := \left[  I_i^{y_0}  \log \Phi((y_0-\bar y_y)a  + X_i' \mu) + \bar I_i^{y_0}  \log \{\Phi(-(y_0-\bar y_y)a  - X_i'\mu)\right],
\end{split}
\end{equation*}
\begin{equation*}
\begin{split}
\hat \alpha_{\bar w_w} & \in \arg\max_{a \in \mathbb{R}}  \En  [\ell_i^{w_0}(a,\hat \nu_{\bar w_w})], \\  
\ell_i^{w_0}(a,\nu) &:= \left[  J_i^{w_0}  \log \Phi((w_0-\bar w_w)a  + X_i'\nu) + \bar J_i^{w_0}  \log \{\Phi(-(w_0-\bar w_w)a  - X_i'\nu)\right],
\end{split}
\end{equation*}
\end{footnotesize}and, for $r \in \{y,w\}$ and $\mR \in \{\mY,\mW\},$ $r_0 \in \RR_n \setminus \bar{\RR}_n$ is such that (i) there are at least $m$ observations between $\bar r$ and $r_0$, and greater than $r_0$ if $r_0 > \bar r_r$ (upper tail) or less than $r_0$ if $r_0 < \bar r_r$ (lower tail), and (ii) $\hat \alpha_{\bar r_r} > 0$. 
\item Estimate $\delta_{yw}$ at $(y,w) \in \bar{\mY}_n \times \bar{\mW}_n$ by restricted BDR, that is,
$$
\hat \delta_{yw} \in \arg\max_{\delta \in \mathbb{R}^{d_x}} \ell^{yw}(\hat \mu_y,\hat \nu_w,\delta) = \En[ \ell_i^{yw}(\hat \mu_y,\hat \nu_w,\delta)],
$$
where
\begin{eqnarray*}
\ell_i^{yw}(\mu,\nu,\delta) &=& I_i^y J_i^w \log \Phi_{2}(X_i'\mu, X_i'\nu; g(X_i'\delta)  + I_i^y \bar J_i^w \log \Phi_{2}(X_i'\mu, -X_i'\nu; -g(X_i'\delta)  \\ &+& \bar I_i^y J_i^w \log \Phi_{2}(-X_i'\mu, X_i'\nu; -g(X_i'\delta)) + \bar I_i^y \bar J_i^w \log \Phi_{2}(-X_i'\mu, -X_i'\nu; g(X_i'\delta)).
\end{eqnarray*}

\item Estimate $\delta_{yw}$ at $(y,w) \in (\mY_n\setminus \bar{\mY}_n) \times (\mW_n\setminus\bar{\mW}_n)$ by imposing the tail restrictions, that is,
$$
\hat \delta_{yw} = \hat \delta_{\bar y_y \bar w_w}.
$$
%where $\bar y_y := \arg \min_{y' \in \bar{\mY}_n} |y-y'|$ and $\bar w_w := \arg \min_{w' \in \bar{\mW}_n} |w-w'|$.
\end{enumerate}
\end{algorithm}

   The counterfactual terms of the decomposition of the joint distribution can be estimated via a plug-in rule. For example, an estimator of the second integral on the right-hand side of \eqref{eq:cdf-decomp} can be constructed as follows:
   $$
   \frac{1}{n_0} \sum_{i=1}^n (1-D_i) \Phi_2(X_i'\hat \mu^1_y,X_i'\hat \nu^1_w;g(X_i'\hat \delta^1_{yw})),
   $$
   where $\hat \mu^1_y, \hat \nu^1_w, \hat \delta^1_{yw}$ are estimators of $\mu^1_y,  \nu^1_w,  \delta^1_{yw}$ in the subsample with $D=1$ and $n_0 := \sum_{i=1}^n (1-D_i)/n$. \footnote{Alternatively, a doubly-robust estimator can be formed as
   $$
   \frac{1}{n_0} \sum_{i=1}^n \frac{(1-D_i)}{1-\hat p(X_i)} 1\{Y_i \leq y, W_i \leq w\} -  \frac{1}{n_0} \sum_{i=1}^n \frac{(D_i-\hat p(X_i))}{1-\hat p(X_i)} \Phi_2(X_i'\hat \mu^1_y,X_i'\hat \nu^1_w;g(X_i'\hat \delta^1_{yw})),
   $$
   where $\hat p(x)$ is an estimator of the propensity score $p(x) = \Pr(D=1 \mid X=x)$.}

\begin{comment}
   The counterfactual terms of the decomposition of the Kendall coefficient can be estimated via a plug-in rule. For example, the average within-group Kendall coefficient $\tau_W$ in group $D=d$ can be estimated by
   \begin{equation}\label{eq:wtauhat}
      \hat \tau_W^{d} = \frac{4}{n_d} \sum_{i=1}^n 1(D_i=d) \Phi_2(X_i'\hat \mu^d_{Y_i}, X_i'\hat \nu^d_{W_i}; g(X_i'\hat \delta^d_{Y_i W_i}))   - 1, 
   \end{equation}
where $\hat \mu^d_y, \hat \nu^d_w, \hat \delta^d_{yw}$ are estimators of $\mu^d_y,  \nu^d_w,  \delta^d_{yw}$ in the subsample with $D=d$ and $n_d := \sum_{i=1}^n 1(D_i=d)$.
   Similarly, the counterfactual average within-group Kendall coefficient $\tau_W^{\pi}$ in \eqref{eq:kendall-counter} can be estimated by
% $$
% \hat \tau_W(\mu^1,\nu^1,\delta^1,F_{X\mid D=0}) = \frac{4}{n_0 n_1} \sum_{i=1}^n (1-D_i) \sum_{j=1}^n D_j   \Phi_2(X_i'\hat \mu_1(Y_j), X_i'\hat \nu_1(W_j); g(X_i'\hat \delta_1(Y_j,W_j))) - 1,
% $$
% where $n_d := \sum_{i=1}^n 1(D_i=d)$.  Alternatively,
$$
\hat \tau_W^{\pi} = \frac{4}{n_1} \sum_{i=1}^n D_i \Phi_2(X_i'\hat \mu^1_{Y_i}, X_i'\hat \nu^1_{W_i}; g(X_i'\hat \delta^1_{Y_i W_i})) \hat \pi_{X_i}   - 1,   \quad \hat \pi_{x} := \frac{(1-\hat p(x))  \hat p}{\hat p(x)(1- \hat p)},
$$
where 
$\hat p = n_1/n$ AND $\hat p(x)$ is an estimator of the propensity score $p(x) = \Pr(D=1 \mid X=x)$.

\end{comment}

\subsection{Inference} We use the exchangeable bootstrap for inference. The following algorithm describes how to obtain counterfactual draws of the BDR estimator.
\begin{algorithm}[Exchangeable Bootstrap Draw of BDR Estimator]\label{alg:bdr_boots} \hfill
\begin{enumerate}
\item Draw a realization of the weights $(\omega_{n1},\ldots,\omega_{nn})$ from a distribution that satisfies Assumption \ref{ass:eb} in Section \ref{sec:theory}. Normalize the weights to add-up to one.
\item Obtain a bootstrap draw of $\hat \mu_y$ at  $y \in \bar{\mY}_n$ and $\hat \nu_w$ at $w \in \bar{\mW}_n$ by weighted DR, that is, 
$$
\hat \mu^*_y \in \arg\max_{\mu  \in \mathbb{R}^{d_x}} \ell_*^y(\mu) =  \En [\omega_{ni} \ell_i^{y}(\mu)], \quad 
\hat \nu^*_w \in \arg\max_{\nu \in \mathbb{R}^{d_x}} \ell_*^w(\mu) =  \En [\omega_{ni} \ell_i^{w}(\nu)].
$$
\item Obtain bootstrap draw of $\hat \mu_y$ at $y \in \mY_n \setminus \bar{\mY}_n$ and $\hat \nu_w$ at $w \in \mW_n \setminus \bar{\mW}_n$ by weighted restricted DR, that is, 
$$
\hat \mu^*_y = (y -\bar y_y)\hat \alpha^*_{\bar y_y}  + x'\hat \mu^*_{\bar y_y} \text{ and } \hat \nu^*_w = (w -\bar w_w)\hat \alpha^*_{\bar w_w}  + x'\hat \nu^*_{\bar w_w},$$ where 
%$\bar y_y := \arg \min_{y' \in \bar{\mY}_n} |y-y'|$, $\bar w_w := \arg \min_{w' \in \bar{\mW}_n} |w-w'|$,  
\begin{equation*}
\hat \alpha^*_{\bar y_y} \in \arg\max_{a \in \mathbb{R}}  \En [ \omega_{ni} \ell_i^{y_0}(a,\hat \mu^*_{\bar y_y})], \quad 
\hat \alpha^*_{\bar w_w} \in \arg\max_{a \in \mathbb{R}}  \En  [\omega_{ni} \ell_i^{w_0}(a,\hat \nu^*_{\bar w_w})], 
\end{equation*}
and, for $r \in \{y,w\}$ and $\mR \in \{\mY,\mW\},$ $r_0 \in \RR_n \setminus \bar{\RR}_n$ is the same as in Algorithm \ref{alg:bdr}(2). 
\item Obtain bootstrap draw of  $\hat \delta_{yw}$ at $(y,w) \in \bar{\mY}_n \times \bar{\mW}_n$ by weighted restricted BDR, that is,
$$
\hat \delta^*_{yw} \in \arg\max_{\delta \in \mathbb{R}^{d_x}} \ell_*^{yw}(\hat \mu_y,\hat \nu_w,\delta) = \En [\omega_{ni} \ell_i^{yw}(\hat \mu_y,\hat \nu_w,\delta)].
$$

\item Obtain bootstrap draw of  $\hat \delta_{yw}$ at $(y,w) \in (\mY_n\setminus \bar{\mY}_n) \times (\mW_n\setminus\bar{\mW}_n)$ by imposing the tail restrictions, that is,
$$
\hat \delta^*_{yw} = \hat \delta^*_{\bar y_y \bar w_w}.
$$
%where $\bar y := \arg \min_{y' \in \bar{\mY}_n} |y-y'|$ and $\bar w := \arg \min_{w' \in \bar{\mW}_n} |w-w'|$
\end{enumerate}
\end{algorithm}

Exchangeable bootstrap draws of the estimators of the functionals of interest can be obtained similarly.

\section{Asymptotic Theory}\label{sec:theory}
To simplify the expressions of some of the assumptions and limit processes, it is convenient to introduce the following notation for the tails. Let $\bar r_r  := \arg \min_{r' \in \bar{\mR}} |r-r'|$ for $r \in \{y,w\}$ and $\bar \mR \in \{\bar \mY,\bar \mW\}$. Note that $\bar r_r = r$ if $r \in \bar \mR$, $\bar r_r = \bar r$ if $r \ge \bar r$ and $\bar r_r = \underline r$ if $r \leq \underline r$, where $\bar r := \sup (\bar \mR)$ and $\underline r := \inf(\bar \mR)$. We use a superscript $d$ to denote variables, parameters and samples from the group $d  \in \{0,1\}$. We also use the following notation for partial derivatives: $\partial_x f(x) := \partial f(x)/\partial x$ and $\partial_{x x} f(x) := \partial^2 f(x)/(\partial x \partial x')$. 
\begin{assumption}[Model and Sampling]\label{ass:fclt} For $d \in \{0,1\}$: (1) Random sampling: $\{Z^d_i :=( Y^d_i,W^d_i,X^d_i)\}_{i=1}^{n_d}$ is a sequence of independent and identically distributed copies of $Z^d := (Y^d,W^d,X^d)$, which is independent across $d$, and $n_d/n \to p_d>0$ for $n := n_0+n_1$.\footnote{Some forms of dependence between the samples across $d$ can be accommodated following the analysis in \cite{Chernozhukov2013inference}. We assume independence for simplicity and because it holds in our application. } 
(2) Model: the joint distribution of $(Y^d,W^d)$ conditional on $X^d$ follows the BDR model \eqref{eq:drmodel} with the tail restrictions. (3) The support of $X^d$, $\mathcal{X}$,  is a  compact set.  (4) The supports of $Y$ and $W$, $\mathcal{Y}$ and $\mathcal{W}$
%support of $Y$ 
are either (i) finite sets or (ii) open intervals. In the second case, the conditional  density functions $f_{Y^d\mid X^d}(y \mid x)$ and $f_{W^d \mid X^d}(w \mid x)$ exist, are uniformly bounded above, and are uniformly continuous in $(y,x)$ on $\mathcal{Y}  \times \mathcal{X}$ and in $(w,x)$ on $\mathcal{W} \times \mathcal{X}$, respectively. (5) Identification and non-degeneracy: the equation $\Ep[\partial_{\delta} \ell^{yw}_i(\mu^d_y,\nu^d_w,\tilde \delta_{yw})] = 0$ possesses a unique solution at $\tilde \delta_{yw} = \delta^d_{yw}$ that lies in the interior of a compact set $\mathcal{D} \subset \mathbb{R}^{d_{\delta}}$ for all $(y,w) \in \bar \mY\bar \mW := \bar \mY \times \bar \mW$; the minimum eigenvalues of the matrices $\Ep[\partial_{\mu \mu}  \ell_i^y(\mu^d_y)]$, $\Ep[\partial_{\alpha \alpha}\ell_i^{y_0}(\alpha^d_{\bar y_y}, \mu_{\bar y_y})]$, $\Ep[\partial_{\nu \nu}  \ell_i^w(\nu^d_w)]$, $\Ep[\partial_{\alpha \alpha}\ell_i^{w_0}(\alpha^d_{\bar w_w}, \mu_{\bar w_w})]$ and $\Ep[\partial_{\delta \delta} \ell^{yw}_i(\mu^d_y,\nu^d_w, \delta^d_{yw})]$  are bounded away from zero uniformly over $y \in \bar \mY$, $y \in \{\underline y, \bar y\}$,  $w \in \bar \mW$,  $w \in \{\underline w, \bar w\}$, and $(y,w) \in \bar \mY \bar \mW$, respectively; and $\Ep\|X\|^2 < \infty$.
%the matrices $H_{1s}$, $H_{2y}$, $H_{3sy}$ and $\Sigma_{\rho}(y,w)$ are nonsingular for each  and $y \in \mathcal{Y}$.
%and $\Sigma_{\nu_y \nu_\tilde y}$ is finite and nonsingular uniformly in $y$ and $\tilde y$ on $\mathcal{Y}$.[FINITE MIGHT FOLLOW FROM THE ASSUMPTION ON THE SUPPORT OF $Z$]
\end{assumption}

Assumption \ref{ass:fclt} implicitly imposes common support for the variables $Y^d$, $W^d$ and $X^d$ across $d$. This condition guarantees that all the counterfactual distributions that we consider are well-defined.

\begin{assumption}[Exchangeable Bootstrap]\label{ass:eb} For each $n_d$ and $d \in \{0,1\}$, $(\omega^d_{n_d1}, ..., \omega^d_{n_dn_d})$ is an exchangeable,\footnote{A sequence of random variables $X_1, X_2, ..., X_n$ is exchangeable if for any finite permutation $\sigma$ of the indices $1,2, ..., n$ the joint
distribution of the permuted sequence $X_{\sigma(1)}, X_{\sigma(2)},
...,X_{\sigma(n)} $ is the same as the joint distribution of the original
sequence.} nonnegative random vector, which is independent of the data and across $d$, such that for some $\epsilon> 0$ 
\begin{equation}  \label{eq: assumptions weighted bootstrap}
\begin{split}
\sup_{n} \Ep[(\omega_{n_d1}^d)^{2+\epsilon}] < \infty, \ \ n_d^{-1}\sum
_{i=1}^{n_d} \left( \omega^d_{n_di} - \bar{\omega}^d_{n_d} \right)^{2} \to_{\Pr} 1, \ \
\bar \omega^d_{n_d} \to_{\Pr} 1,
\end{split}
\end{equation}
where $\bar \omega^d_{n_d} = n_d^{-1} \sum_{i=1}^{n_d} \omega^d_{n_di} $. 
\end{assumption}

Let  $\hat \theta^d_{yw} := ( \hat \mu_y^{d'}, \hat \nu_w^{d'}, \hat \delta_{yw}^{d'})'$ and $\hat \theta^{d*}_{yw} := ( \hat \mu^{d*'}_y, \hat \nu^{d*'}_w, \hat \delta^{d*'}_{yw})'$  be the estimator and corresponding bootstrap draw of the parameter $\theta^d_{yw} := ( \mu_y^{d'},\nu_w^{d'},\delta_{yw}^{d'})'$, for $y \in \bar \mY$ and $w \in \bar \mW$. Similarly, let  $\hat \xi_y^d := ( \hat \alpha_y^d,\hat \mu_y^{d'})'$ and $\hat \xi_w^d := ( \hat \alpha_w^d, \hat \nu_w^{d'})'$ be the estimators of the tail parameters $\xi_y^d := ( \alpha_y^d, \mu_y^{d'})'$ and $\xi_w^d := ( \alpha_w^d, \nu_w^{d'})'$, for $y \in\{\underline{y} , \bar y\}$ and $w \in\{\underline{w} , \bar w\}$,  and $\hat \xi^{d*}_y := ( \hat \alpha^{d*}_y,\hat \mu^{d*'}_y)'$ and $\hat \xi^{d*}_w := ( \hat \alpha^{d*}_w, \hat \nu^{d*'}_w)'$ be the corresponding bootstrap draws of the estimators. 
Let $Z_n \leadsto  Z $ in $\mathbb{D}$ denote weak convergence of a
stochastic process $Z_n$ to a random element $Z$ in a normed space $\mathbb{D
}$, as defined in \cite{van1996weak}. For example, $\mathbb{D
} = \ell^{\infty}(\bar \mY \bar \mW)$, the set of measurable and bounded functions on $\bar \mY \bar \mW$.

In order to state the results about bootstrap validity formally, we follow the notation and
definitions in \cite{van1996weak}. Let $D_{n}$ denote the data
vector and $M_{n}$ be the vector of random variables used to generate
bootstrap draws given $D_{n}$. Consider the random element $%
\mathbb{Z}^{*}_{n} = \mathbb{Z}_{n}(D_{n}, M_{n})$ in a normed space $%
\mathbb{E}$. We say that the bootstrap law of $\mathbb{Z}^{*}_{n}$
consistently estimates the law of some tight random element $\mathbb{Z}$ and
write $\mathbb{Z}^{*}_{n} \leadsto_{\Pr} \mathbb{Z} $ in $\mathbb{D}$
if: 
\begin{equation}  \label{boot1}
\begin{array}{r}
\sup_{h \in\text{BL}_{1}(\mathbb{D})} \left| \Ep_{M_{n}} h \left( \mathbb{Z}%
^{*}_{n}\right) - \Ep h(\mathbb{Z})\right| \rightarrow_{\Pr} 0,%
\end{array}%
\end{equation}
where $\text{BL}_{1}(\mathbb{D})$ denotes the space of functions with
Lipschitz norm at most 1 and $\Ep_{M_{n}}$ denotes the conditional expectation
with respect to $M_{n}$ given the data $D_{n}$; and $ \rightarrow_{\Pr}$ denotes
convergence in (outer) probability.

%\green{Not sure but it seems that theta was not defined yet. I assume it is the vector of mu, nu and delta?}

\begin{lemma}[FCLT and Bootstrap FCLT for $(\hat \theta_{yw}^d,\hat \xi_{\bar y}^d,\hat \xi_{\underline y}^d,\hat \xi_{\bar w}^d,\hat \xi_{\underline w}^d)$]\label{lemma:fclt} Assume that Assumption \ref{ass:fclt} holds. Then, for $d \in \{0,1\}$, (i)
$$
\sqrt{n}(\hat \theta^d_{yw} -  \theta^d_{yw}) \leadsto Z_{yw}^{d\theta} := \sqrt{p_d} \mathbb{G}[\psi_i(\theta^d_{yw})] \text{ in } \ell^{\infty}(\bar \mY \bar \mW)^{d_{\theta}},
 $$
where $\mathbb{G}$ is a Brownian bridge and $\psi_i(\theta^d_{yw}) := H^{yw}(\theta^d_{yw})^{-1} S^{yw}_i(\theta^d_{yw})$ with 
\begin{equation}\label{eq:score}
S^{yw}_i(\theta_{yw}^d) = \left[\begin{array}{c}
     \partial_{\mu} \ell_i^y(\mu_y^d) \\
     \partial_{\nu} \ell_i^w(\nu_w^d) \\
     \partial_{\delta} \ell_i^{yw}(\mu_y^d,\nu_w^d,\delta_{yw}^d)
     \end{array}\right],    
\end{equation}
and
\begin{equation}\label{eq:hessian}
H^{yw}(\theta_{yw}^d) = \Ep \left[\begin{array}{ccc}
     \partial_{\mu \mu}\ell_i^y(\mu_y^d) & 0 & 0 \\
     0 & \partial_{\nu \nu}\ell_i^w(\nu_w^d) & 0 \\
     \partial_{\delta \mu}\ell_i^{yw}(\mu_y^d,\nu_w^d,\delta_{yw}^d) & \partial_{\delta \nu}\ell_i^{yw}(\mu_y^d,\nu_w^d,\delta_{yw}^d) & \partial_{\delta \delta}\ell_i^{yw}(\mu_y^d,\nu_w^d,\delta_{yw}^d)
\end{array}\right];
\end{equation}
and (ii) 
$\sqrt{n}(\hat \xi_r^d - \xi_r^d) \leadsto Z^{d \xi_r}$ in $\mathbb{R}^{d_{\xi_r}}$, for $r \in \{\bar r, \underline r\}$ and $r \in \{y,w\}$, where 
$$
Z^{d \xi_r} :=  \sqrt{p_d} \Ep \left[\begin{array}{cc}
     \partial_{\alpha \alpha}\ell_i^{r_0}(\alpha_{\bar r_r}^d, \mu_{\bar r_r}^d) & \partial_{\alpha \mu}\ell_i^{r_0}(\alpha_{\bar r_r}^d, \mu_{\bar r_r}^d)
\\    0 &  \partial_{\mu \mu}\ell_i^y(\mu_{\bar r_r}^d)  
\end{array}\right]^{-1}
\mathbb{G} \left[ \begin{array}{c}
     \partial_{\alpha}\ell_i^{r_0}(\alpha_{\bar r_r}^d, \mu_{\bar r_r}^d) 
\\    \partial_{\mu}\ell_i^r(\mu_{\bar r_r}^d)  
\end{array} \right],
$$ 
has a zero-mean normal distribution.
% ; and (iii) $\sqrt{n}(\hat \xi_w - \xi_w) \leadsto Z^{\xi_w}$ in $\mathbb{R}^{d_{\xi_w}}$, for $w \in \{\bar w, \underline w\}$, where 
%  $$
% Z^{\xi_w} :=  \Ep \left[\begin{array}{cc}
%      \partial_{\alpha \alpha}\ell_i^{w_0}(\alpha_{\bar w_w}, \mu_{\bar w_w}) & \partial_{\alpha \mu}\ell_i^{w_0}(\alpha_{\bar w_w}, \mu_{\bar w_w})
% \\    0 &  \partial_{\mu \mu}\ell_i^w(\mu_{\bar w_w})  
% \end{array}\right]^{-1}
% \mathbb{G} \left[ \begin{array}{c}
%      \partial_{\alpha}\ell_i^{w_0}(\alpha_{\bar w_w}, \mu_{\bar w_w}) 
% \\    \partial_{\mu}\ell_i^w(\mu_{\bar w_w})  
% \end{array} \right],
% $$ 
% has a zero-mean normal distribution. 
Moreover, (i) and (ii)  hold jointly, and $(Z_{yw}^{0\theta},Z^{0 \xi_{\bar r}},Z^{0 \xi_{\underline r}})$ and $(Z_{yw}^{1\theta},Z^{1 \xi_{\bar r}},Z^{1 \xi_{\underline r}})$ are independent.  If in addition Assumption \ref{ass:eb} holds, for $d \in \{0,1\}$, jointly (i$^*$)
$$
\sqrt{n}(\hat \theta^{d*}_{yw} -  \hat \theta^d_{yw}) \leadsto_{\Pr} Z_{yw}^{d\theta} \text{ in } \ell^{\infty}(\bar \mY \bar \mW)^{d_{\theta}},
$$
and (ii$^*$) $\sqrt{n}(\hat \xi^{d*}_r - \hat \xi^d_r) \leadsto_{\Pr} Z^{d\xi_r}$ in $\mathbb{R}^{d_{\xi_r}}$, for $r \in \{\bar r, \underline r\}$ and $r \in \{y,w\}$.
% , and (iii$^*$) $\sqrt{n}(\hat \xi^*_w - \hat \xi_w) \leadsto_{\Pr} Z^{\xi_w}$ in $\mathbb{R}^{d_{\xi_w}}$, for $w \in \{\bar w, \underline w\}$.
\end{lemma}

The quantities of interest are functionals of the joint distribution of $(Y,W)$ conditional on $X$ and the marginal distribution of $X$. To state the results about these functionals it  is convenient to introduce some notation. The following definitions apply to all $j,k,l,m \in \{0,1\}$. Let $F^m_X$, $\hat F^m_X$ and $\hat F^{m*}_X$ be the joint distribution of $X^m$,  empirical counterpart and bootstrap draw;  and $F^{(jkl)}_{ywx} := \Phi_2(x'\mu^j_y,x'\nu^k_w;g(x'\delta^l_{yw}))$, $\hat F^{(j,k,l)}_{ywx} := \Phi_2(x'\hat \mu^j_y,x'\hat \nu^k_w;g(x'\hat \delta^l_{yw}))$ and $\hat F^{(jkl)*}_{ywx} := \Phi_2(x'\hat \mu^{j*}_y,x'\hat \nu^{k*}_w;g(x'\hat \delta^{l*}_{yw}))$ be the counterfactual conditional distribution,  its estimator and bootstrap draw. Consider the empirical processes and their bootstrap draws, $(y,w,x) \mapsto \hat Z^{(jkl)}_{ywx} := \sqrt{n}(\hat F^{(jkl)}_{ywx} - F^{(jkl)}_{ywx})$, $(y,w,x) \mapsto \hat Z^{(jkl)*}_{ywx} := \sqrt{n}(\hat F^{(jkl)*}_{ywx} - \hat F^{(jkl)}_{ywx})$,  $f \mapsto \hat G^m_X(f) := \sqrt{n} \int f \mathrm{d} (\hat F^m_X - F^m_X)$ and $f \mapsto \hat G^{m*}_X(f) := \sqrt{n} \int f \mathrm{d} (\hat F^{m*}_X - \hat F^m_X)$, for $f \in \mF$, where $\mF$ is a class of measurable functions that (i) includes $F^{(jkl)}_{ywx}$,  the indicators of all the rectangles in $\bar{\mathbb{R}}^{d_x+2}$, for $\bar{\mathbb{R}} = \mathbb{R} \cup \{-\infty,+\infty\},$ the extended real line, and (ii) is totally bounded under the metric:
$$
\lambda(f; \tilde f) = \left[\int (f - \tilde f)\mathrm{d} F_X \right]^{1/2}, \quad f,\tilde f \in \mF.
$$
For $d \in \{0,1\}$, the selection matrix $\vS_d^{(jkl)}$  picks up the elements of $Z_{ywx}^{d\theta}$ corresponding to the components used in the construction of the counterfactual conditional distribution. For example, $\vS_0^{(010)} = \diag(\boldsymbol{I}_{d_{\mu}},\boldsymbol{0}_{d_{\nu}},\boldsymbol{I}_{d_{\delta}})$ and $\vS_1^{(010)} = \diag(\boldsymbol{0}_{d_{\mu}},\boldsymbol{I}_{d_{\nu}},\boldsymbol{0}_{d_{\delta}})$, where $\boldsymbol{0}_{p}$ is a $p\times p$ matrix of zeros and $\boldsymbol{I}_{p}$ is the identity matrix of size $p$.

\begin{theorem}[FCLT and Bootstrap CLT for $\hat F^{(jkl)}_{ywx}$ and $\hat G^m_X(f)$]\label{thm:clt-cdf} (i) Assume that Assumption \ref{ass:fclt} holds. Then, jointly in $j,k,l,m \in \{0,1\}$,
$$
(\hat Z_{ywx}^{(jkl)},  \hat G^m_X(f)) \leadsto (Z^{(jkl)}_{ywx},   G^m_X(f)) \text{ in } \ell^{\infty}(\mY\mW\mX\mF),
 $$
where $G^m_X(f) := \sqrt{p_m} \mathbb{G}^m(f)$, with $\mathbb{G}^0$ and $\mathbb{G}^1$ independent, and
$$
Z_{ywx}^{(jkm)} :=  \partial_{\theta} F^{(jkl)}_{\bar y_y \bar w_w x} \sum_{d \in \{0,1\}}  \vS_d^{(jkl)} Z_{\bar y_y \bar w_w}^{d\theta} + \partial_{\alpha_y} F^{(jkl)}_{\bar y_y \bar w_w x} Z^{j \alpha_{\bar y_y}} 1(y \not\in \bar \mY) + \partial_{\alpha_w} F^{(jkl)}_{\bar y_y \bar w_w x} Z^{k \alpha_{\bar w_w}} 1(w \not\in \bar \mW),
$$
with $\partial_{\theta} F^{(jkl)}_{\bar y_y \bar w_w x} := \partial_{\theta} \Phi_2(x'\mu^j_{\bar y_y},x'\nu^k_{\bar w_w};g(x'\delta^l_{\bar y_y \bar w_w}))$, $\partial_{\alpha_y} F^{(jkl)}_{\bar y_y w x} := \partial_{\alpha_{\bar y_y} }\Phi_2(\alpha^j_{\bar y_y}(y - \bar y_y) + x'\mu^j_{\bar y_y},x'\nu^k_w;g(x'\delta^l_{yw}))
$, $
\partial_{\alpha_w} F^{(jkl)}_{y \bar w_w x} := \partial_{\alpha_{\bar w_w} }\Phi_2( x'\mu^j_{y},\alpha^l_{\bar w_w}(w - \bar w_w) + x'\nu^l_{\bar w_w};g(x'\delta^l_{yw}))
$,  $Z^{d\alpha_{r}} := e_1'Z^{d\xi_{r}}$ for $r \in \{y,w\}$ and $d \in \{0,1\}$, $e_1$ is a unitary vector with a one in the first position, 
 and 
$Z_{yw}^{d\theta}$, $Z^{d\xi_y}$ and $Z^{d\xi_w}$ are defined in Lemma \ref{lemma:fclt}. (ii) If in addition Assumption \ref{ass:eb} holds,   then, jointly in $j,k,l,m \in \{0,1\}$,
$$
(\hat Z^{(jkl)*}_{ywx},  \hat G^{m*}_Z(f)) \leadsto_{\Pr} (Z^{(jkl)}_{ywx},   G^m_Z(f)) \text{ in } \ell^{\infty}(\mY\mW\mX\mF).
 $$
\end{theorem}

Let
$$
\hat F_{Y,W}^{(j,k,l,m)}(y,w) = \dfrac{1}{n_m} \sum_{i=1}^{n_m}  \Phi_2(X_i^{m'}\hat \mu_y^j, X_i^{m'}\hat \nu_w^k;g(X_i^{m'}\hat \delta^l_{yw})),
$$
be an estimator of the counterfactual joint distribution in \eqref{eq:counter-cdf} and
$$
\hat F_{Y,W}^{*(j,k,l,m)}(y,w) = \dfrac{1}{n_m} \sum_{i=1}^{n_m}  \omega_{in_m}^m \Phi_2(X_i^{m'}\hat \mu_y^{j*}, X_i^{m'}\hat \nu_w^{k*};g(X_i^{m'}\hat \delta^{l*}_{yw})),
$$
be the corresponding bootstrap draw. 
The following result follows from Theorem \ref{thm:clt-cdf}, Lemma D.1 of \cite{Chernozhukov2013inference}, which establish the Hadamard differentiability of the counterfactual map, and the functional delta method.

\begin{corollary}[FCLT and Bootstrap FCLT for $\hat F_{Y,W}^{(j,k,l,m)}$] Under the assumptions of Theorem \ref{thm:clt-cdf},  jointly for $i,j,k,l \in \{0,1\}$,
$$
\sqrt{n}(\hat F_{Y,W}^{(j,k,l,m)}(y,w) - F_{Y,W}^{(j,k,l,m)}(y,w)) \leadsto Z_{yw}^{(j,k,l,m)} := \int Z_{ywx}^{(jkl)} \dd F^m_{X}(x) +  G^m_X(F^{(jkl)}_{ywx}) \text{ in $\ell^{\infty}(\mY\mW)$},
$$
and
$$
\sqrt{n}(\hat F_{Y,W}^{*(j,k,l,m)}(y,w) - \hat F_{Y,W}^{(j,k,l,m)}(y,w)) \leadsto_{\Pr} Z_{yw}^{(j,k,l,m)}  \text{ in $\ell^{\infty}(\mY\mW)$}.
$$
\end{corollary}

\section{Empirical Illustration}

\label{s:empirical}
We now employ BDR to study intergenerational income mobility. BDR is a particularly useful and powerful tool for this area of research since it can model the joint distribution of incomes for different generations of family members. %This allows one to model the transition probabilities of movements across the income distribution conditional on covariates. By employing BDR estimates we can evaluate if the observed patterns at various points of the distribution are driven by different effects.  This provides a more complete description of mobility although it is more difficult to draw conclusions regarding overall mobility. Accordingly, we also produce and decompose the conditional Kendall's $\tau$ to provide an overall measure of correlation. 
We focus on the joint distribution of a child’s and their father's labor incomes. We begin by documenting the raw data. This provides some insight into the presence of intergenerational mobility in these data. We then employ BDR to estimate the joint conditional distribution of these two outcomes conditional on covariates for both the child and the father.
We examine the role of the estimated local dependence parameter in generating these observed outcomes. Finally, we decompose the observed differences in dependence between fathers/sons and fathers/daughters into composition, sorting and marginals effects.\footnote{We acknowledge the use of the term "sorting" in this potential empirical context is somewhat inappropriate given that one would not interpret the process by which fathers and children are paired is best characterized by a sorting process. However, we retain this description to avoid the introduction of additional terminology.}

\subsection{Data}

\label{s:data}
Our analysis examines data from the Panel Study of Income Dynamics (PSID), a rich longitudinal dataset including family and individual-level income data for individuals in the United States since 1968. The survey was conducted yearly until 1997 and biannually thereafter. The PSID has been employed in previous examiniations of  intergenerational mobility (see, for example, \cite{Callaway2019} or \cite{landerso2017scandinavian}). 

Although the PSID data has 904,796 observations, the number for which we can construct father/children pairs is 6,6767 comprising 3,424 daughters and 3,243 sons.
% frequency of observations is the same for both father-daughter and father-son pairs.
Our empirical analysis is based on this smaller sample. A pair is included if both the father and the child are observed at least once between the ages of 25 and 50. The outcome variables are their respective  average annual labor earnings. This encompasses wages, business income, bonuses, and overtime payments. Labor earnings have been standardized to 1982 dollars. Note that we exclude observations with zero labor income earnings for that year. The average frequency of observations is the same for both father-daughter and father-son pairs. That is, 9 years for children and 15 years for fathers. The equal frequency for males and females is puzzling given the higher labor force participation rates of males. This might reflect the construction of the data set but probably also be due to our relatively weak requirement that the child only needs to be observed with earnings in one year. The higher average annual work hours for sons (2011) than daughters (1540) is consistent with males having a relatively higher level of labor market engagement than females.

The inclusion of only pairs for which both the father and child report labor income may introduce some form of selection bias. Moreover, not accounting for the frequency that the pair is observed may also introduce an additional form of bias. While each of these issues is interesting, we delay their treatment to future work as they are beyond the scope of this paper.

% Although the PSID data covers 904,796 observations, the number of observations for which we can construct father/children pairs is  

%The average frequency of observations is the same for both father-daughter and father-son pairs. That is, 9 years for children and 15 years for fathers.  
%Note that we exclude observations with zero labor income earnings for that year. %\%textcolor{red}{Jonas, can we add something what observations are included and which are excluded? What proportion of females and males are working....That is, do we have more daughters buecause there are more in the the data } \textcolor{orange}{I have no good answer. 
%The average hours worked for sons is higher, 2011 vs 1540.
%Once we condition on the children and fathers being linked, we have more daughters, but that could be just a random artefact of the data.} We do so as it is not clear how to interpret these observations. We also do not account for the frequency at which we observe the individual with positive earnings. We acknowledge that each of these issues may result in some form of selection bias but a treatment of this issue is beyond the scope of this paper.

%esvalues to account for the wide time span that is covered by the data. 
Studies in the literature rely on various income measures to study intergenerational mobility. These include, for example, labor earnings, tax records, or lifetime income imputed over the life cycle.  \cite{Mogstad2021} note that the estimated degree of mobility can vary considerably depending on the measure used. We focus on labor income, noting that intergenerational persistence tends to be stronger for broader income concepts \citep{landerso2017scandinavian}. % ???In this regard, our results can be interpreted as a conservative lower bound on intergenerational mobility. 

%Note that we exclude observations with zero earnings. We do so as it is not clear how to interpret these observations. Nor do we account for the frequency at which we observe the individual with positive earnings. We acknowledge that this may result in some form of selection bias but a treatment of this issue here is beyond this paper.

\begin{table}[h!]
\begin{center}
\begin{threeparttable}[b]
\setlength{\tabcolsep}{0pt}
\caption{Descriptive Statistics PSID Data}   \label{t:desc_s}
\begin{tabular*}{15cm}{ @{\extracolsep{\fill}}lrrrrr} %
\toprule
& & \multicolumn{2}{c}{Education}  & \multicolumn{2}{c}{Race} \\
\cmidrule{3-4}  \cmidrule{5-6} 
Sons & All & No Coll. & College & Other & White \\
\midrule
Labor Income Children (1000 USD) &   17.84 &   16.76 &   22.09 &   13.58 &   19.98 \\ 
Labor Income Father (1000 USD) &   21.32 &   18.84 &   31.07 &   15.11 &   24.43 \\ 
Age Children &   31.71 &   31.93 &   30.86 &   31.51 &   31.81 \\ 
Age Father &   39.41 &   39.75 &   38.07 &   39.33 &   39.45 \\ 
White Children &    0.67 &    0.63 &    0.85 &    0.06 &    0.98 \\ 
White Father &    0.67 &    0.62 &    0.84 &    0.00 &    1.00 \\ 
College Children &    0.30 &    0.22 &    0.61 &    0.17 &    0.36 \\ 
College Father &    0.20 &    0.00 &    1.00 &    0.10 &    0.26 \\ 
HH Size Children &    2.58 &    2.65 &    2.32 &    2.52 &    2.61 \\ 
HH Size Father &    4.70 &    4.86 &    4.08 &    5.33 &    4.39 \\ 
Children Origin Urb. 1 &    0.07 &    0.09 &    0.02 &    0.07 &    0.07 \\ 
Children Origin Urb. 2 &    0.52 &    0.50 &    0.60 &    0.37 &    0.60 \\ 
Children Origin Urb. 3 &    0.37 &    0.38 &    0.34 &    0.53 &    0.30 \\ 
Children Origin Urb. 4 &    0.03 &    0.03 &    0.04 &    0.03 &    0.03 \\ 
Decade Children & 2004.63 & 2003.65 & 2008.51 & 2005.66 & 2004.12 \\ 
Decade Parents & 1983.55 & 1983.06 & 1985.48 & 1984.94 & 1982.86 \\ 

\vspace{.1cm} \\
Daughters \\
\midrule
Labor Income Children (1000 USD) &   12.18 &   11.04 &   16.95 &   10.72 &   13.10 \\ 
Labor Income Father (1000 USD) &   21.14 &   18.78 &   30.99 &   14.99 &   24.98 \\ 
Age Children &   32.00 &   32.20 &   31.17 &   32.16 &   31.90 \\ 
Age Father &   39.37 &   39.69 &   38.02 &   39.75 &   39.13 \\ 
White Children &    0.63 &    0.57 &    0.85 &    0.06 &    0.98 \\ 
White Father &    0.62 &    0.56 &    0.84 &    0.00 &    1.00 \\ 
College Children &    0.37 &    0.29 &    0.68 &    0.24 &    0.45 \\ 
College Father &    0.19 &    0.00 &    1.00 &    0.08 &    0.27 \\ 
HH Size Children &    2.93 &    3.02 &    2.53 &    2.98 &    2.89 \\ 
HH Size Father &    4.79 &    4.95 &    4.14 &    5.52 &    4.34 \\ 
Children Origin Urb. 1 &    0.06 &    0.06 &    0.03 &    0.07 &    0.05 \\ 
Children Origin Urb. 2 &    0.55 &    0.54 &    0.57 &    0.39 &    0.64 \\ 
Children Origin Urb. 3 &    0.37 &    0.38 &    0.35 &    0.52 &    0.28 \\ 
Children Origin Urb. 4 &    0.03 &    0.02 &    0.05 &    0.02 &    0.03 \\ 
Decade Children & 2005.60 & 2004.60 & 2009.76 & 2005.57 & 2005.61 \\ 
Decade Parents & 1984.04 & 1983.51 & 1986.27 & 1984.31 & 1983.88 \\ 

\bottomrule

\end{tabular*}
\begin{tablenotes}[flushleft]
\small 
\item \textit{ \tiny Notes: Number of observations 3426 for daughters and 3255 for sons. All statistics are means. Child origin refers to a categorical variable with 4 distinct values measuring the degree of urbanization where the child grew up.}  
\end{tablenotes}
\end{threeparttable}
\end{center}
\end{table}

The variables displayed in Table \ref{t:desc_s} are all available since 1968 with the exception of ``college'', which indicates that the individual has a college degree,  which is available from 1975. Observations with missing entries for any of the variables are excluded. Table \ref{t:desc_s} indicates that the father's information is generally collected during the 1980s, while the children's information is typically gathered in the early 2000s. Fathers are observed later in life, mostly between the ages of 34 and 47, while children are observed earlier between the ages of 27 and 36. This also explains the large difference in labor income in favor of fathers. Unsurprisingly, we find higher levels of labor income for fathers with a college degree. The same is true for their children. This can be partially explained by the high level of persistence in college education. For example, 30 percent of sons have a college degree, but that number rises to 61 percent among sons whose fathers have a college degree. Similar figures are found for daughters. We also find that white fathers, and their children, generally have higher labor income. Household size is smaller for children, which can be attributed to their younger age, although it may also capture broader demographic changes and trends. 
%Given the strong relationship between age and labor income, controlling for these differences is crucial for the analysis. 

%\textcolor{red}{Jonas, can we start with a figure like figure 1 which shows the raw data so we can get sense  of level of dependence in raw data?}
%\textcolor{orange}{I have added two graphs at the end of the document: one like Figure 1 with the estimated $\rho$ for the raw data and one for the transition matrices where once we set $\rho=0$ to see how large the effect is. For the $\rho$ figure: we see that the values tend to be closer to zero ones we include covariates for the marginals, but not always. I am not sure its helpful, what do you think? For the transition matrices: We may want to use this, what do you think Ivan?}

\subsection{Unconditional Mobility} 
\begin{figure}[h]
\caption{Transition matrices, raw data (bootstrapped standard errors are shown in parentheses.)} \label{f:tm_raw}
	\begin{subfigure}{.45\textwidth}
		\centering
		\includegraphics[scale=0.36]{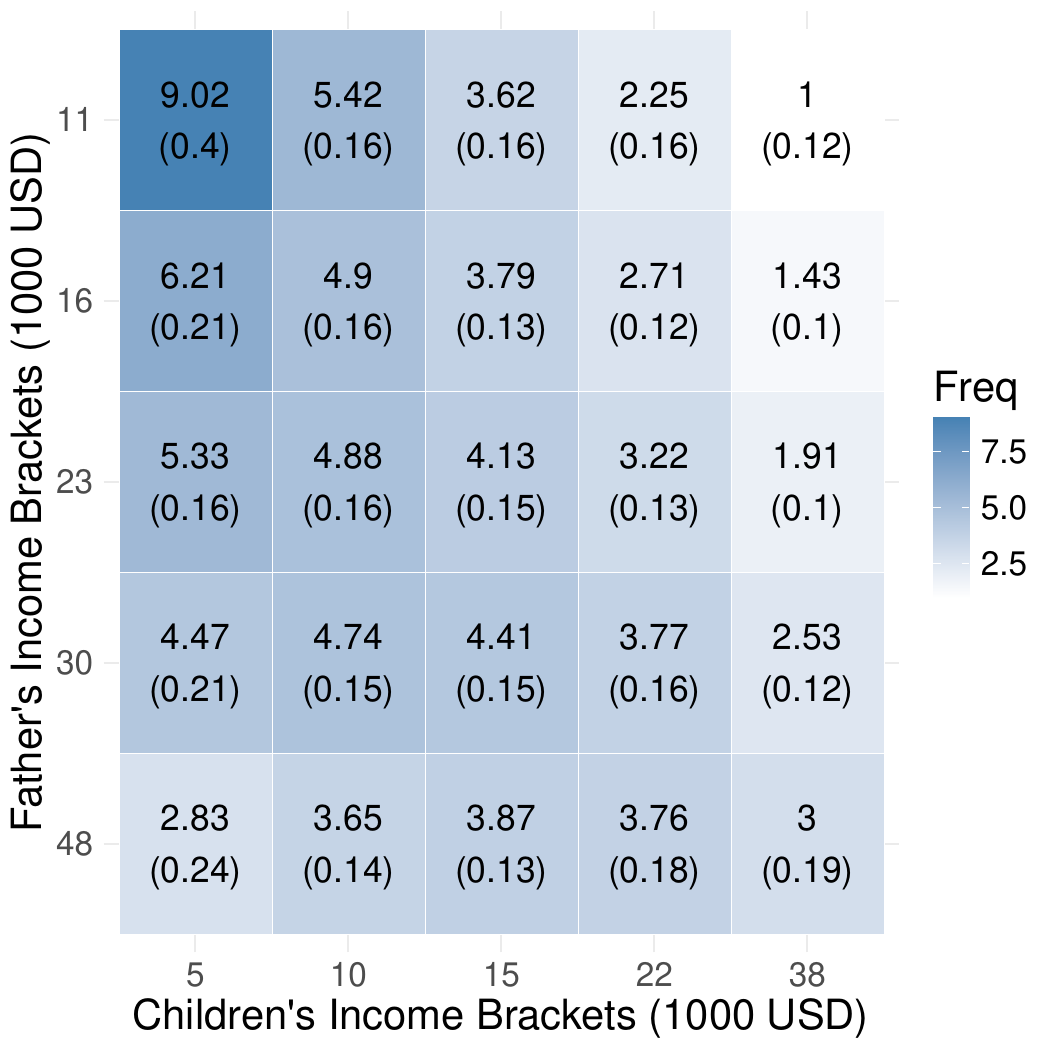}
		\caption{Daughters}
	\end{subfigure}%
	\begin{subfigure}{.45\textwidth}
		\centering
		\includegraphics[scale=0.36]{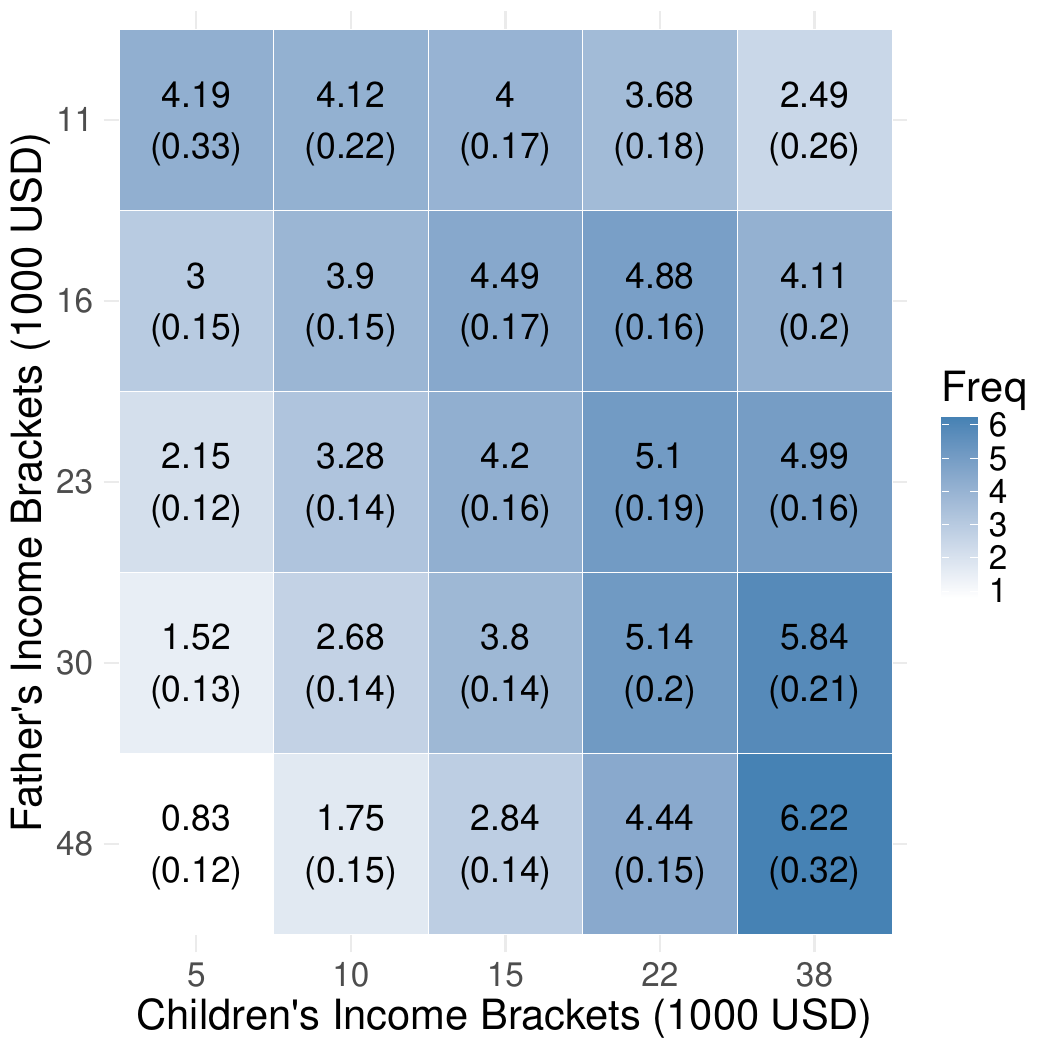}
		\caption{Sons}
	\end{subfigure} 
	\caption*{\footnotesize \textit{Notes: The results in Figure \ref{f:tm_raw} show the probability mass function for the raw data separately for sons and daughters. Reported standard errors equal the interquartile range of the bootstrap estimates divided by the interquartile range of the standard normal distribution.  } }
\end{figure}

Figure \ref{f:tm_raw} plots the transition matrices for the outcome variables. 
The cells of the transition matrices are based on the quintiles of the fathers in USD on the y-axis and the quintiles of the children in USD on the x-axis. The values within these cells represent the percentage of observations observed with a father-daughter or father-son combination within these cells. Due to the overall wage gap between women and men, sons earn more than daughters and the percentages of sons are relatively higher on the right side of the figure, while they are relatively higher on the left side of the figure for daughters. As relative darkness of the squares capture larger probabilities, it is immediately apparent that father-daughter pairs are relatively more frequently located in the upper left corner and the father-son pairs are more relatively frequent in the lower right corner. A high degree of mobility would imply equal probabilities across all cells within each row. Therefore,  the observed deviation from this pattern illustrates the limited mobility in the United States.  

\vspace{1cm}

\subsection{BDR Estimation} \label{ss:estimation} We now focus on the estimates of the BDR model. We first estimate the model using the covariates described below for each of the marginals while also allowing $\rho$ to be a function of the covariates. We also estimate the model in which $\rho$ is not a function of the covariates although these latter results are relegated to the appendix. 

The empirical results reported in the main text uses the following variables. For the marginal distribution of father's labor income we employ  age, age$^2$ and decade fixed effects for both fathers and children, father's and child's race, father's and child's household size and indicators for a college degree of the father and the child's origin (measuring the degree of urbanization where the child was raised). We use the same set of variables for the marginal distribution of the children. When $\rho$ is a function of the covariates we employ all of the covariates listed above but exclude the decade fixed effects. We acknowledge that some covariates are potentially endogenous. Accounting for this endogeneity is beyond the scope of this paper and we defer this important issue to further work. However, in the appendix we provide the corresponding results when we exclude some variables from the conditioning set. We note that their exclusion does not have substantial implication for the results reported.

\subsection{Model predictions and Counter Factual}

\begin{figure}[!h]
\vspace{-.5cm}
\caption{Counterfactual distributions (bootstrapped standard errors are shown in parentheses.)} \label{f:tm_count}
\vspace{.2cm}
	\begin{subfigure}{.45\textwidth}
		\centering
		\includegraphics[scale=0.29]{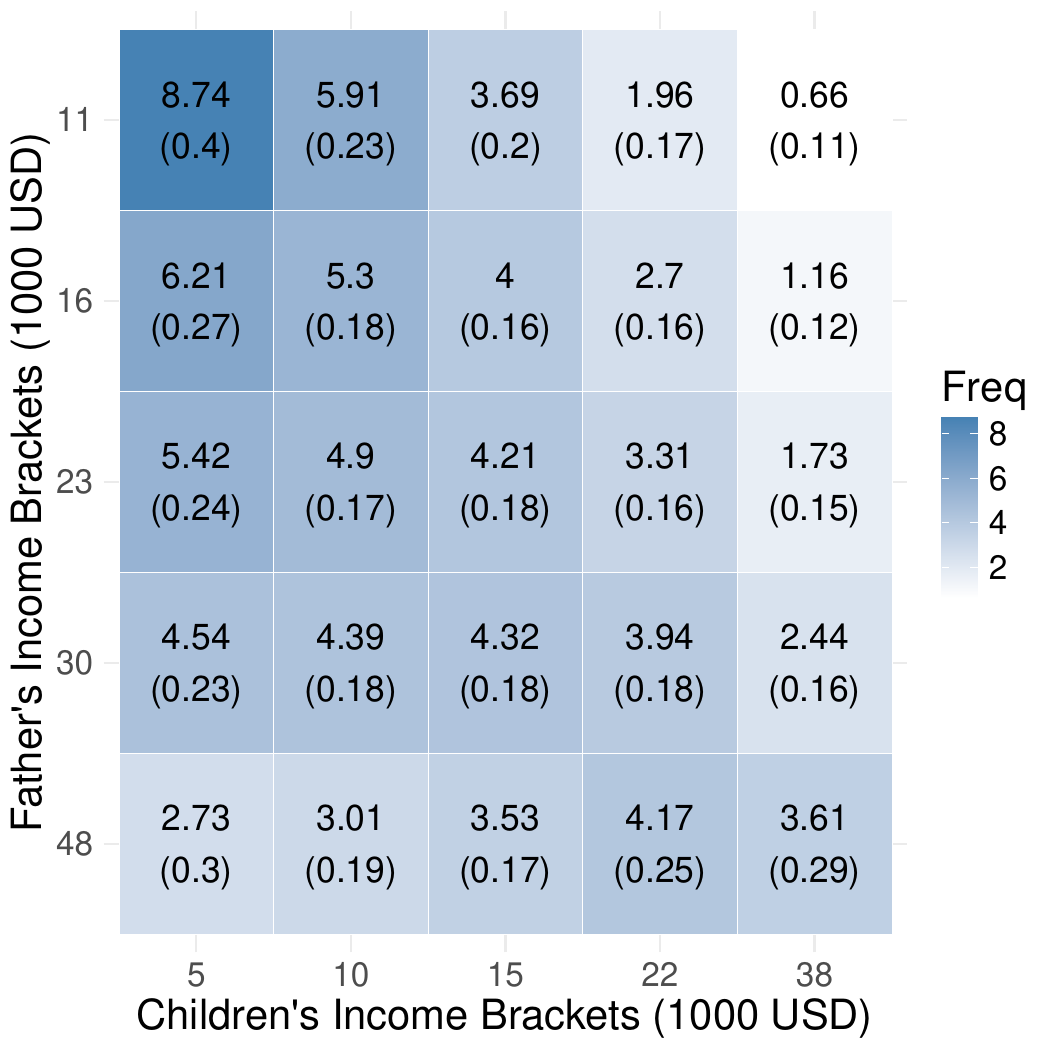}
		\caption{Daughters}
	\end{subfigure}%
	\begin{subfigure}{.45\textwidth}
		\centering
		\includegraphics[scale=0.29]{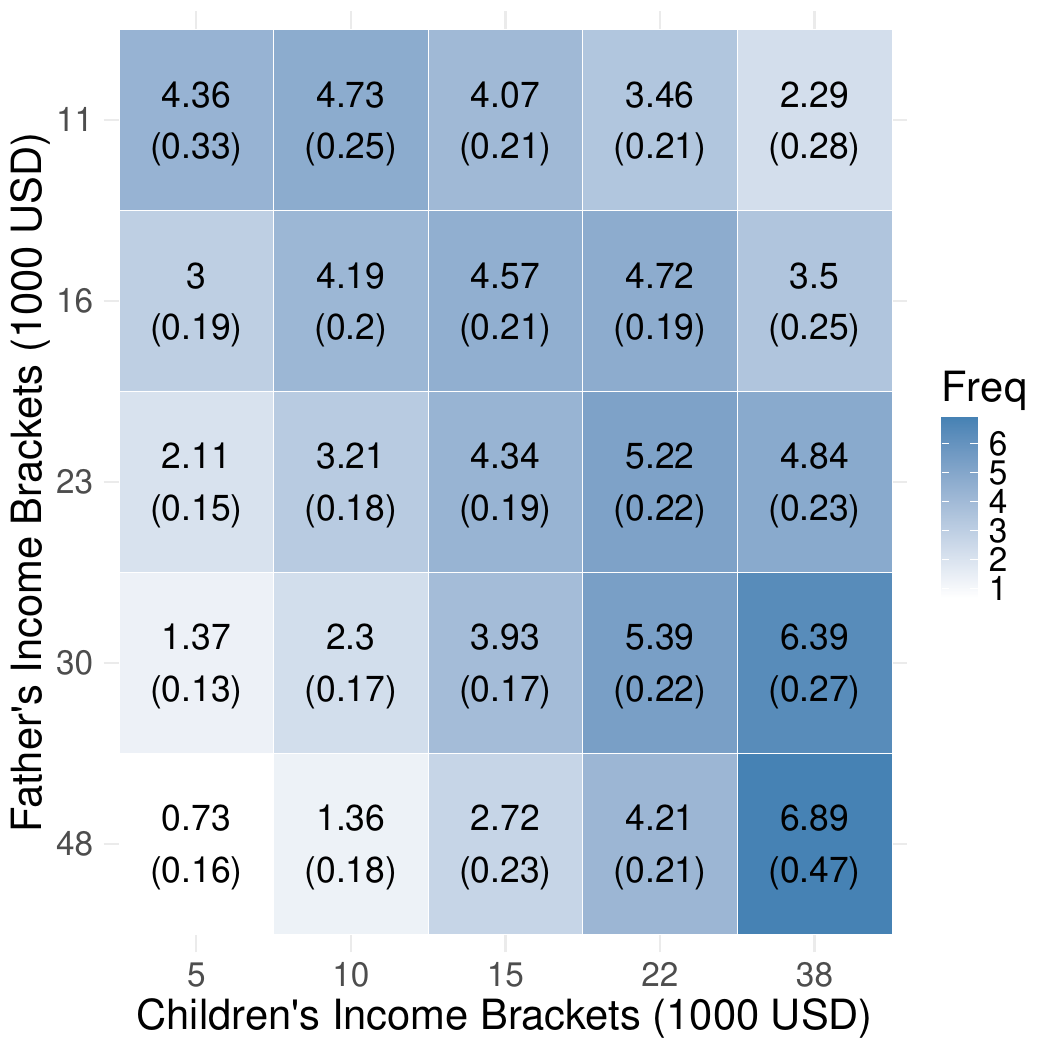}
		\caption{Sons}
	\end{subfigure} \\ 
        \vspace{.5cm}
	\begin{subfigure}{.45\textwidth}
		\centering
		\includegraphics[scale=0.29]{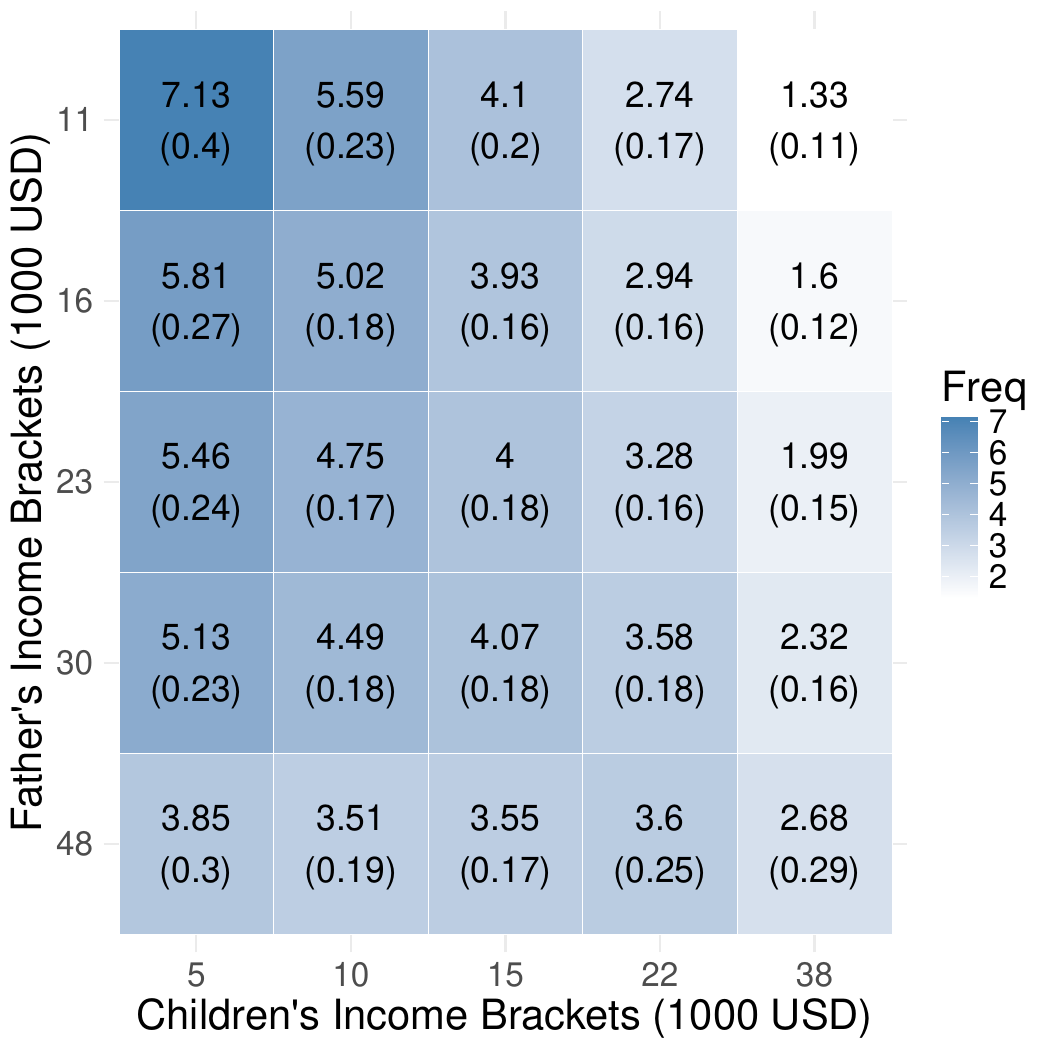}
		\caption{Daughters,  $\rho = 0$}
	\end{subfigure}%
	\begin{subfigure}{.45\textwidth}
		\centering
		\includegraphics[scale=0.29]{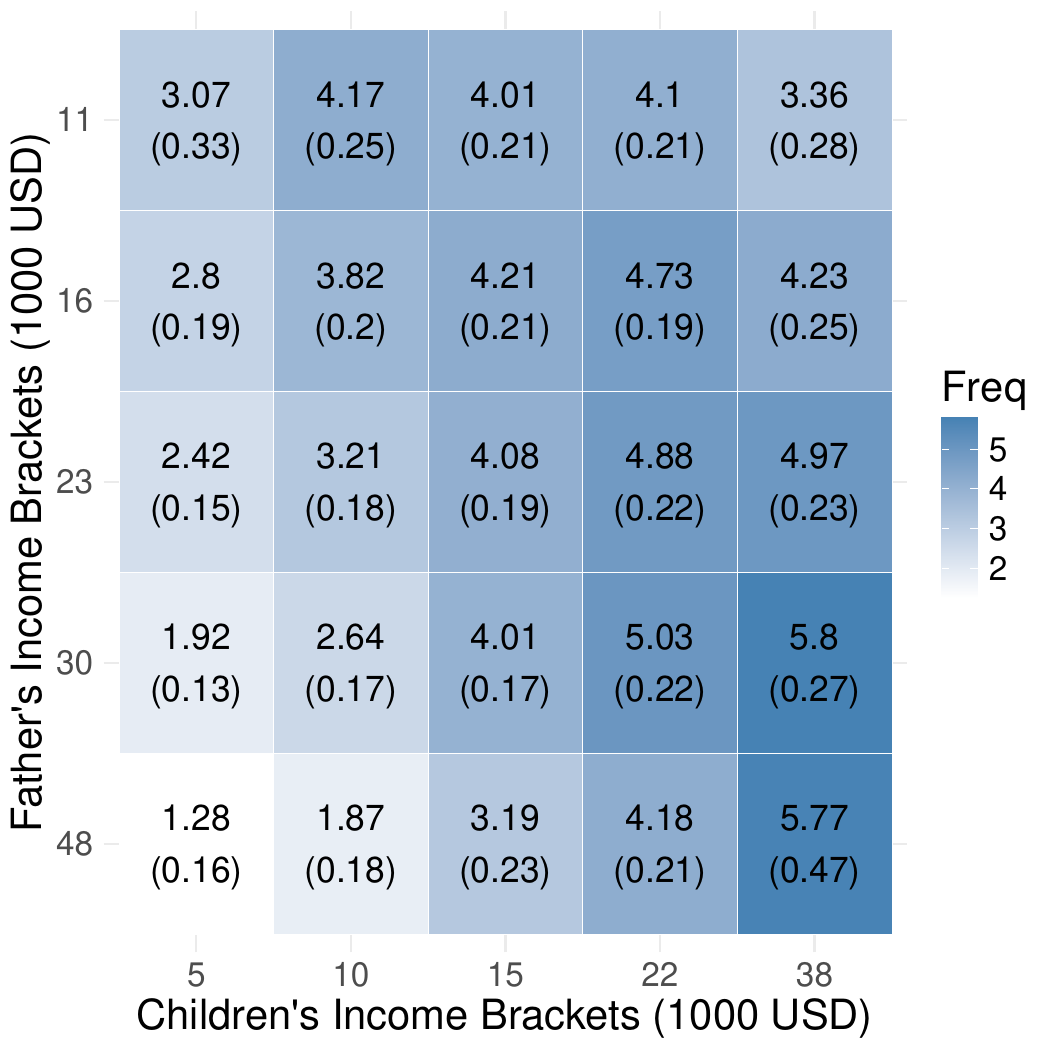}
		\caption{Sons,  $\rho = 0$}
	\end{subfigure} \\ 
        \vspace{.5cm}
	\begin{subfigure}{.45\textwidth}
		\centering
		\includegraphics[scale=0.29]{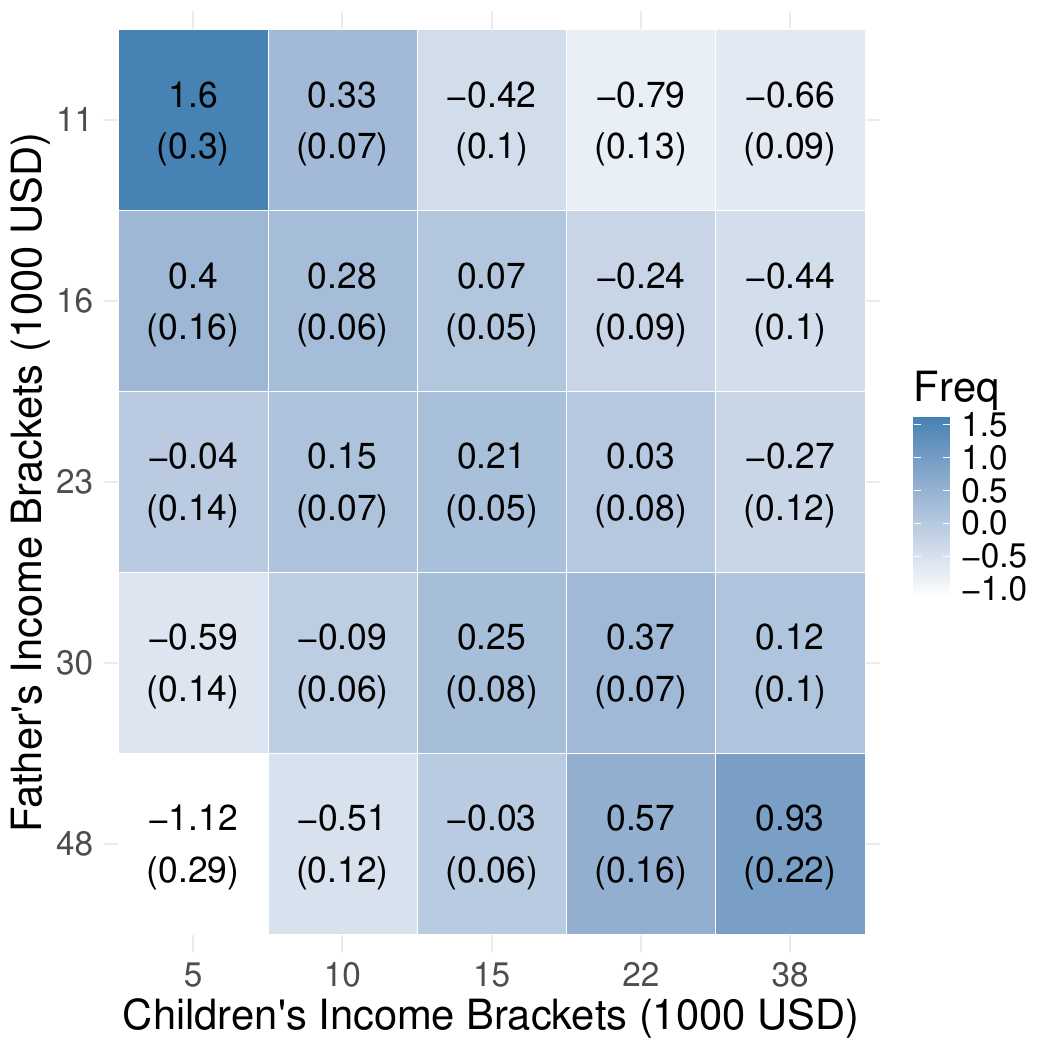}
		\caption{Daughters, difference due to $\rho$}
	\end{subfigure}
	\begin{subfigure}{.45\textwidth}
		\centering
		\includegraphics[scale=0.29]{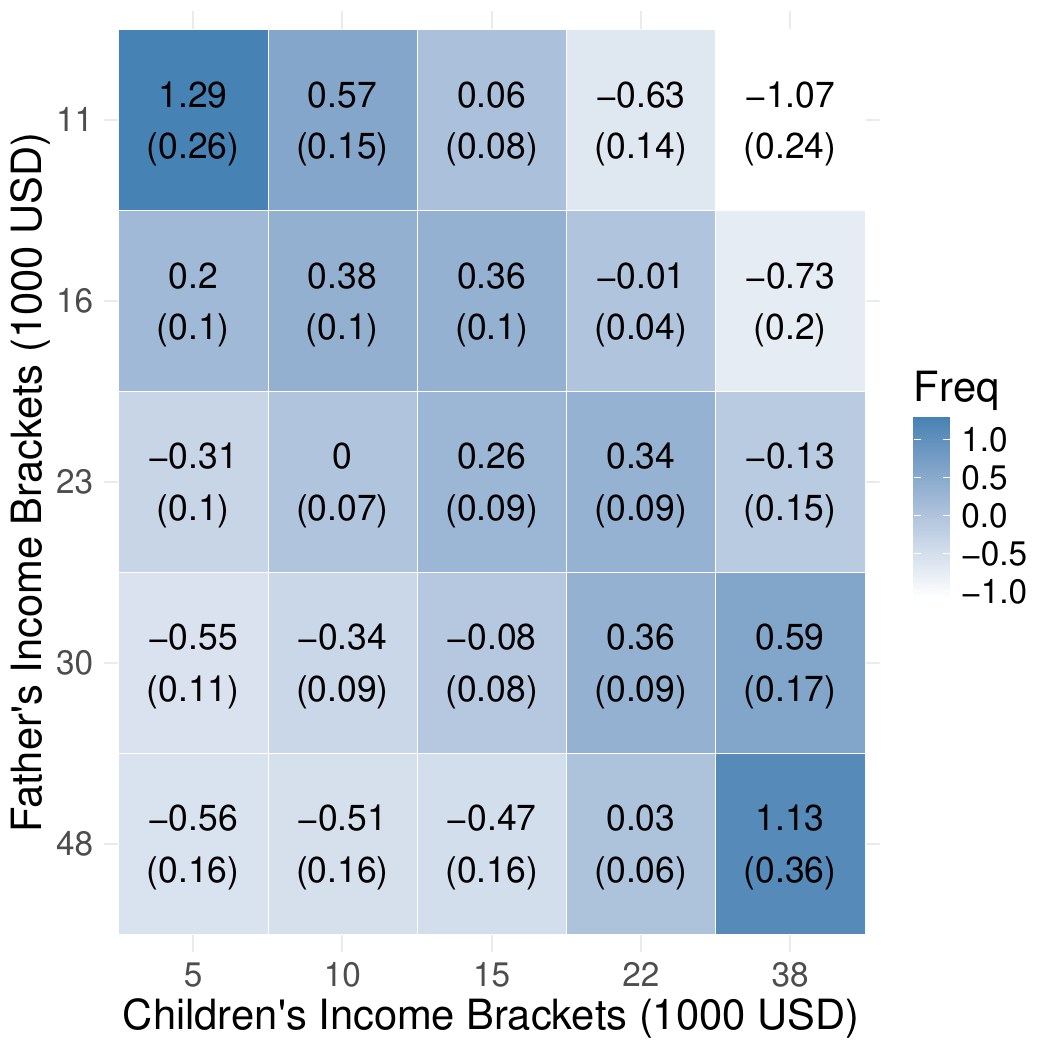}
		\caption{Sons, difference due to $\rho$}
	\end{subfigure}
	\caption*{\tiny \textit{Notes: Panel (A) and (B) show the predicted probability mass functions from the full model. For panels (C) and (D), $\hat{\rho}$ is artificially set to zero. The probability mass functions capture the dependence due to the marginals. Panel (E) and (F) show the differences between (A) and (C) and (B) and (D). Standard errors calculated as for Figure \ref{f:tm_raw}.}}
\end{figure}

We first report the capacity of the BDR estimates to reproduce the observed joint distribution of fathers' and children's earnings and examine the impact on this predicted joint distribution when we eliminate the dependence operating through $\rho$. Figure \ref{f:tm_count} presents the estimated transition matrices based on the BDR model. To construct these matrices, we estimate the joint distribution of father-child earnings using the set of covariates listed above. We then integrate over the empirical distribution of the observed covariates to obtain the average joint distribution.%, which is subsequently interpolated to evaluate the distribution at the specified quintile value.

Panels (A) and (B) display the resulting transition matrices for daughters and sons, respectively. These matrices correspond to the estimated joint distribution. The values of these matrices are calculated based on equation \eqref{eq:tm}. Panels (C) and (D) show the same average distribution, but under the hypothetical assumption that the local correlation is set to zero. This counterfactual allows us to isolate the contribution of local dependence to intergenerational mobility. Panels (E) and (F) present the difference between the two distributions and quantifies the effect of local dependence.

 It is useful to consider two issues before focusing on the results from this exercise. The first is the interpretation of $\rho$ in this context and the second is the resulting estimates of $\rho$. Given the nature of the outcomes, the parameter $\rho$ captures how the outcomes move together after conditioning out the impact of the covariates. If the estimated value of $\rho$ was zero, this implies no dependence between the outcomes after conditioning out the covariates. If the estimated $\rho$ is positive (negative) this implies that,  conditional on the covariates, a higher value for one outcome is associated with a higher (lower) outcome for the other. In this particular context this dependence  may capture some local correlation between unobservables affecting the respective outcomes. 

As noted above an attractive feature of BDR is its capacity to flexibly model the local dependence. Although we do not report our estimates of $\rho$,  a notable feature of the results is the substantial heterogeneity in these estimates across different points on the grid. This reflects that the level of dependence varies greatly at different points of the joint distribution. This highlights the importance of adopting a flexible approach to its estimation. 

Several insights emerge from these results. First, the estimated transition matrices in Panels (A) and (B) closely resemble those obtained from the raw data (Figure \ref{f:tm_raw}), indicating that the model successfully captures the observed patterns of mobility. While this is reassuring it is not surprising given the nature of the estimator and the large number of parameters estimated. %\textcolor{red}{are the right shares imposed in estimation?}\textcolor{orange}{I am not sure i understand this question. All models are estimated separately for sons and daughters. }
Second, setting the local correlation parameter ($\rho$) to zero leads to a considerable flattening of the dependence structure. This effect is evident for both sons and daughters and is particularly pronounced in the upper tail of the income distribution, where the level of  dependence is relatively strong in the raw data. The influence of local dependence is quantitatively meaningful. For example, the probability that a son reaches the top decile is 6.9\% under local dependence, compared to 5.8\% under the counterfactual with $\rho = 0$. This difference is economically significant and explains a large fraction of the departure from the benchmark probability of 4\%, which would be expected under full independence and in the absence of gender wage gaps and compositional effects. This highlights the role of local correlation in capturing some aspect of the observed persistent intergenerational advantages. Comparable patterns are observed for daughters, for whom the impact of $\rho$ is most pronounced in the upper-left region of the figure.

Note that we also estimated the model predictions and the counterfactuals with a specification that does not take account of the covariates in $\rho$. These results are presented in Figure \ref{f:tm_count_1}. The results are very similar to the results presented in Figure \ref{f:tm_count}. This suggests that the results in this particulalr context are not sensitive to the treatment of the local dependence.

\subsection{Decomposition of Joint Conditional CDF}

\begin{figure}[!h]
\vspace{-.5cm}
\caption{Decomposition of difference in transition matrices between sons and daughters (bootstrapped standard errors in parentheses).} \label{f:tm_decom}
\vspace{.2cm}
	\begin{subfigure}{.45\textwidth}
		\centering
		\includegraphics[scale=0.29]{figures_new/pmf_d_bdr.pdf}
		\caption{Daughters}
	\end{subfigure} 
    \begin{subfigure}{.45\textwidth}
		\centering
		\includegraphics[scale=0.29]{figures_new/pmf_s_bdr.pdf}
		\caption{Sons}
	\end{subfigure} \\
        \begin{subfigure}{.45\textwidth}
		\centering
		\includegraphics[scale=0.29]{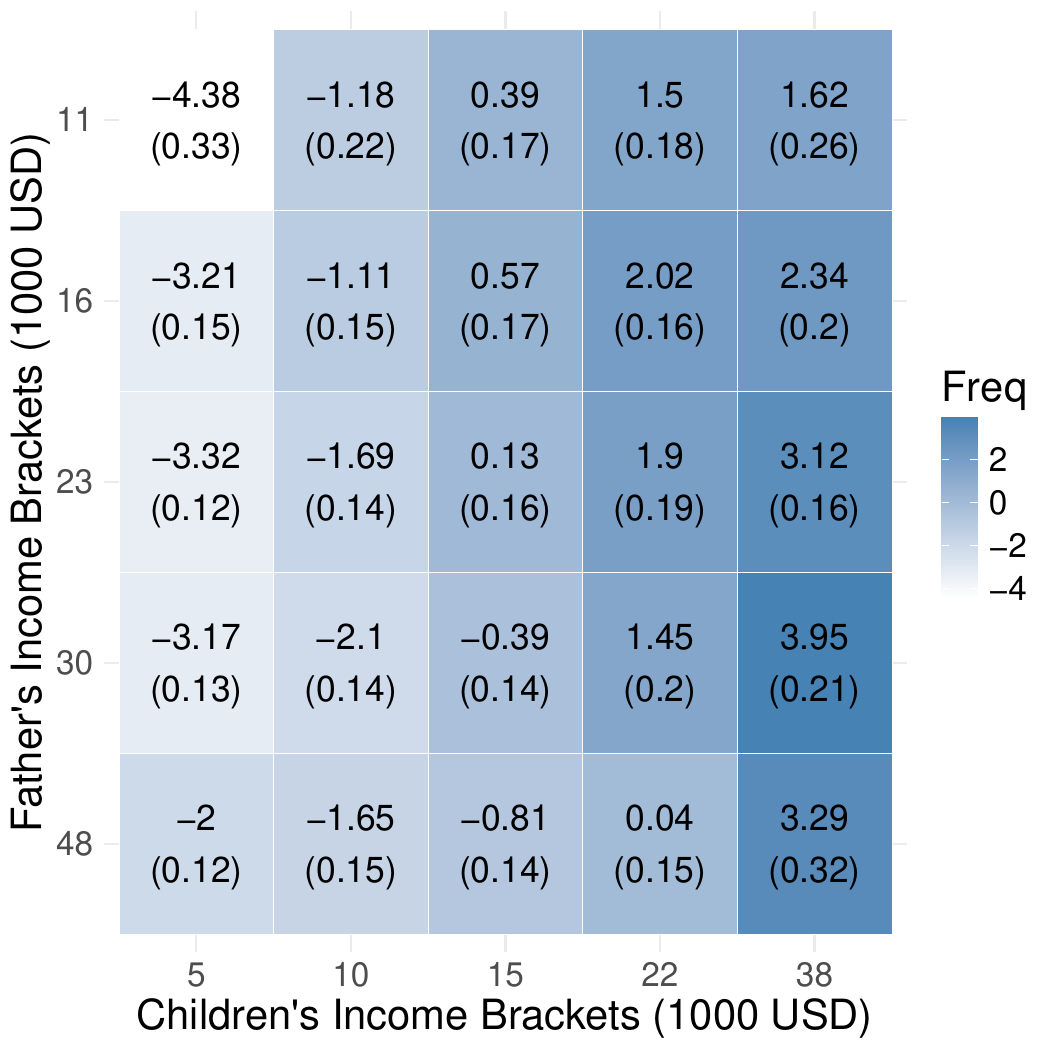}
		\caption{Total Difference (B-A)}
	\end{subfigure} 
	\begin{subfigure}{.45\textwidth}
		\centering
		\includegraphics[scale=0.29]{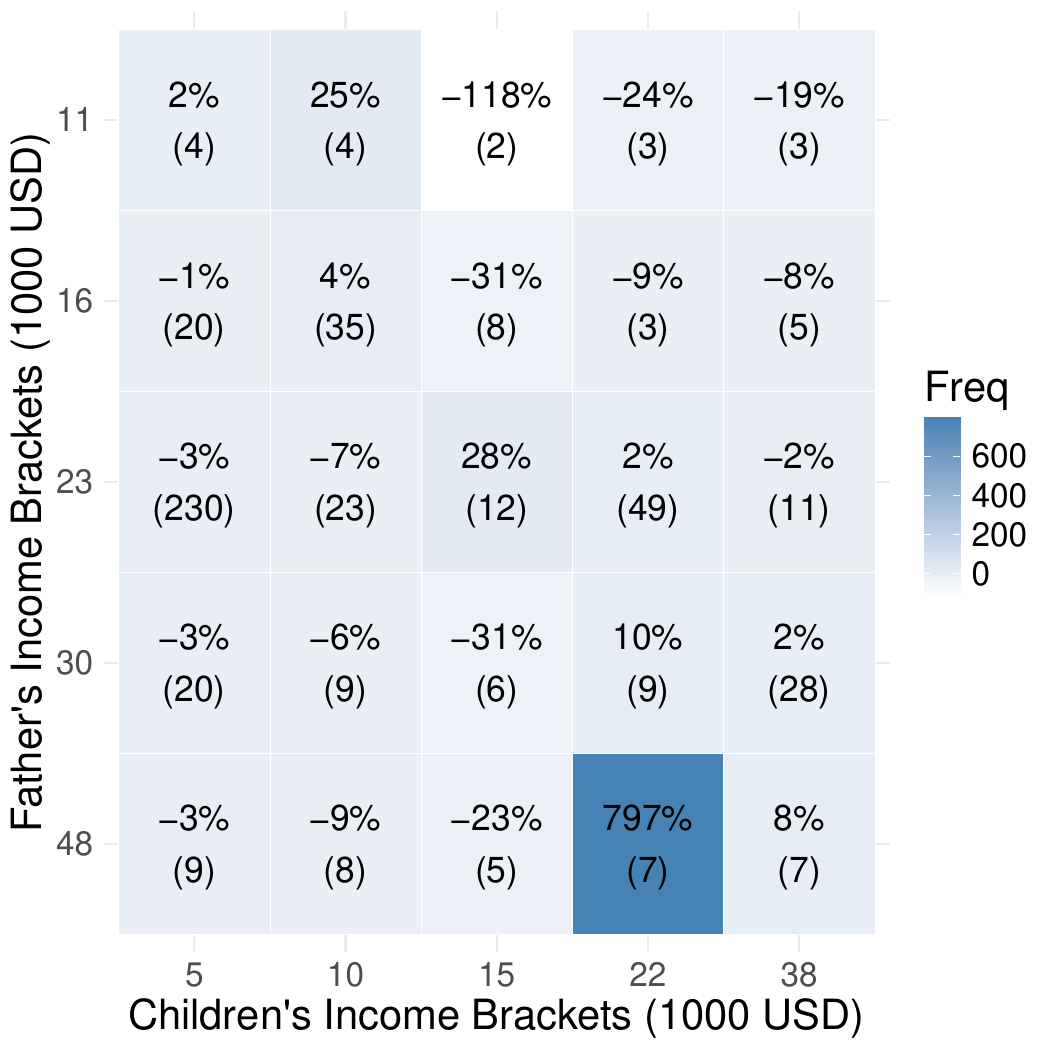}
		\caption{Composition}
	\end{subfigure} \\ 
        \vspace{.5cm}
	\begin{subfigure}{.45\textwidth}
		\centering
		\includegraphics[scale=0.29]{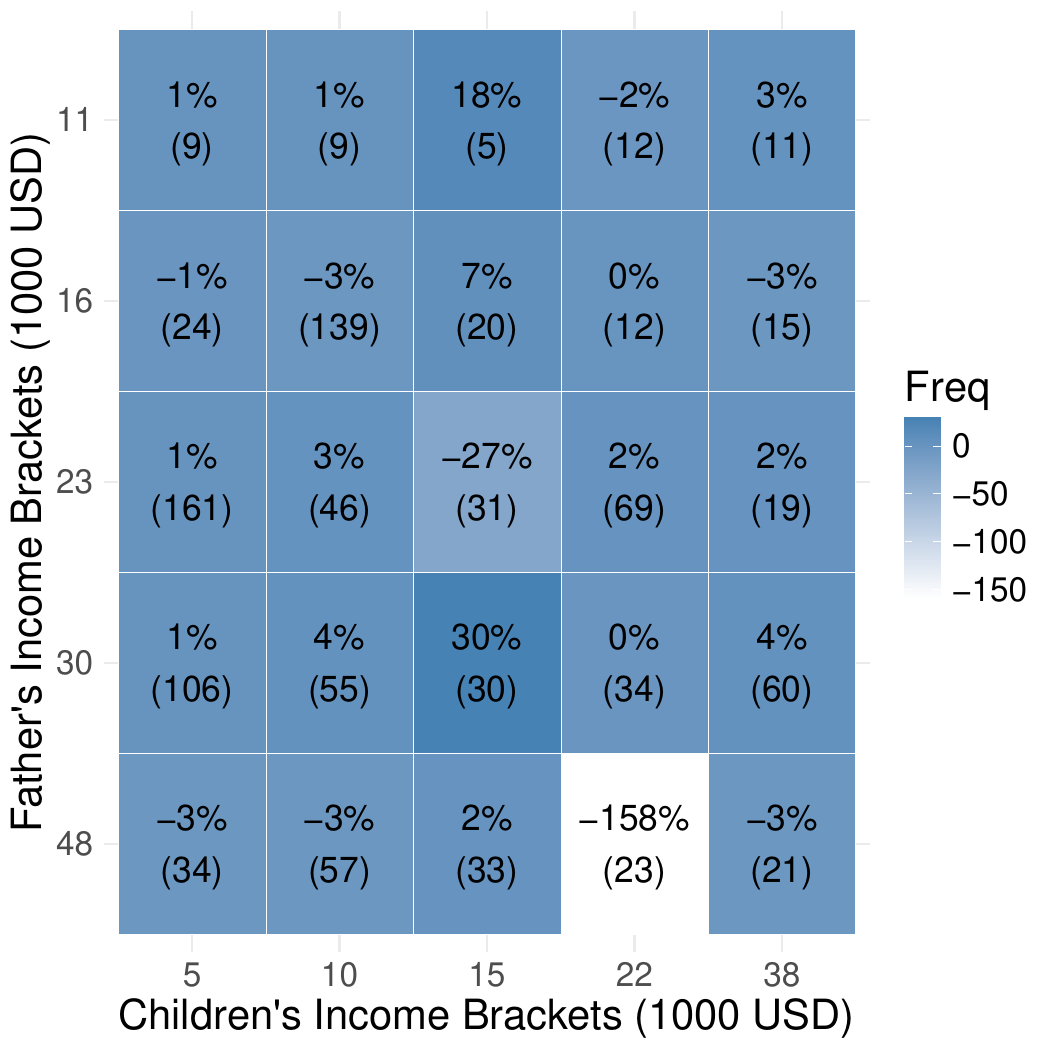}
		\caption{Sorting}
	\end{subfigure}%
        \begin{subfigure}{.45\textwidth}
		\centering
		\includegraphics[scale=0.29]{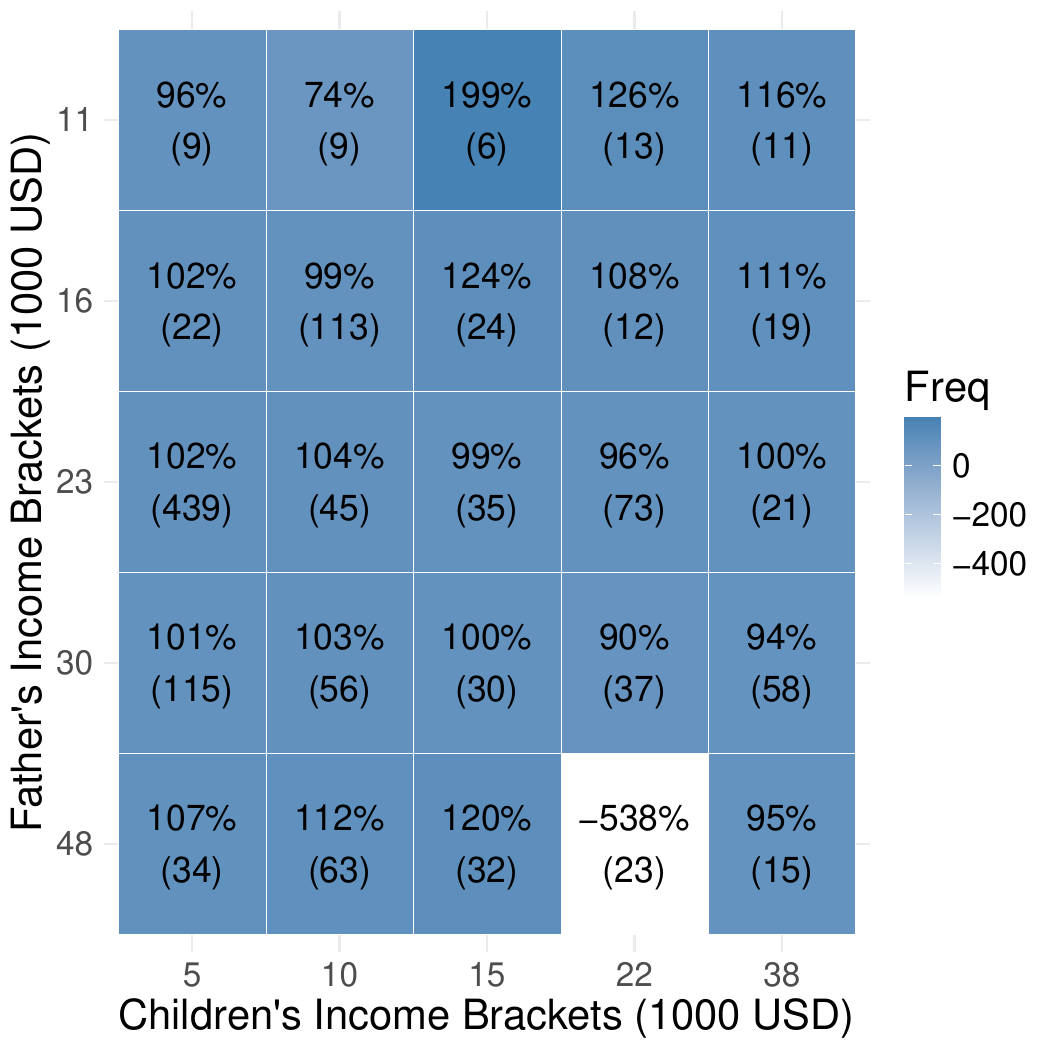}
		\caption{Marginals}
	\end{subfigure} \\ 
	\caption*{\tiny \textit{Notes: Figure \ref{f:tm_decom} shows the decomposition of the joint distribution between sons and daughters into three terms. Values in Panels (D), (E), and (F) are in percentage differences of the total difference. Note that there is not an equal number of observations in all rows as sons and daughters have different marginal distributions. For this figure, we evaluated the joint at the common labor income levels. Standard errors calculated as in Figure \ref{f:tm_raw}.}}
\end{figure}

Figure \ref{f:tm_decom} presents a decomposition of the differences in transition matrices between sons and daughters. The decomposition follows equation \eqref{eq:cdf-decomp} and separates the overall difference in the transition matrices into three distinct components: (i) Composition effects, which capture differences in observable characteristics (e.g. education, region, cohort); (ii) sorting effects, reflecting differences in the estimated local dependence parameter $\rho$; and (iii) marginal effects, which account for gender-specific differences in how covariates relate to the marginal earnings distributions.

Panels (A) and (B) display the estimated transition matrices for sons and daughters, respectively, using the BDR model with the set of covariates as described in Section \ref{ss:estimation}. Panel (C) shows the element-wise difference between the two matrices. Panels (D), (E), and (F) then report how much of this difference can be explained by composition, sorting, and marginal effects, respectively, expressed as a percentage of the absolute difference in each cell.\footnote{Note that the difference in Panel (C) is, in some instances, very close to zero, which means that the composition, sorting, and marginal effects are divided by a number near zero. This partially explains the occasional occurrence of very large values in Panels (D)–(F).}

Several key insights emerge from this analysis. Marginal effects are by far the most important driver of the observed gender differences in intergenerational earnings mobility. In many cells, they account for the majority of the variation across gender, highlighting the importance of differential returns to characteristics in shaping the marginal earnings distributions of sons and daughters. These marginal effects capture the differences in the labor market experiences, reflecting both supply and demand factors, for females and males. Sorting effects, driven by differences in the estimated local correlation, explain roughly 10\% of the overall difference. This suggests that sons and daughters differ not only in their marginal earnings prospects, but also in how tightly their outcomes are linked to their fathers’ incomes, even after controlling for observed characteristics. Composition effects contribute very little to the observed differences. This indicates that the sons and daughters in our sample are, on average, very similar in terms of the included covariates. This suggests that the large difference in intergenerational earnings mobility for daughters and sons cannot be explained by differences in observable characteristics. These results emphasize that gender differences in intergenerational earnings mobility arise primarily from structural differences in earnings determination. These arise through the impact of observables on earnings and the nature of dependence between father's and child's outcomes.

We also provide a robustness check that excludes the children's race and household size. The results are presented in Table \ref{f:tm_decom_1}. We find no substantial deviations from the results shown in Figure \ref{f:tm_decom}. While we do not consider this an exhaustive examination of the sensitivity of our results, it does not provide some indication of their sensitivity to the exclusion of variables which are potentially important.

\subsection{Discussion}
Our empirical findings highlight the importance of modeling both the marginal distributions and the local dependence structure when analyzing intergenerational mobility. The BDR framework enables a flexible and interpretable decomposition of mobility, and highlights that both observable characteristics and residual local association patterns contribute to the overall structure. Importantly, one can isolate the channels through which gender differences in income dependence emerge. The stronger association observed in father-son pairs is largely due to the differences in the marginal distributions. However, accounting for differences in sorting, capturing local dependence, is also important. Overall, the evidence suggests that policies aiming to promote intergenerational equity should consider not just differences in endowments, but also structural differences in how income is transmitted across generations.

\section{Conclusion} \label{s:conclusion}
We employ bivariate distribution regression to estimate the joint distribution of two outcome variables conditional on set(s) of covariates. Our approach 
%to modeling conditional bivariate distributions 
is particularly valuable when the dependence between the outcomes persists even after controlling for observed covariates.  
%Our analysis relies on a result from \cite{Chernozhukov2018distribution} which shows that any conditional joint distribution has a local Gaussian representation. 
We describe how BDR can be implemented and present some associated functionals of interest. The functionals of primary interest in this setting are those involving  the unexplained dependence. We provide a decomposition of the joint distributions for different groups into composition, marginal and sorting effects. We provide algorithms for estimation and inference, along with the associated theory. 
We also provide a similar decomposition for the associated transition matrices
%We also showed bootstrap validity.
Using data from the Panel Survey of Income Dynamics, we model the joint distribution of parents' and children's earnings and find that the local dependence structure is an important contributor in explaining intergenerational mobility.

%\section{Bivariate Distribution Regression}

\newpage
\appendix

\section{Proofs}\label{app:proof}

\subsection{Notation and Auxiliary Result for Lemma \ref{lemma:fclt}}\label{app:notation}
Let 
$$\mathbb{G}_{n}[f]:=\mathbb{G}_{n}[f(W)]:=\dfrac{1}{\sqrt{n}} \sum_{i=1}^{n}(f(W)-\Ep[f(W)]).$$
When the function $\hat{f}$ is estimated, the notation should be interpreted as
$$\mathbb{G}_{n}[\hat{f}]=\mathbb{G}_{n}[f]|_{f=\hat{f}}.$$

% We also follow the notation and definitions in \cite{van1996weak}
% of bootstrap consistency. Let $D_{n}$ denote the data vector and $E_{n}$ be
% the vector of bootstrap weights. Consider the random element $%
% Z_{n}^{b}=Z_{n}(D_{n},E_{n})$ in a normed space $\mathbb{Z}$. We say that
% the bootstrap law of $Z_{n}^{b}$ consistently estimates the law of some
% tight random element $Z$ and write $Z_{n}^{b}\leadsto_{\Pr}Z$ in $%
% \mathbb{Z}$ if 
% \begin{equation}
% \begin{array}{r}
% \sup_{h\in\text{BL}_{1}(\mathbb{Z})}\left|\Ep^{b}h%
% \left(Z_{n}^{b}\right)- \Ep h(Z)\right|\rightarrow_{\Pr^{*}}0,%
% \end{array}
% \label{boot1}
% \end{equation}
% where $\text{BL}_{1}(\mathbb{Z})$ denotes the space of functions with
% Lipschitz norm at most 1, ${\Ep}^{b}$ denotes the conditional
% expectation with respect to $E_{n}$ given the data $D_{n}$, and $%
% \rightarrow_{\Pr^{*}}$ denotes convergence in (outer) probability.

We use the $Z$-process framework from Appendix E.1 of \citet{Chernozhukov2013inference}.  The score vector in \eqref{eq:score} has the analytical form:
\begin{footnotesize}{\begin{equation}
\begin{split}\label{eq:score2}
        & S^{yw}_i(\theta^d_{yw}) =  \\ & \left[\begin{array}{c}
     \dfrac{\phi(X_i^{d'}\mu_y^d)}{\Phi(X_i^{d'}\mu_y^d)\Phi(-X_i^{d'}\mu_y^d)} \left[I_i^{d y} - \Phi(X_i^{d'}\mu_y^d)\right] \\
     \dfrac{\phi(X_i^{d'}\nu_w^d)}{\Phi(X_i^{d'}\nu_w^d)\Phi(-X_i^{d'}\nu_w^d)} \left[J_i^{dw} - \Phi(X_i^{d'}\nu_w^d)\right] \\
     \left( \dfrac{I_i^{dy} J_i^{dw}}{\Phi^d_{2i}(y,w)} - \dfrac{I_i^{dy} \bar J_i^{dw}}{\Phi^d_{2i}(y,-  w)} - \dfrac{\bar I_i^{dy} J_i^w}{\Phi^d_{2i}(-  y,w)} + \dfrac{\bar I_i^{dy} \bar J_i^{dw}}{\Phi^d_{2i}(-  y,-  w)}  \right) \phi_2(X_i^{d'}\mu_y,X_i^{d'}\nu_w;g(X_i^{d'}\delta_{yw}^d)) \dot g(X_i^{d'}\delta_{yw}^d)
     \end{array}\right] \otimes X_i^d,
\end{split}
\end{equation}}\end{footnotesize}where
$\Phi^d_{2i}(y,w) := \Phi_2(X_i^{d'}\mu_y^d,X_i^{d'}\nu_w^d;g(X_i^{d'}\delta_{yw}^d))$, $\Phi^d_{2i}(y,-  w) := \Phi_2(X_i^{d'}\mu_y^d,- X_i^{d'}\nu_w^d; - g(X_i^{d'}\delta_{yw}^d))$, $\Phi^d_{2i}(-  y,w) := \Phi_2(- X_i^{d'}\mu_y^d,X_i^{d'}\nu_w^d; - g(X_i^{d'}\delta_{yw}^d))$, $\Phi^d_{2i}(-  y, -  w) := \Phi_2(- X_i^{d'}\mu_y^d, - X_i^{d'}\nu_w^d;g(X_i^{d'}\delta_{yw}^d))$,  and $\dot g(u) := \dd g(u)/\dd u$.
The non-zero components of the expected Hessian matrix in \eqref{eq:hessian} have the analytical form
\begin{equation*}
\begin{split}
\Ep \left[ \partial_{\mu \mu}\ell_i^y(\mu_y^d) \right] = - \Ep \left[ \dfrac{\phi(X_i^{d'}\mu_y^d)^2}{\Phi(X_i^{d'}\mu_y^d)\Phi(-X_i^{d'}\mu_y^d)} X_i^d X_i^{d'}\right], \\\ \Ep \left[ \partial_{\nu \nu}\ell_i^w(\nu_w^d) \right] = - \Ep \left[ \dfrac{\phi(X_i^{d'}\nu_w^d)^2}{\Phi(X_i^{d'}\nu_w^d)\Phi(-X_i^{d'}\nu_y^d)} X_i^d X_i^{d'}\right],
\end{split}
\end{equation*}
\begin{align*}    
\Ep&\left[\partial_{\delta  \mu} \ell_{i}^{yw}(\theta_{yw}^d)\right]\\
&=-\Ep\Bigg[\left\{ \dfrac{\Phi_{2i}^{d\mu}(y,w)}{\Phi^d_{2i}(y,w)}-\dfrac{\Phi_{2i}^{d\mu}(y,-  w)}{\Phi^d_{2i}(y,-  w)} + \dfrac{\Phi_{2i}^{d\mu}(y,w)}{\Phi^d_{2i}(- y,w)}-\dfrac{\Phi_{2i}^{d\mu}(y,-  w)}{\Phi^d_{2i}(-  y,-  w)}\right\} \Phi_{2i}^{d\rho}(y,w)\dot g(X^{d'}\delta_{yw}^d) X_i^d X_i^{d'}\Bigg],
\end{align*}

\begin{align*}    
\Ep&\left[\partial_{\delta  \nu} \ell_{i}^{yw}(\theta_{yw}^d)\right]\\
&=-\Ep\Bigg[\left\{ \dfrac{\Phi_{2i}^{d\nu}(y,w)}{\Phi^d_{2i}(y,w)}+\dfrac{\Phi_{2i}^{d\nu}(y,  w)}{\Phi^d_{2i}(y,-  w)} - \dfrac{\Phi_{2i}^{d\nu}(-y,w)}{\Phi^d_{2i}(-  y,w)}-\dfrac{\Phi_{2i}^{d\nu}(-y,  w)}{\Phi^d_{2i}(-  y,-  w)}\right\} \Phi_{2i}^{d\rho}(y,w)\dot g(X_i^{d'}\delta^d_{yw}) X_i^d X_i^{d'}\Bigg],
\end{align*}

\begin{align*}    
\Ep&\left[\partial_{\delta  \delta} \ell_{i}^{yw}(\theta_{yw}^d)\right]\\
&= - \Ep\Bigg[\left\{ \dfrac{1}{\Phi^d_{2i}(y,w)}+\dfrac{1}{\Phi^d_{2i}(y,-  w)} + \dfrac{1}{\Phi^d_{2i}(-  y,w)}+\dfrac{1}{\Phi^d_{2i}(-  y,-  w)}\right\} \Phi_{2i}^{d\rho}(y,w)^2\dot g(X_i^{d'}\delta^d_{yw})^2 X_i^d X_i^{d'}\Bigg],
\end{align*}
where 
\begin{align*}
\Phi_{2i}^{d\mu}(\pm y,\mp w) & = \partial_{\mu} \left. \Phi_2(\pm X_i^{d'}\mu_y^d,\mp X_i^{d'}\nu_w^d;g(X_i^{d'}\delta_{yw}^d))\right|_{\mu = \pm X_i^{d'}\mu_y^d} \\ & = \Phi\left( \frac{\mp X_i^{d'}\nu_w^d - g(X_i^{d'}\delta_{yw}^d) (\pm X_i^{d'}\mu_y^d) }{\sqrt{1-g(X_i^{d'}\delta_{yw}^d)^2}}\right)\phi(X_i^{d'}\mu_y^d),
\end{align*}
\begin{align*}
\Phi_{2i}^{d\nu}(\pm y,\mp w) & = \partial_{\nu} \left. \Phi_2(\pm X_i^{d'}\mu_y^d,\mp  X_i^{d'}\nu_w^d;g(X_i^{d'}\delta_{yw}^d))\right| _{\nu = X_i^{d'}\nu_w^d} \\ & = \Phi\left( \frac{\pm X_i^{d'}\mu_y^d - g(X_i^{d'}\delta_{yw}^d)(\mp X_i^{d'}\nu_w^d) }{\sqrt{1-g(X_i^{d'}\delta_{yw}^d)^2}}\right)\phi(X_i^{d'}\nu_w^d),
\end{align*}
and
$$
\Phi_{2i}^{d\rho}(y,w) = \partial_{\rho} \left. \Phi_2(X_i^{d'}\mu_y^d,X_i^{d'}\nu_w^d;\rho)\right|_{\rho = g(X_i^{d'}\delta_{yw}^d)} = \phi_2(X_i^{d'}\mu_y^d,X_i^{d'}\nu_w^d;g(X_i^{d'}\delta_{yw}^d)).
$$

Also, by the inverse of the partitioned matrix formula, the inverse of $H(\theta^d_{yw})$ is
\begin{tiny}{\begin{equation*}\label{eq:inv-hess}
\left[\begin{array}{ccc}
    \Ep[ \partial_{\mu \mu}\ell_i^y(\mu_y^d)]^{-1} & 0 & 0 \\
     0 & \Ep[\partial_{\nu \nu}\ell_i^w(\nu_w^d)]^{-1} & 0 \\
     - \Ep[\partial_{\delta \delta}\ell_i^{yw}(\theta_{yw}^d)]^{-1} \Ep[\partial_{\delta \mu}\ell_i^{yw}(\theta_{yw}^d)] \Ep[\partial_{\mu \mu}\ell_i^y(\mu_y^d)]^{-1} & -\Ep\partial_{\delta \delta}\ell_i^{yw}(\theta_{yw}^d)]^{-1} \Ep\partial_{\delta \nu}\ell_i^{yw}(\theta_{yw}^d)] \Ep[\partial_{\nu \nu}\ell_i^w(\nu_w^d)]^{-1} & \partial_{\delta \delta}\ell_i^{yw}(\theta_{yw}^d)^{-1}
\end{array}\right].    
\end{equation*}}\end{tiny}

The following lemma, which combines results from \citet{Chernozhukov2013inference}  and \cite{Chernozhukov2018distribution}, provides sufficient conditions to verify Condition $Z$ in \citet{Chernozhukov2013inference} on the compact set $\bar \mY \times \bar \mW$. Let $(\theta,y,w)\mapsto\Psi(\theta,y,w):=\Ep[S^{yw}_i(\theta)]$ and $\Theta = \mathcal{M} \times \mathcal{N} \times \mD$ be the parameter space for $\theta_{yw}^d = (\mu_y^{d'},\nu_w^{d'},\delta_{yw}^{d'})'$.
%and $(\theta,y,w)\mapsto\hat \Psi(\theta,y,w):=\En[S^{yw}_i(\theta_{yw})]$. %The proof of Lemma \ref{lem:conz} is provided in CFL.

\begin{lemma}[Sufficient Condition for $Z$]\label{lem:conz}
Suppose that $\mathcal{M} = \mathcal{N} = \R^{d_x}$, $\mD$ is a compact subset of $\mathbb{R}^{d_{x}}$ and $\mathcal{\bar Y \bar W}:= \mathcal{\bar Y}\times\mathcal{\bar W}$ is a compact set in $\mathbb{R}^{2}$. Let $\mathcal{I}$ be an open set containing $\mathcal{\bar Y \bar W}$. Suppose that (a) $\Psi:\Theta\times\mathcal{I}\mapsto\mathbb{R}^{d_{\theta}}$ is continuous, (b) $\mu \mapsto \Ep[] \partial_{\mu} \ell_i^y(\mu)]$ and $\nu \mapsto \Ep [\partial_{\nu} \ell_i^w(\nu)]$ are the gradients of convex functions for each $y \in \bar \mY$ and $w \in \bar \mW$, respectively, and $\delta \mapsto \Ep[\partial_{\delta} \ell_i^{yw}(\mu_y^d,\nu_w^d,\delta)]$ possesses a unique zero at $\delta_{yw}$
that is in the interior of $\mD$ for each $(y,w)\in\mathcal{\bar Y \bar W}$ and $d \in \{0,1\}$, (c) $\partial\Psi(\theta,y,w)/\partial(\theta',y,w)$ exists at $(\theta^d_{yw},y,w)$, and is continuous at $(\theta_{yw}^d,y,w)$ for each
$(y,w)\in\mathcal{\bar Y \bar W}$ and $d\in\{0,1\}$, and $\dot{\Psi}_{\theta^d_{yw},y,w}=\partial\Psi(\theta,y,w)/\partial\theta^{'} |_{\theta_{yw}^d}$ obeys $\inf_{(y,w)\in\mathcal{\bar Y \bar W}}\inf_{||h||=1}\left\Vert \dot{\Psi}_{\theta_{yw}^d,y,w}h\right\Vert >c_{0}>0$ for each $d \in \{0,1\}$. Then Condition Z of \citet{Chernozhukov2013inference} holds and $(y,w)\mapsto\theta^d_{yw}$ is continuously
differentiable for each $d \in \{0,1\}$.
\end{lemma}

\subsection{Proof of Lemma \ref{lemma:fclt}} 
We drop the group index $d$ to lighten the notation whenever it does not cause confusion. An identical proof applies to both groups and the joint convergence follows from the marginal convergence  by independence. 

We first show the result of the unrestricted estimator $\hat \theta_{yw}$  and then for the estimator of the the tail parameters.

\paragraph{\textbf{FCLT and Bootstrap FCLT for $\hat \theta_{yw}^d$}} 
The proof follows the same steps as the proof of Theorem 5.2 of \citet{Chernozhukov2013inference}  and Theorem C.1 of \cite{Chernozhukov2018distribution}. Let $\Psi(\theta,y,w):=\Pr\left[S^{yw}_i(\theta)\right]$ and $\hat{\Psi}(\theta,y,w):=\Pr_n\left[S^{yw}_i(\theta)\right]$, where $\Pr_{n}$ represents the empirical measure and $\Pr$ represents the corresponding probability measure. From the first-order conditions,  $\hat{\theta}_{yw}=\phi\left(\hat{\Psi}(\cdot,y,w),0\right)$
for each $(y,w)\in\mathcal{\bar Y \bar W}$, where $\phi$ is the $Z$-map defined
in Appendix E.1 of \citet{Chernozhukov2013inference}. The random vector $\hat{\theta}_{yw}$ is the estimator of $\theta_{yw}=\phi\left(\Psi(\cdot,y,w),0\right)$ in the notation of this framework. Applying Step 1 below, we obtain
$$
\sqrt{n}\left(\hat{\Psi}-\Psi\right)\leadsto Z^{\Psi}\text{ in }\ell^{\infty}\left(\mathcal{\bar Y \bar W}\times\mathbb{R}^{d_{\theta}}\right)^{d_{\theta}},
$$
where $Z_{\Psi}(y,w,\theta)=\mathbb{G}S^{yw}(\theta)$, $\mathbb{G}$ is a $P$-Brownian bridge, and $Z_{\Psi}$ has continuous paths a.s.

Step 2 verifies the conditions of Lemma \ref{lem:conz} for $\Psi$,
which also implies that $(y,w)\mapsto\theta_{yw}$ is continuously differentiable
in the set $\mathcal{\bar Y \bar W}$. Then, by Lemma E.2 of \citet{Chernozhukov2013inference}, the
map $\phi$ is Hadamard differentiable with derivative map $(\psi,0)\mapsto-\dot{\Psi}^{-1}\psi$
at $(\Psi,0)$. Therefore, we can conclude  that
\begin{align}\label{eq:fclt_coeffs}
\sqrt{n}\left(\hat{\theta}^d_{yw}-\theta^d_{yw}\right)\leadsto Z^{d\theta}_{yw} :=-\sqrt{p_d} \dot{\Psi}_{\theta^d_{yw},y,w}^{-1} Z^{\Psi}(y,w,\theta^d_{yw})\text{ in }\ell^{\infty}\left(\mathcal{\bar Y \bar W}\right)^{d_{\theta}},
\end{align}
where $(y,w)\mapsto Z^{d\theta}_{yw}$ has continuous paths a.s., because
$$
\label{eq:fclt_coeffs}
\sqrt{n_d}\left(\hat{\theta}^d_{yw}-\theta^d_{yw}\right)\leadsto - \dot{\Psi}_{\theta^d_{yw},y,w}^{-1} Z^{\Psi}(y,w,\theta^d_{yw})\text{ in }\ell^{\infty}\left(\mathcal{\bar Y \bar W}\right)^{d_{\theta}},
$$
by the functional delta
method and 
$$
\sqrt{n_d}/\sqrt{n} \to_{\Pr} \sqrt{p_d}.
$$

\begin{step}[Donskerness] We verify that $\mathcal{G}=\left\{ S_i^{yw}(\theta):(y,w,\theta)\in\mathcal{\bar Y \bar W}\times\mathbb{R}^{d_{\theta}}\right\} $
is a $\Pr$-Donsker with a square-integrable envelope. By examining the expression  
 in \eqref{eq:score2}, we see that $ S_i^{yw}(\theta)$ is a Lipschitz transformation of VC functions with Lipschitz coefficient bounded by $c\left\Vert X_i \right\Vert $ for some constant $c$ and envelope function $c\left\Vert X_i \right\Vert $,
which is square-integrable. Hence $\mathcal{G}$ is $\Pr$-Donsker by Example 19.9 in \citet{van1998asymptotic}.
\end{step}

\begin{step}[Verification of the Conditions of Lemma \ref{lem:conz}]
Conditions (a) and (b) follow directly from Assumption \ref{ass:fclt}. To verify (c),
note that for $(\tilde{\theta}_{yw},\tilde{y},\tilde{w})$ in the neighborhood
of $(\theta_{yw},y,w)$,
$$
\dfrac{\partial\Psi(\tilde{\theta}_{yw},\tilde{y},\tilde{w})}{\partial(\tilde{\theta}_{yw}',\tilde{y},\tilde{w})}=\left[H^{\tilde y \tilde w}(\tilde{\theta}_{yw}),J^{\tilde y \tilde w}(\tilde{\theta}_{yw}),K^{\tilde y \tilde w}(\tilde{\theta}_{yw})\right].
$$
To characterize the analytical expressions of  $H^{\tilde y \tilde w}(\tilde{\theta}_{yw})$, $J^{\tilde y \tilde w}(\tilde{\theta}_{yw})$ and $K^{\tilde y \tilde w}(\tilde{\theta}_{yw})$, we use the notation and expressions introduced in Appendix \ref{app:notation}. Let $\tilde \Phi_{2i}(y,w)$, $\tilde \Phi_{2i}(y,-w)$, $\tilde \Phi_{2i}(-y,w)$, $\tilde \Phi_{2i}(-y,-w))$, $\tilde \Phi_{2i}^{\mu}(\pm y,\mp w)$, $\tilde \Phi_{2i}^{\nu}(\pm y,\mp w)$ and $\tilde \Phi_{2i}^{\rho}( y, w)$ be the same expressions as $ \Phi_{2i}(y,w)$, $ \Phi_{2i}(y,-w)$, $ \Phi_{2i}(-y,w)$, $ \Phi_{2i}(-y,-w))$, $ \Phi_{2i}^{\mu}(\pm y,\mp w)$, $ \Phi_{2i}^{\nu}(\pm y,\mp w)$ and $ \Phi_{2i}^{\rho}( y, w)$ in Appendix \ref{app:notation} but evaluated at $\tilde{\theta}_{yw}=(\tilde{\mu}_{y},\tilde{\nu}_{w},\tilde{\delta}_{yw})$.\footnote{Note that $\tilde \Phi_{2i}(y,w)$ is different from $\Phi_{2i}(\tilde y,\tilde w)$, because the former is evaluated at $\tilde{\theta}_{yw}$, a value near $\theta_{yw}$, whereas the latter is evaluated at $\theta_{\tilde y \tilde w}$, the true value of $\theta_{yw}$ at $(y,w) = (\tilde y, \tilde w)$.} 
Then,
$$
J^{\tilde y \tilde w}(\tilde{\theta}_{yw}):=\dfrac{\partial\Psi(\tilde{\theta}_{yw},\tilde{y},\tilde{w})}{\partial\tilde{y}}=\Ep\begin{bmatrix}
\dfrac{\phi(X_i'\tilde \mu_y)}{\Phi(X_i'\tilde \mu_y)\Phi(-X_i'\tilde \mu_y)} f_{Y \mid X}(\tilde y \mid X) \\
0\\
J_3^{\tilde y \tilde w}(\tilde{\theta}_{yw})
\end{bmatrix},
$$
where 
\begin{footnotesize}{\begin{multline*}
J_3^{\tilde y \tilde w}(\tilde{\theta}_{yw}) = \left( \dfrac{1}{\tilde \Phi_{2i}(y,-  w)} - \dfrac{1}{\tilde \Phi_{2i}(-  y,-  w)}  \right)  f_{Y \mid X}(\tilde y \mid X) \tilde \Phi^{\rho}_{2i}(y, w) \dot g(X_i'\tilde \delta_{yw})
     X_i \\
      + \left( \dfrac{1}{\tilde \Phi_{2i}(y,w)} - \dfrac{1}{\tilde \Phi_{2i}(y,-  w)} - \dfrac{1}{\tilde \Phi_{2i}(-  y,w)} + \dfrac{1}{\tilde \Phi_{2i}(-  y,-  w)}  \right) F_{Y \mid W,X}(\tilde y \mid \tilde w, X) f_{W \mid X}(\tilde w \mid X) \tilde \Phi^{\rho}_{2i}(y, w) \dot g(X_i'\tilde \delta_{yw}) X_i;
\end{multline*}}\end{footnotesize}
and
$$
K^{\tilde y \tilde w}(\tilde{\theta}_{yw}):=\dfrac{\partial\Psi(\tilde{\theta}_{yw},\tilde{y},\tilde{w})}{\partial\tilde{w}}=\Ep\begin{bmatrix}
0 \\
\dfrac{\phi(X_i'\tilde \nu_w)}{\Phi(X_i'\tilde \nu_w)\Phi(-X_i'\tilde \nu_w)} f_{W \mid X}(\tilde w \mid X) \\
K_3^{\tilde y \tilde w}(\tilde{\theta}_{yw})
\end{bmatrix},
$$
where 
\begin{footnotesize}{\begin{multline*}
K_3^{\tilde y \tilde w}(\tilde{\theta}_{yw}) = \left( \dfrac{1}{\tilde \Phi_{2i}(- y, w)} - \dfrac{1}{\tilde \Phi_{2i}(-  y,-  w)}  \right)  f_{W \mid X}(\tilde w \mid X) \tilde \Phi^{\rho}_{2i}(y, w) \dot g(X_i'\tilde \delta_{yw})
     X_i \\
      + \left( \dfrac{1}{\tilde \Phi_{2i}(y,w)} - \dfrac{1}{\tilde \Phi_{2i}(y,-  w)} - \dfrac{1}{\tilde \Phi_{2i}(-  y,w)} + \dfrac{1}{\tilde \Phi_{2i}(-  y,-  w)}  \right) F_{W \mid Y,X}(\tilde w \mid \tilde y, X) f_{Y \mid X}(\tilde y \mid X) \tilde \Phi^{\rho}_{2i}(y, w) \dot g(X_i'\tilde \delta_{yw})X_i.
\end{multline*}}\end{footnotesize}
Also, 
\begin{equation*}
H^{\tilde y \tilde w}(\tilde \theta_{yw}) :=\dfrac{\partial\Psi(\tilde y,\tilde{s},\tilde{\theta}_{yw})}{\partial \tilde{\theta}_{yw}'}= \Ep \left[\begin{array}{ccc}
     H^{\tilde y \tilde w}_{11}(\tilde \theta_{yw}) & 0 & 0 \\
     0 & H^{\tilde y \tilde w}_{22}(\tilde \theta_{yw}) & 0 \\
     H^{\tilde y \tilde w}_{31}(\tilde \theta_{yw}) & H^{\tilde y \tilde w}_{32}(\tilde \theta_{yw}) & H^{\tilde y \tilde w}_{33}(\tilde \theta_{yw})
\end{array}\right].    
\end{equation*}
where
$$
H^{\tilde y \tilde w}_{11}(\tilde \theta_{yw}) = \left\{ G(X_i'\tilde \mu_y) \left[\Phi(X_i'\mu_{\tilde y}) - \Phi(X_i'\tilde \mu_y)\right] - \dfrac{\phi(X_i'\mu_y)^2}{\Phi(X_i'\mu_y)\Phi(-X_i'\mu_y)} \right\} X_i X_i',
$$
$$
H^{\tilde y \tilde w}_{22}(\tilde \theta_{yw}) = \left\{ G(X_i'\tilde \nu_w) \left[\Phi(X_i'\nu_{\tilde w}) - \Phi(X_i'\tilde \nu_w)\right] - \dfrac{\phi(X_i'\nu_w)^2}{\Phi(X_i'\nu_w)\Phi(-X_i'\nu_w)} \right\} X_i X_i',
$$
\begin{footnotesize}{\begin{multline*}
H^{\tilde y \tilde w}_{31}(\tilde \theta_{yw}) = \\ -\left\{ \dfrac{\tilde \Phi_{2i}^{\mu}(y,w)\Phi_{2i}(\tilde y,\tilde w)}{\tilde \Phi_{2i}(y,w)^2}-\dfrac{\tilde \Phi_{2i}^{\mu}(y,-  w)\Phi_{2i}(\tilde y,-\tilde w)}{\tilde \Phi_{2i}(y,-  w)^2} + \dfrac{\tilde \Phi_{2i}^{\mu}(y,w)\Phi_{2i}(-\tilde y,\tilde w)}{\tilde \Phi_{2i}(- y,w)^2}-\dfrac{\tilde \Phi_{2i}^{\mu}(y,-  w)\Phi_{2i}(-\tilde y,-\tilde w)}{\tilde \Phi_{2i}(-  y,-  w)^2}\right\} \times \\
\tilde \Phi_{2i}^{\rho}(y,w)\dot g(X_i^{'}\tilde \delta_{yw}) X_i X_i^{'}  \\
-\left\{ \dfrac{\Phi_{2i}(\tilde y,\tilde w)}{\tilde \Phi_{2i}(y,w)}-\dfrac{\Phi_{2i}(\tilde y,-\tilde w)}{\tilde \Phi_{2i}(y,-  w)} - \dfrac{\Phi_{2i}(-\tilde y,\tilde w)}{\tilde \Phi_{2i}(- y,w)}+\dfrac{\Phi_{2i}(-\tilde y,-\tilde w)}{\tilde \Phi_{2i}(-  y,-  w)}\right\} \tilde \Phi_{2i}^{\rho\mu}(y,w)\dot g(X_i^{'}\tilde \delta_{yw}) X_i X_i^{'}, 
\end{multline*}}\end{footnotesize}
\begin{footnotesize}{\begin{multline*}
H^{\tilde y \tilde w}_{32}(\tilde \theta_{yw}) = -\left\{ \dfrac{\Phi_{2i}(\tilde y,\tilde w)}{\tilde \Phi_{2i}(y,w)}-\dfrac{\Phi_{2i}(\tilde y,-\tilde w)}{\tilde \Phi_{2i}(y,-  w)} - \dfrac{\Phi_{2i}(-\tilde y,\tilde w)}{\tilde \Phi_{2i}(- y,w)}+\dfrac{\Phi_{2i}(-\tilde y,-\tilde w)}{\tilde \Phi_{2i}(-  y,-  w)}\right\} \tilde \Phi_{2i}^{\rho\nu}(y,w)\dot g(X_i^{'}\tilde \delta_{yw}) X_i X_i^{'} \\ -\left\{ \dfrac{\tilde \Phi_{2i}^{\nu}(y,w)\Phi_{2i}(\tilde y,\tilde w)}{\tilde \Phi_{2i}(y,w)^2}+\dfrac{\tilde \Phi_{2i}^{\nu}(y,  w)\Phi_{2i}(\tilde y,-\tilde w)}{\tilde \Phi_{2i}(y,-  w)^2} - \dfrac{\tilde \Phi_{2i}^{\nu}(-y,w)\Phi_{2i}(-\tilde y,\tilde w)}{\tilde \Phi_{2i}(- y,w)^2}-\dfrac{\tilde \Phi_{2i}^{\nu}(- y, w)\Phi_{2i}(-\tilde y,-\tilde w)}{\tilde \Phi_{2i}(-  y,-  w)^2}\right\} \times \\
\tilde \Phi_{2i}^{\rho}(y,w)\dot g(X_i^{'}\tilde \delta_{yw}) X_i X_i^{'}, 
\end{multline*}}\end{footnotesize}
\begin{footnotesize}{\begin{multline*}
H^{\tilde y \tilde w}_{33}(\tilde \theta_{yw}) = -\left\{ \dfrac{\Phi_{2i}(\tilde y,\tilde w)}{\tilde \Phi_{2i}(y,w)}-\dfrac{\Phi_{2i}(\tilde y,-\tilde w)}{\tilde \Phi_{2i}(y,-  w)} - \dfrac{\Phi_{2i}(-\tilde y,\tilde w)}{\tilde \Phi_{2i}(- y,w)}+\dfrac{\Phi_{2i}(-\tilde y,-\tilde w)}{\tilde \Phi_{2i}(-  y,-  w)}\right\} \times \\ \left[\tilde \Phi_{2i}^{\rho\rho}(y,w)\dot g(X_i^{'}\tilde \delta_{yw}) + \tilde \Phi_{2i}^{\rho}(y,w)\ddot g(X_i^{'}\tilde \delta_{yw}) \right] X_i X_i^{'} \\ -\left\{ \dfrac{\Phi_{2i}(\tilde y,\tilde w)}{\tilde \Phi_{2i}(y,w)^2}+\dfrac{\Phi_{2i}(\tilde y,-\tilde w)}{\tilde \Phi_{2i}(y,-  w)^2} + \dfrac{\Phi_{2i}(-\tilde y,\tilde w)}{\tilde \Phi_{2i}(- y,w)^2}+\dfrac{\Phi_{2i}(-\tilde y,-\tilde w)}{\tilde \Phi_{2i}(-  y,-  w)^2}\right\} \tilde \Phi_{2i}^{\rho}(y,w)^2 \dot g(X_i^{'}\tilde \delta_{yw}) X_i X_i^{'}, 
\end{multline*}}\end{footnotesize}
$$
G(u) = \frac{\dd}{\dd u} \left[ \frac{\phi(u)}{\Phi(u)\Phi(-u)}\right] = - \frac{\phi(u)}{\Phi(u)\Phi(-u)}\left[u + \frac{1-2\Phi(u)}{\Phi(u)\Phi(-u)} \right],
$$
$$
\tilde \Phi_{2i}^{\rho \mu} = \partial^2_{\rho, \mu} \left. \Phi_2(\mu,X_i'\tilde \nu_w;\rho)\right|_{\mu = X_i'\tilde \mu_y, \rho = g(X_i'\tilde \delta_{yw})} = - \frac{X_i'\tilde \mu_y - g(X_i'\tilde \delta_{yw}) X_i'\tilde \nu_w}{1-g(X_i'\tilde \delta_{yw})^2}\phi_2(X_i'\tilde \mu_y,X_i'\tilde \nu_w;g(X_i'\tilde \delta_{yw})),
$$
$$
\tilde \Phi_{2i}^{\rho \nu} = \partial^2_{\rho, \nu} \left. \Phi_2(X_i'\tilde \mu_y,\nu;\rho)\right|_{\nu = X_i'\tilde \nu_w, \rho = g(X_i'\tilde \delta_{yw})} = - \frac{X_i'\tilde \nu_w - g(X_i'\tilde \delta_{yw}) X_i'\tilde \mu_y}{1-g(X_i'\tilde \delta_{yw})^2}\phi_2(X_i'\tilde \mu_y,X_i'\tilde \nu_w;g(X_i'\tilde \delta_{yw})),
$$
and
$$
\tilde \Phi_{2i}^{\rho \rho} = \partial^2_{\rho, \rho} \left. \Phi_2(X_i'\tilde \mu_y,X_i'\tilde \nu_w;\rho)\right|_{\rho = g(X_i'\tilde \delta_{yw})} = \partial_{\rho} \left. \phi_2(X_i'\tilde \mu_y,X_i'\tilde \nu_w;\rho)\right|_{\rho = g(X_i'\tilde \delta_{yw})}.
$$

We want to show that $(\tilde{\theta}_{yw},\tilde{y},\tilde{w})\mapsto H^{\tilde y \tilde w}(\tilde{\theta}_{yw})$, $(\tilde{\theta}_{yw},\tilde{y},\tilde{w})\mapsto J^{\tilde y \tilde w}(\tilde{\theta}_{yw})$, and $(\tilde{\theta}_{yw},\tilde{y},\tilde{w})\mapsto K^{\tilde y \tilde w}(\tilde{\theta}_{yw})$ are continuous at $(\theta_{yw},y,w)$ for each $(y,w)\in\mathcal{\bar Y \bar W}$. The verification of the continuity follows from using the expressions, the dominated convergence theorem, and the following ingredients: (1) a.s. continuity of the map $(\tilde{\theta}_{yw},\tilde{s},\tilde{y})\mapsto \partial_{\theta} S_i^{\tilde y \tilde w}(\tilde \theta_{yw})$, (2) domination of $\left\Vert \partial_{\theta} S_i^{\tilde y \tilde w}(\tilde \theta_{yw})\right\Vert$  by a square-integrable function $c\left\Vert X_i \right\Vert$  for some constant $c>0$, (3) a.s. continuity and uniform boundedness of the conditional density functions $y\mapsto f_{Y \mid X}(y \mid x)$ and $w \mapsto f_{W \mid X}(w \mid x)$ by Assumption \ref{ass:fclt}, and (4) $\tilde \Phi_{2i}(y,w)$, $\tilde \Phi_{2i}(y,-w)$, $\tilde \Phi_{2i}(-y,w)$, and $\tilde \Phi_{2i}(-y,-w))$ being bounded away from zero, and $\tilde \Phi_{2i}^{\mu}(\pm y,\mp w)$, $\tilde \Phi_{2i}^{\nu}(\pm y,\mp w)$ and $\tilde \Phi_{2i}^{\rho}( y, w)$ being bounded, uniformly on $\tilde{\theta}_{yw}\in\mathbb{R}^{d_{\theta}}$ a.s. Finally, the last part of (c) follows because the minimum eigenvalue of $ \dot{\Psi}_{\theta_{yw},y,w} = H^{yw}(\theta_{wy})$ is bounded away from zero uniformly on $(y,w)\in\mathcal{\bar Y \bar W}$ by Assumption \ref{ass:fclt}.
\end{step}

% The bootstrap FCLT
% \begin{align*}\sqrt{n}\left(\hat{\theta}_{yw}-\theta_{yw}\right)\leadsto_{\Pr}  Z^{\theta}_{yw} :=-H(\theta_{yw})^{-1} Z^{\Psi}(y,w,\theta_{yw})\text{ in }\ell^{\infty}\left(\mathcal{\bar Y \bar W}\right)^{d_{\theta}},
% \end{align*}
% follows from Assumption \ref{ass:eb} by the same steps as in the proof of Theorem D.2 in CFL.

\paragraph{\textbf{CLT and Bootstrap CLT for $(\hat \xi^d_{\bar y},\hat \xi^d_{\underline y},\hat \xi^d_{\bar w},\hat \xi^d_{\underline w})$}} We use the notation $R$ to refer to either $Y$ and $W$, as the proof is identical in both cases. In what follows it is convenient to analyze the estimator of the lower tail, $\underline r$,  the analysis for estimator of upper tail follows exactly the same steps, switching the sign of the dependent variable, $ R \in \{ Y, W, -Y, -W\}$. 

The estimator $\hat \xi_{\underline r}$  can be seen as a Z-estimator with moment function
$$
\varphi_i(\alpha_{\underline r}, \mu_{\underline r}) = \left(
     \partial_{\alpha}\ell_i^{r_0}(\alpha_{\underline r}, \mu_{\underline r}),
      \partial_{\mu}\ell_i^r(\mu_{\underline r})' \right)'.
$$
Invoking Z-process theory for the simple case of finite-dimensional space $\R^{d_x+1}$,
we have that jointly in $R \in \{Y, -Y, W, -W\}$,
\begin{equation}
 \sqrt{n/n_d} \sqrt{n_d}\left(\hat \xi^d_{\underline r}-\xi^d_{\underline r} \right)  
 \leadsto  \sqrt{p_d} \Ep \left[\begin{array}{cc}
     \partial_{\alpha \alpha}\ell_i^{r_0}(\alpha^d_{\underline r}, \mu^d_{\underline r}) & \partial_{\alpha \mu}\ell_i^{r_0}(\alpha^d_{\underline r}, \mu^d_{\underline r})
\\    0 &  \partial_{\mu \mu}\ell_i^r(\mu^d_{\underline r})  
\end{array}\right]^{-1} \mathbb{G}(\varphi_i(\alpha^d_{\underline r}, \mu^d_{\underline r})).
\label{eq:tail_coeffs}\end{equation}
In fact using Hadamard differentiability results for Z-processes given in \cite{Chernozhukov2013inference}, we conclude that convergence results (\ref{eq:fclt_coeffs}) and (\ref{eq:tail_coeffs})  hold jointly.\footnote{Of course, it is cumbersome to put this joint convergence statement into one display, so we state this verbally.}

The Hadamard differentiability results for Z-processes also imply
that the bootstrap analogs of the results (\ref{eq:fclt_coeffs}) and (\ref{eq:tail_coeffs}) are valid and hold jointly as well. We omit writing the formulas for these convergence results, since they are analogous to (\ref{eq:fclt_coeffs}) and (\ref{eq:tail_coeffs}). \qed

\subsection{Proof of Theorem \ref{thm:clt-cdf}}
Let $UC(\bar \mY \bar \mW,\rho)$ denote the set of functions mapping $\bar \mY \bar \mW$ to the real line that are uniformly continuous with respect to $\rho$, the standard metric in $\R^2$.  The proof follows similar steps to the proof of Theorem 5.2 in \cite{Chernozhukov2013inference}, suitably modified to extend the process to the tails. The main differences are highlighted in Steps 1, 2, and 3 below.

\textsc{Step 1.}(Results for coefficients and empirical measures). 
Application of the Hadamard differentiability results for Z-processes in \cite{Chernozhukov2013inference}, together with Lemma \ref{lemma:fclt}, gives that, 
\begin{equation}\label{eq:fclt_coeffs2}
\sqrt{n/n_m} \sqrt{n_m} \ \hat G^m_X(f) \leadsto G^m_X(f)  \text { in  } \ell^{\infty}(\mathcal{F}), 
\end{equation}
jointly with  (\ref{eq:fclt_coeffs}) and (\ref{eq:tail_coeffs}) for all $d,m \in \{0,1\}$. 
% \footnote{\cite{chernozhukov+13inference} gives detailed arguments on how H-differentiability of Z-processes implies that $
% (\sqrt{n}(\hat{\beta}_{Y}(y)-\beta_{Y}(y)),\hat G_{X}(f))\rightsquigarrow(H_Y(y),G_X(f)) 
% $ in $\ell^{\infty}({\bar{\mathcal{Y}}}%
% )^{d_x}\times\ell^{\infty}(\mathcal{F})$, where $\hat G_{X}(f)$ is the empirical process induced by the marginal distribution of $X$. The extension to stacking another Z-process is straightforward, implying the result \eqref{eq:fclt_coeffs}.
% }

\textsc{Step 2.}(Main: Results for conditional cdfs).
Here we show that, for all $j,k,l,m \in \{0,1\}$,
\begin{align*}
& (\hat Z_{ywx}^{(jkl)}, \hat G^m_X(f)) \leadsto (Z_{ywx}^{(jkl)},   G^m_X(f)) \text{ in } \ell^{\infty}(\mY\mW\mX\mF),\\
& (\hat Z^{(jkl)*}_{ywx}, \hat G^{m*}_X(f)) \leadsto_{\Pr} (Z_{ywx}^{(jkl)},  G^m_X(f)) \text{ in } \ell^{\infty}(\mY\mW\mX\mF).    
\end{align*}

As in the proof of Lemma \ref{lemma:fclt}, we drop the group indexes $i$, $j$ and $k$ to lighten the notation, whenever it does not cause confusion.

It is convenient to extend the processes $\hat \mu_y$, $\hat \nu_w$ and $\hat \delta_{yw}$ to the tails as $$\hat \mu_y = \hat \mu_{\bar y_y} + (y-\bar y_y)\hat \alpha_{\bar y _y} e_1, \ \hat \nu_w = \hat \nu_{\bar w_w} + (w-\bar w_w)\hat \alpha_{\bar w _w} e_1, \ \hat \delta_{yw} = \hat \delta_{\bar y_y \bar w_w}, \quad y \in \mY\setminus \bar{\mY}, w \in \mW\setminus \bar{\mW},$$ where $e_1$ is a unitary $d_x$-vector with a one in the first component.  Likewise, the coefficient functions are given by 
$$  \mu_y =   \mu_{\bar y_y} + (y-\bar y_y)  \alpha_{\bar y _y} e_1, \   \nu_w =   \nu_{\bar w_w} + (w-\bar w_w)  \alpha_{\bar w _w} e_1, \   \delta_{yw} =   \delta_{\bar y_y \bar w_w}, \quad y \in \mY\setminus \bar{\mY}, w \in \mW\setminus \bar{\mW},$$
by assumption.  

%The uniform $\epsilon$-covering numbers for this class
%can be bounded independently of $F_{X_{}}$ by the previous argument and ..., and
%so the Pollard's entropy integral is finite. Hence we can construct a class
%of functions $\mathcal{F}$ containing the union of all the families %$%
%\mathcal{F}_Y$, $\mathcal{F}_W$ and the indicators of
%all rectangles in $\overline{\mathbb{R}}^{d_x+2}$. Note that these indicators
%form a VC class. The final set $\mathcal{F}$ therefore is a DKP class.[VICTOR: THIS STEP NEEDS TO BE UPDATED]

For the body part, $(y,w) \in \bar \mY \times \bar \mW$,
consider the mapping $ \phi: \mathbb{D}_ \phi \subset  \ell^{\infty}(\bar \mY \bar \mW)^{d_{\theta}} \to \ell^{\infty}(\bar \mY \bar \mW \mathcal{X})$, defined as $$t \mapsto  \phi(t), \quad  \phi(t)(y,w,x):= \Phi_2 \left(x'm_y, x'n_w, g(x'd_{yw})\right), \quad t_{yw} := (m_y,n_w,d_{yw}).$$ 
It is straighforward to deduce that this map is Hadamard differentiable at $t_{yw} = \theta_{yw}$ tangentially to $UC(\bar \mY \bar \mW,\rho)^{d_{\theta}}$ with the derivative map given by: $$v \mapsto  \phi_{}^{'}(v), \quad  \phi_{}^{'}(v)(y, w, x)= \partial_{\theta} \Phi_2(x'\mu_{y},x'\nu_{w};g(x'\delta_{yw}))  \cdot  v_{yw}.$$

The tail part can be divided in 3 regions: (I) $(\mY \setminus \bar \mY) \times (\mW \setminus \bar \mW)$, (II) $ \bar \mY \times (\mW \setminus \bar \mW)$, and (III) $(\mY \setminus \bar \mY) \times  \bar \mW$. 
For  $(y,w) \in I$, we consider the mapping
$\phi_I: \mathbb{R}^{3d_x+2} \to \ell^\infty((\mY \setminus \bar \mY) (\mW \setminus \bar \mW) \mX)$ defined by:
\begin{multline*}
(t,a,b) \mapsto \phi_I(t,a,b), \quad  \phi_I(t,a,b)(y,w,x) := \Phi_2(x'm + (y-\bar y_y) a, x'n + (w-\bar w_w) b, g(x'd)), \\ t:=(m,n,d).    
\end{multline*}
This map is also Hadamard differentiable  at $(t,a,b) =
(\theta_{\bar y_y,\bar w_w},\alpha_{\bar y_y},\alpha_{\bar w_w})$ tangentially to the entire domain with the derivative $(v,h,u) \mapsto \phi_I'(v,h, u)$, where
\begin{multline*}
    \phi_I'(v,h, u)(y,w,x) = \partial_{\theta} \Phi_2 (x'\mu_{\bar y_y} + (y-\bar y_y) \alpha_{\bar y_y}, x'\nu_{\bar w_w} + (w-\bar w_w) \alpha_{\bar w_w}, g(x'\delta_{\bar y_y \bar w_w})) v_{\bar y_y \bar w_w} \\ + \partial_{\alpha_{\bar y_y}} \Phi_2 (x'\mu_{\bar y_y} + (y-\bar y_y) \alpha_{\bar y_y}, x'\nu_{\bar w_w} + (w-\bar w_w) \alpha_{\bar w_w}, g(x'\delta_{\bar y_y \bar w_w})) )(y - \bar y_y) h \\ + \partial_{\alpha_{\bar w_w}} \Phi_2 (x'\mu_{\bar y_y} + (y-\bar y_y) \alpha_{\bar y_y}, x'\nu_{\bar w_w} + (w-\bar w_w) \alpha_{\bar w_w}, g(x'\delta_{\bar y_y \bar w_w})) )(w - \bar w_w) u.
\end{multline*}
The derivative is a bounded (continuous) linear operator (note that as $y \to \pm \infty$ or $w \to \pm \infty$, the derivative vanishes, with the linear growth factors $y - \bar y_w$ or $w - \bar w_w$ being dominated by the term $\partial_{\alpha_{\bar y_y}} \Phi_2$ or $\partial_{\alpha_{\bar w_w}} \Phi_2$ with exponential tails). Similarly, for $(y,w) \in II$, we consider the mapping
$\phi_{II}: \ell^{\infty}(\bar \mY)^{2d_x}\times \mathbb{R}^{d_x+1} \to \ell^\infty(\bar \mY  (\mW \setminus \bar \mW) \mX)$ defined by:
$$
(t,b) \mapsto \phi_{II}(t,b), \quad  \phi_{II}(t,b)(y,w,x) := \Phi_2(x'm_y, x'n + (w-\bar w_w) b, g(x'd_y)), \ t_y:=(m_y,n,d_y).
$$
This map is Hadamard differentiable  at $(t_{y},b) =
(\theta_{y,\bar w_w},\alpha_{\bar w_w})$ tangentially to the entire domain with the derivative $(v,u) \mapsto \phi_{II}'(v,u)$, where
\begin{multline*}
    \phi_{II}'(v, u)(y,w,x) = \partial_{\theta} \Phi_2 (x'\mu_{y}, x'\nu_{\bar w_w} + (w-\bar w_w) \alpha_{\bar w_w}, g(x'\delta_{y \bar w_w})) v_{y \bar w_w} \\  + \partial_{\alpha_{\bar w_w}} \Phi_2 (x'\mu_{y}, x'\nu_{\bar w_w} + (w-\bar w_w) \alpha_{\bar w_w}, g(x'\delta_{y \bar w_w})) )(w - \bar w_w) u.
\end{multline*}
The derivative is a bounded (continuous) linear operator. Finally, for $(y,w) \in III$, we consider the mapping
$\phi_{III}: \ell^{\infty}(\bar \mW)^{2d_x}\times \mathbb{R}^{d_x+1} \to \ell^\infty((\mY \setminus \bar \mY) \bar \mW \mX)$ defined by:
$$
(t,a) \mapsto \phi_{III}(t,a), \quad  \phi_{III}(t,a)(y,w,x) := \Phi_2(x'm + (y-\bar y_y) a, x'n_w, g(x'd_w)), \ t_w:=(m,n_w,d_w).
$$
This map is Hadamard differentiable  at $(t_{w},a) =
(\theta_{\bar y_y,w},\alpha_{\bar y_y})$ tangentially to the entire domain with the derivative $(v,h) \mapsto \phi_{III}'(v,h)$, where
\begin{multline*}
    \phi_{III}'(v, h)(y,w,x) = \partial_{\theta} \Phi_2 (x'\mu_{\bar y_y} + (y-\bar y_y) \alpha_{\bar y_y}, x'\nu_{w}, g(x'\delta_{\bar y_y w})) v_{\bar y_y w} \\  + \partial_{\alpha_{\bar y_y}} \Phi_2 (x'\mu_{\bar y_y} + (y-\bar y_y) \alpha_{\bar y_y}, x'\nu_{w}, g(x'\delta_{\bar y_y w}))(y - \bar y_y) h.
\end{multline*}
The derivative is a bounded (continuous) linear operator.

We can now define the "extended map" that combines the body and tail pieces $(t, a, b) \mapsto \bar  \phi(t,a,b)$, where
\begin{multline*}
\bar  \phi(t,a,b)(y,w,x):=  \phi(t)(y,w,x) 
1( (y,w) \in \bar \mY \times \bar \mW)  + \phi_I(t,a,b)(y,w,x) 
1( (y,w) \in I) \\ + \phi_{II}(t,b)(y,w,x) 
1( (y,w) \in II) + \phi_{III}(t,a)(y,w,x) 
1( (y,w) \in III).    
\end{multline*}
By combining the four differentiability results above, we can deduce that this map is Hadamard differentiable with the derivative map $(v, h, u) \mapsto \bar  \phi_{}^{'}(v,h,u)$, where
\begin{multline*}
 \bar  \phi_{}^{'}(v,h,u) (y,w,x) :=
 \phi_{}^{'}(v)(y,w,x) 1( (y,w) \in \bar \mY \times \bar \mW) + \phi_I(v,u,h)(y,w,x) 
1( (y,w) \in I) \\ + \phi_{II}(v,u)(y,w,x) 
1( (y,w) \in II)  + \phi_{III}(v,h)(y,w,x) 
1( (y,w) \in III).        
\end{multline*}

Then, the claim for the conditional process $\hat Z^{(ijk)}_{ywx}$ follows by \eqref{eq:fclt_coeffs}, \eqref{eq:tail_coeffs} and the functional delta method, and expression of  the limit process can be obtained after some algebra.

\textsc{Step 3.} (Auxiliary: Donskerness). One key ingredient for the result is to show that $\mathcal{F}$ is a Dudley-Koltchinskii-Pollard (DKP) class, namely it has bounded uniform covering entroy integral and obeys standard measurability condition (Dudley's image-admissible Suslin condition). We omit any discussion of measurability in this paper, but we note that it trivially holds.  The proof in \cite{Chernozhukov2013inference} relies on compactness of the set $\mathcal{YW}\mathcal{X}$ and applies immediately only to $\bar \mY \bar \mW\mathcal{X}$. We extend the result to $\mathcal{YWX}$. Note that $\mathcal{F} = \{F_{Y,W \mid X}(y,w \mid \cdot): y \in\mathcal{Y}, w \in \mW\}$
is a uniformly bounded ``parametric" family indexed by $y \in \mY$ and  $w \in \mW$
that obeys $|F_{Y,W \mid X_{}}(y,w \mid \cdot) - F_{Y,W \mid X_{}}(y',w' \mid \cdot)| \leq
L(| y - y^{'}| + | w - w^{'}|)$, given the assumption that the density functions $f_{Y \mid X_{}}$ and $f_{W \mid X_{}}$ are uniformly bounded by some constant $L$. 
This is enough to bound the covering numbers for the index set $\bar \mY \times \bar \mW$, but is not enough
to bound the covering number over the unbounded set $\mY \times \mW$.

Under our modelling hypotheses, there exists a small enough constant $C>0$ such that
$$
F_{Y,W \mid X}(y,w \mid \cdot) \leq \exp(C y); \ \ y <0; \ \   \quad 1-F_{Y,W \mid X}(y,w \mid \cdot) \leq \exp(- C y ); \ \ y>0; 
$$
and
$$
F_{Y,W \mid X}(y,w \mid \cdot) \leq \exp(C w); \ \ w <0; \ \   \quad 1-F_{Y,W \mid X}(y,w \mid \cdot) \leq \exp(- C w ); \ \ w>0. 
$$
Let $R_{j} = -M(\eps) - \eps/(2L) + j(\eps/L)$, with $j = 0,..., J$, where $J= \lceil 2 M(\eps) L/\eps \rceil +1 $,  $M(\eps) = \log (1/\eps)/C$ and $\lceil \cdot \rceil$ is the ceiling function. Let $R_{-1} = -\infty$ and $R_{J+1}=+ \infty$. The sets $B_{jk} = \{ F_{Y,W \mid X} (y,w \mid X):  R_{j} \leq  y \leq R_{j+1}, R_{k} \leq  w \leq R_{k+1} \}$
for $j,k \in \{-1,..., J\}$ have the $L^2$ diameter of at most $\sqrt{2} \eps$ independently of the distribution of $F_X$:

\begin{itemize}

\item Indeed by the previous paragraph, if $j,k \in \{0,..., J-1\}$, then the diameter of the set $B_{j,k}$ is at most $\sqrt{2} L (\eps/L)  = \sqrt{2} \eps$.

\item For $j \in \{-1, J\}$ or $k \in \{-1, J\}$, then any pair of conditional cdfs in the same ball obey:
$$
|F_{Y,W \mid X_{}}(y,w \mid \cdot) - F_{Y,W \mid X_{}}(y',w' \mid \cdot)| \leq \exp (- M(\eps)C) = \exp ( -[\log (1/\eps)/C] C ) \leq \epsilon.
$$
\end{itemize}

The number of sets is at most $[2 \log (1/\eps)L/(C\eps) + 5]^2$. It follows that the uniform covering number of the function set $\mathcal{F} = \{F_{Y,W \mid X}(y,w \mid X): y \in \mathcal{Y}, w \in \mathcal{W}\}$
is bounded by $(1+ (1/\eps^2))^2$ up to a constant that does not depend on the distribution of $X$. The set  $\mathcal{F}$ is therefore a DKP class. \qed
\bibliographystyle{aer}
\bibliography{bdr}
%\end{singlespacing}

\newpage
\section{Robustness checks}\label{app:robustness}
\begin{comment}
\begin{figure}[h!]
\vspace{-.5cm}
\caption{Estimated Local Dependence $\rho$} \label{f:rho_all}
\vspace{.2cm}
\begin{scriptsize}
\begin{center}
        \begin{subfigure}[t]{.45\textwidth}
		\centering
		\includegraphics[scale=0.25]{figures/rho_bdr_d.pdf}
		\caption{Daughters, \textit{main} $X$}
	\end{subfigure}%
	%\hspace{2.5cm}
	\begin{subfigure}[t]{.45\textwidth}
		\centering
		\includegraphics[scale=0.25]{figures/rho_bdr_s.pdf}
		\caption{Sons, \textit{main} $X$}
	\end{subfigure} \\
        \vspace{.5cm}
        \begin{subfigure}[t]{.45\textwidth}
		\centering
		\includegraphics[scale=0.25]{figures/rho_bdrall_d.pdf}
		\caption{\centering Daughters, \textit{main} $X$ +\newline  children's race and HH size}
	\end{subfigure}%
	%\hspace{2.5cm}
	\begin{subfigure}[t]{.45\textwidth}
		\centering
		\includegraphics[scale=0.25]{figures/rho_bdrall_s.pdf}
		\caption{\centering Sons, \textit{main} $X$ + \newline children's race and HH size}
	\end{subfigure} 
	\caption*{\tiny \textit{Notes: Panel (A) and (B) show the average local correlation if \textit{main} covariates are used to estimate the marginals and \textit{main} covariates are used to estimate $\rho$. Panel (C) and (D) show the same object when childrens race and household size are added as covariates. The BDR model was estimated using a grid of $15 \times 15$ values. To ensure readability, this figure shows a $5 \times 5$ matrix of $\hat{\rho}$}. }
\end{center}
\end{scriptsize}
\end{figure}
\end{comment}

\begin{figure}[h]
\vspace{-0.5cm}
\caption{Counterfactual transition matrices using no covariates for $\rho$ (boostrapped standard errors in parentheses).} \label{f:tm_count_1}
\vspace{.2cm}
	\begin{subfigure}{.4\textwidth}
		\centering
		\includegraphics[scale=0.28]{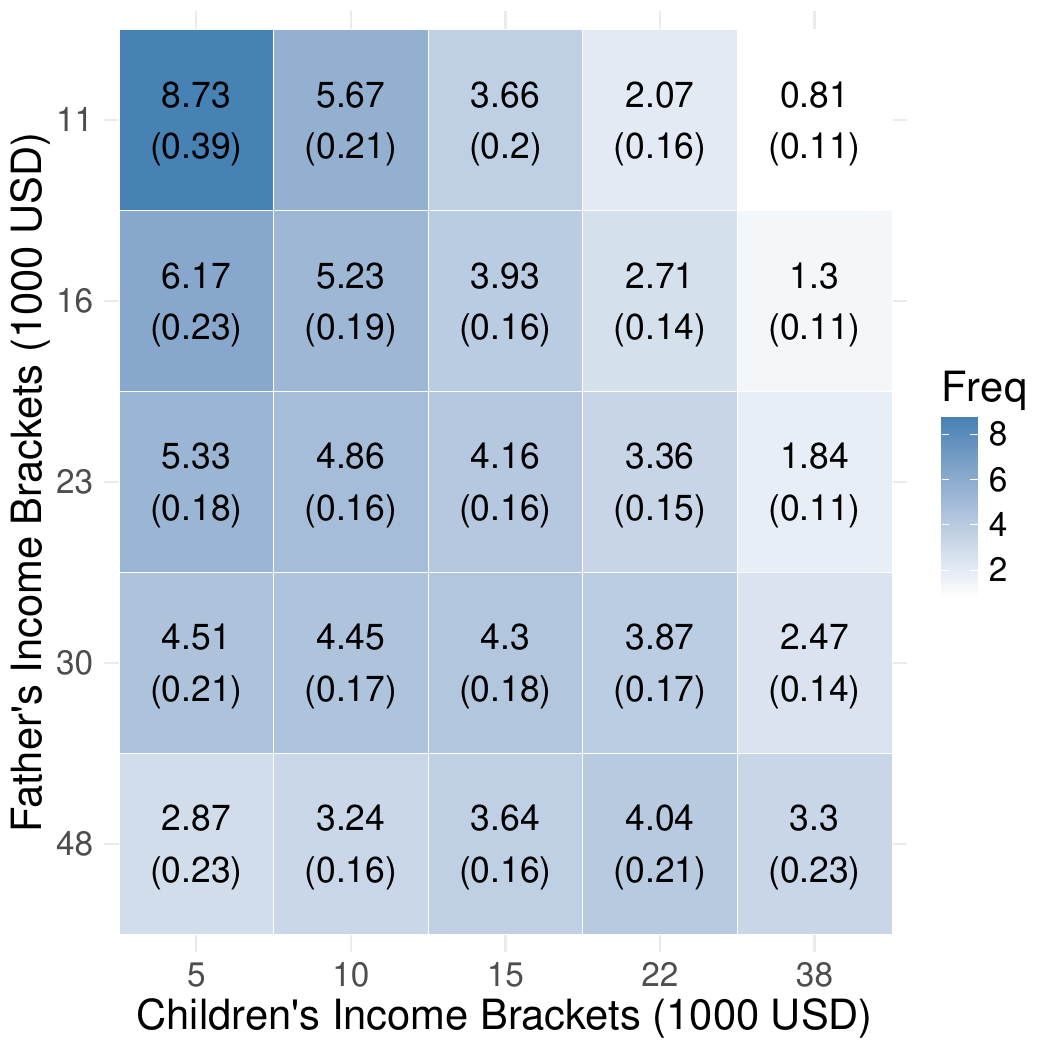}
		\caption{Daughters}
	\end{subfigure}%
	\begin{subfigure}{.4\textwidth}
		\centering
		\includegraphics[scale=0.28]{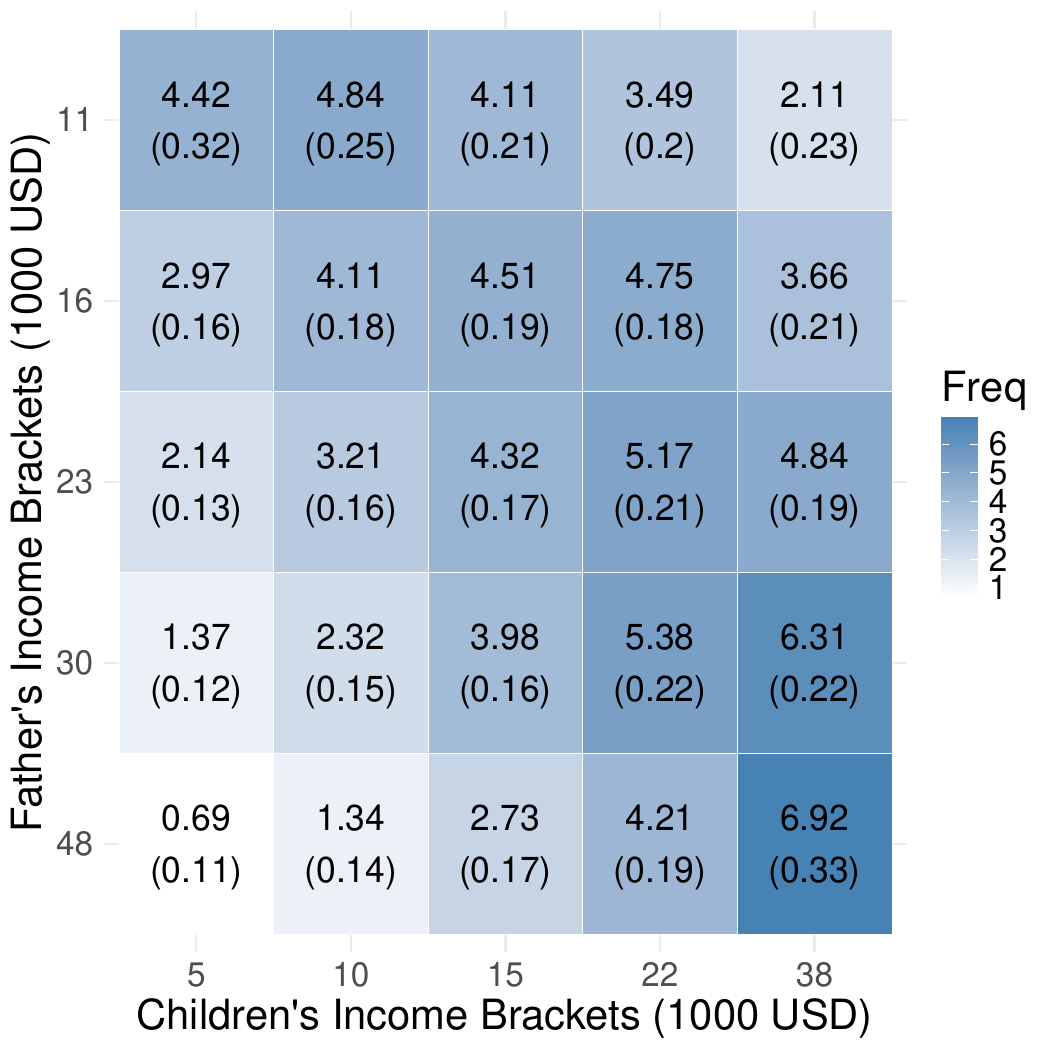}
		\caption{Sons}
	\end{subfigure} \\ 
        \vspace{.5cm}
	\begin{subfigure}{.4\textwidth}
		\centering
		\includegraphics[scale=0.28]{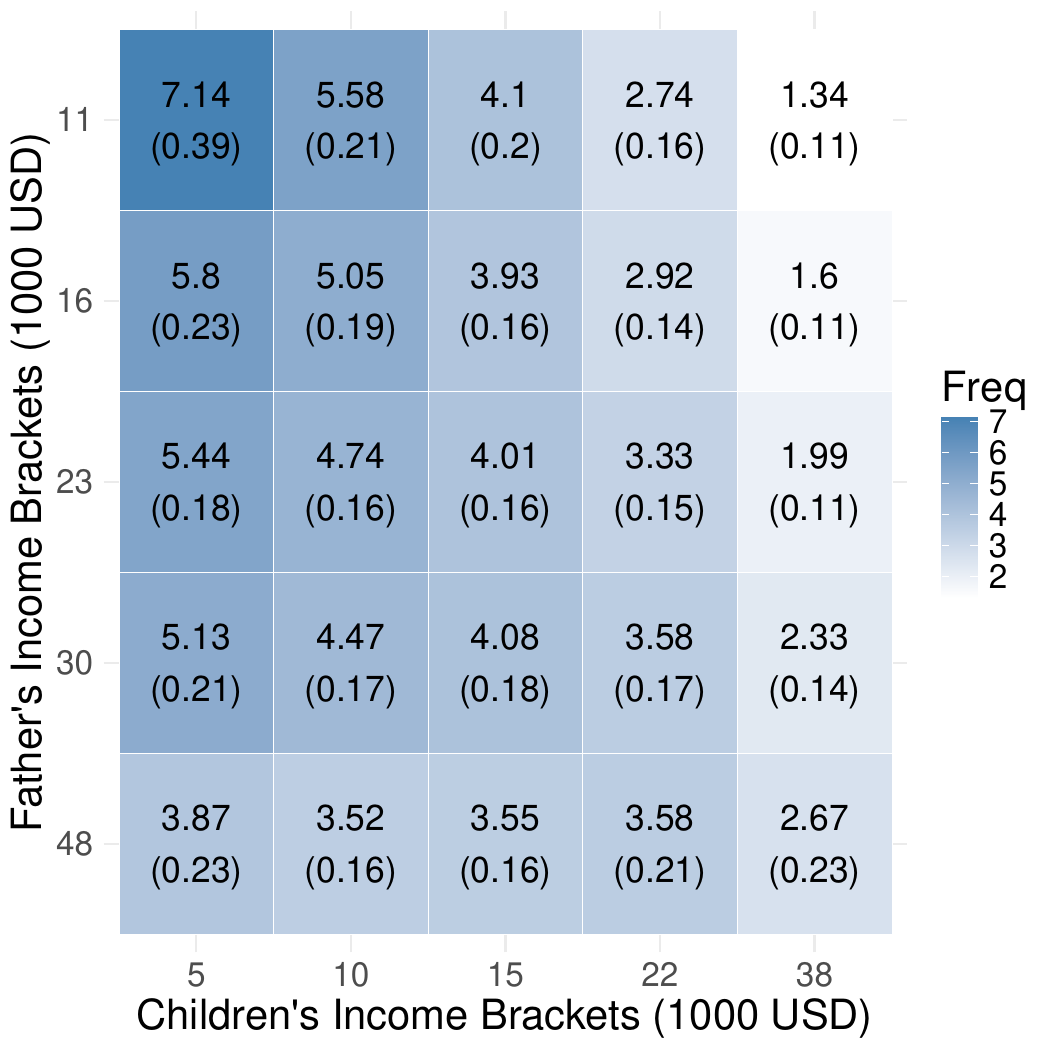}
		\caption{Daughters,  $\rho = 0$}
	\end{subfigure}%
	\begin{subfigure}{.4\textwidth}
		\centering
		\includegraphics[scale=0.28]{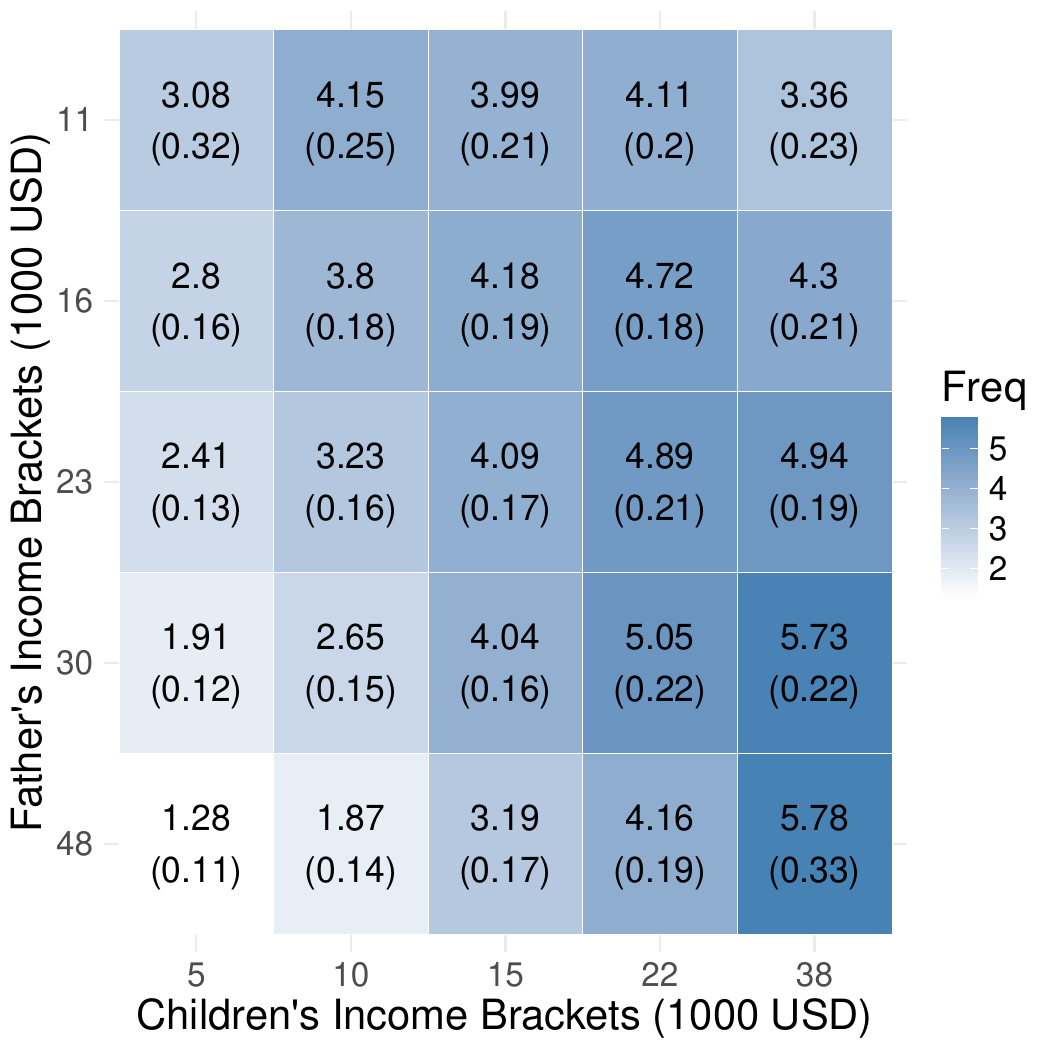}
		\caption{Sons,  $\rho = 0$}
	\end{subfigure} \\ 
        \vspace{.5cm}
	\begin{subfigure}{.4\textwidth}
		\centering
		\includegraphics[scale=0.28]{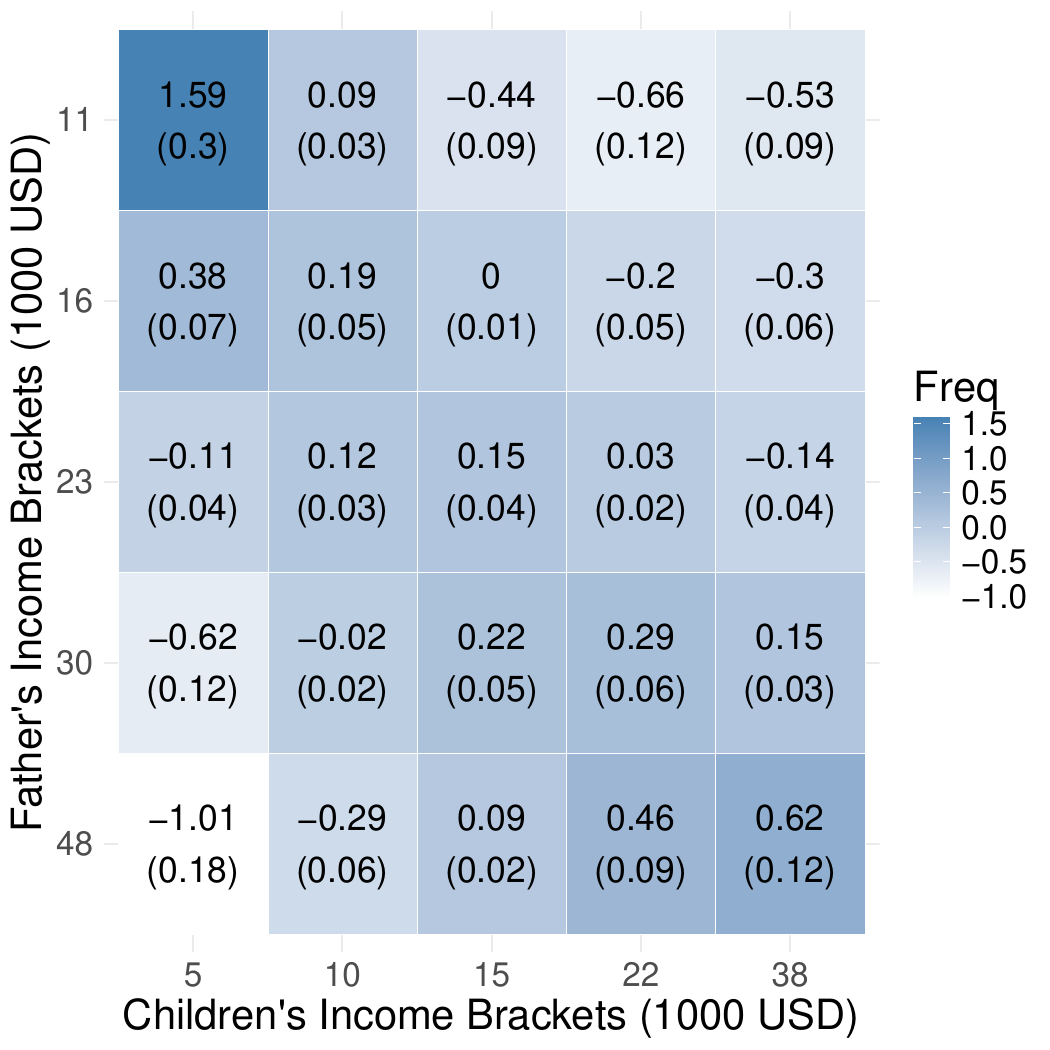}
		\caption{Daughters, difference due to $\rho$}
	\end{subfigure}
	\begin{subfigure}{.4\textwidth}
		\centering
		\includegraphics[scale=0.29]{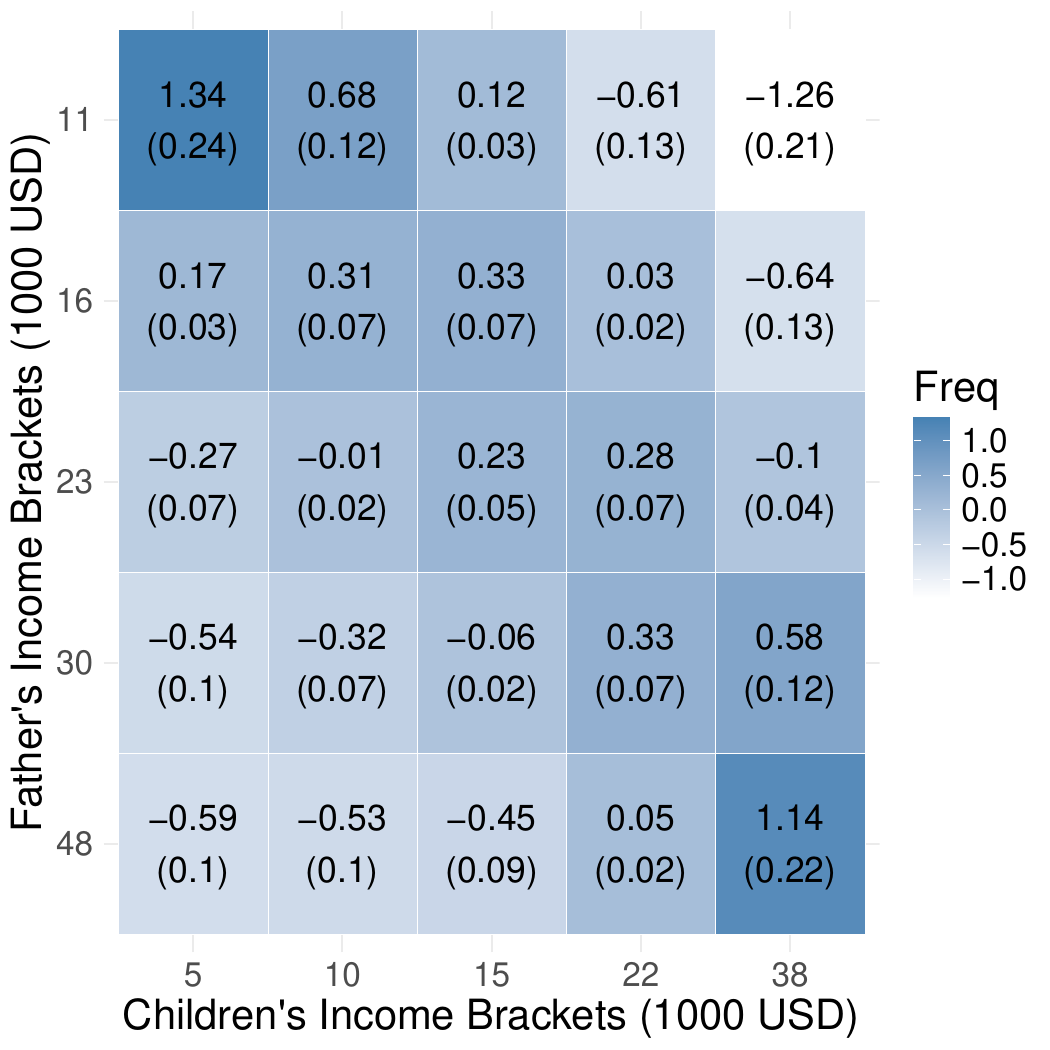}
		\caption{Sons, difference due to $\rho$}
	\end{subfigure}
	\caption*{\tiny \textit{Notes: Panel (A) and (B) show the predicted transition matrices from the full model. For panels (C) and (D), $\hat{\rho}$ is artificially set to zero. The transition matrices capture the dependence due to the marginals. Panel (E) and (F) show the differences between (A) and (C) and (B) and (D). Standard errors calculated as in Figure \ref{f:tm_raw}.}}
\end{figure}

\begin{figure}[h]
\vspace{-.5cm}
\caption{Decomposition of difference in transition matrices between sons and daughters excluding race and household size for children.} \label{f:tm_decom_1}
\vspace{.2cm}
	\begin{subfigure}{.45\textwidth}
		\centering
		\includegraphics[scale=0.25]{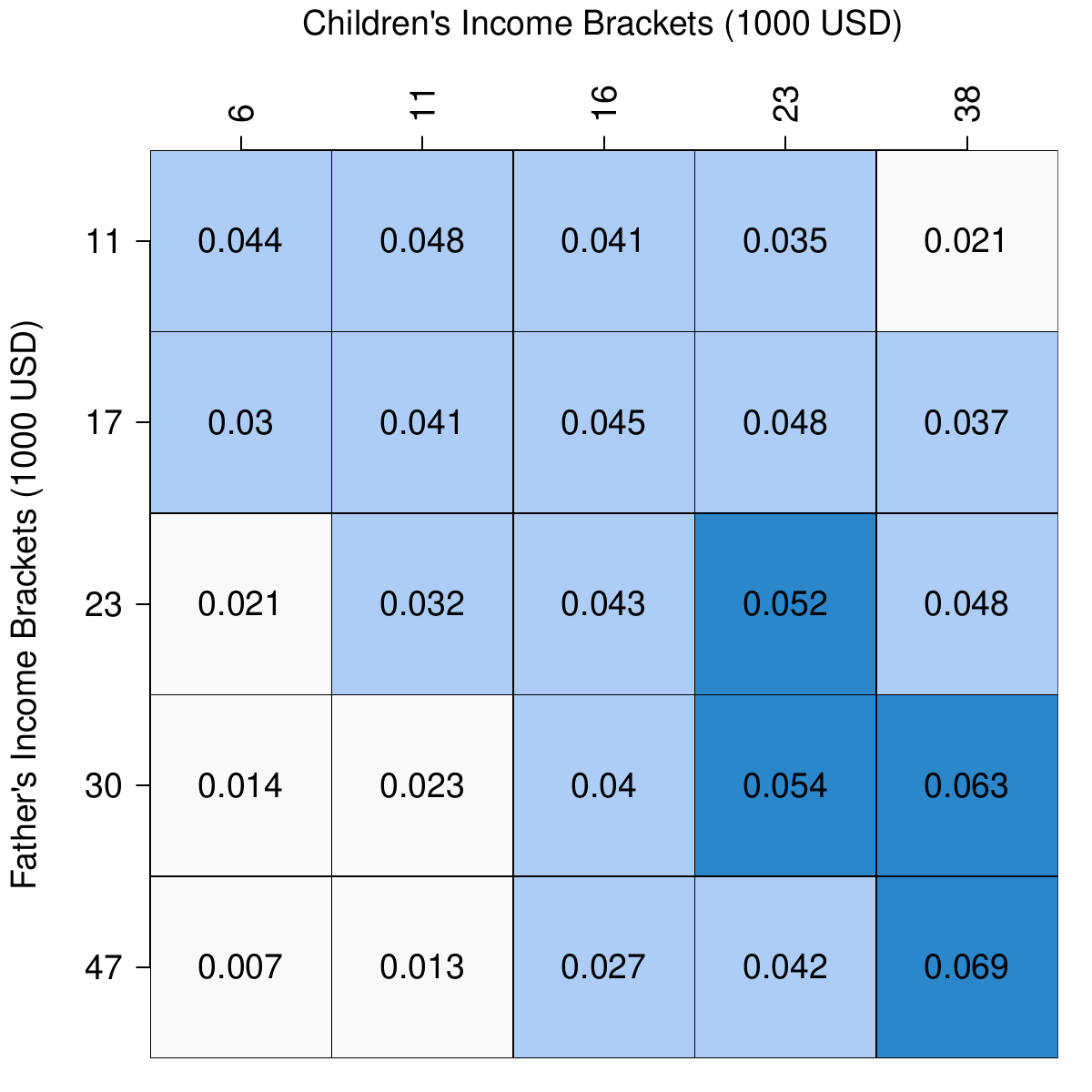}
		\caption{Sons}
	\end{subfigure} 
	\begin{subfigure}{.45\textwidth}
		\centering
		\includegraphics[scale=0.25]{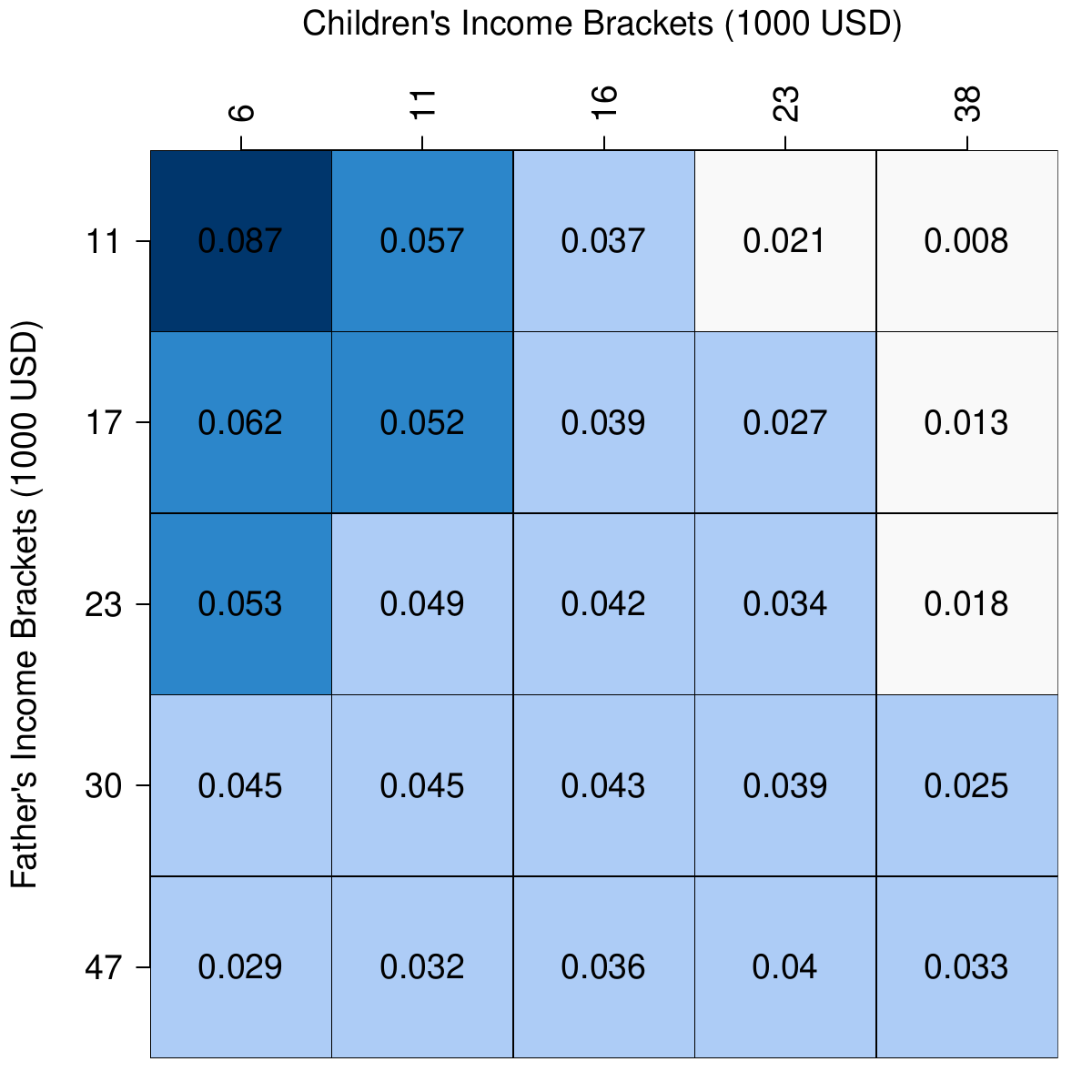}
		\caption{Daughters}
	\end{subfigure} \\ 
        \begin{subfigure}{.45\textwidth}
		\centering
		\includegraphics[scale=0.25]{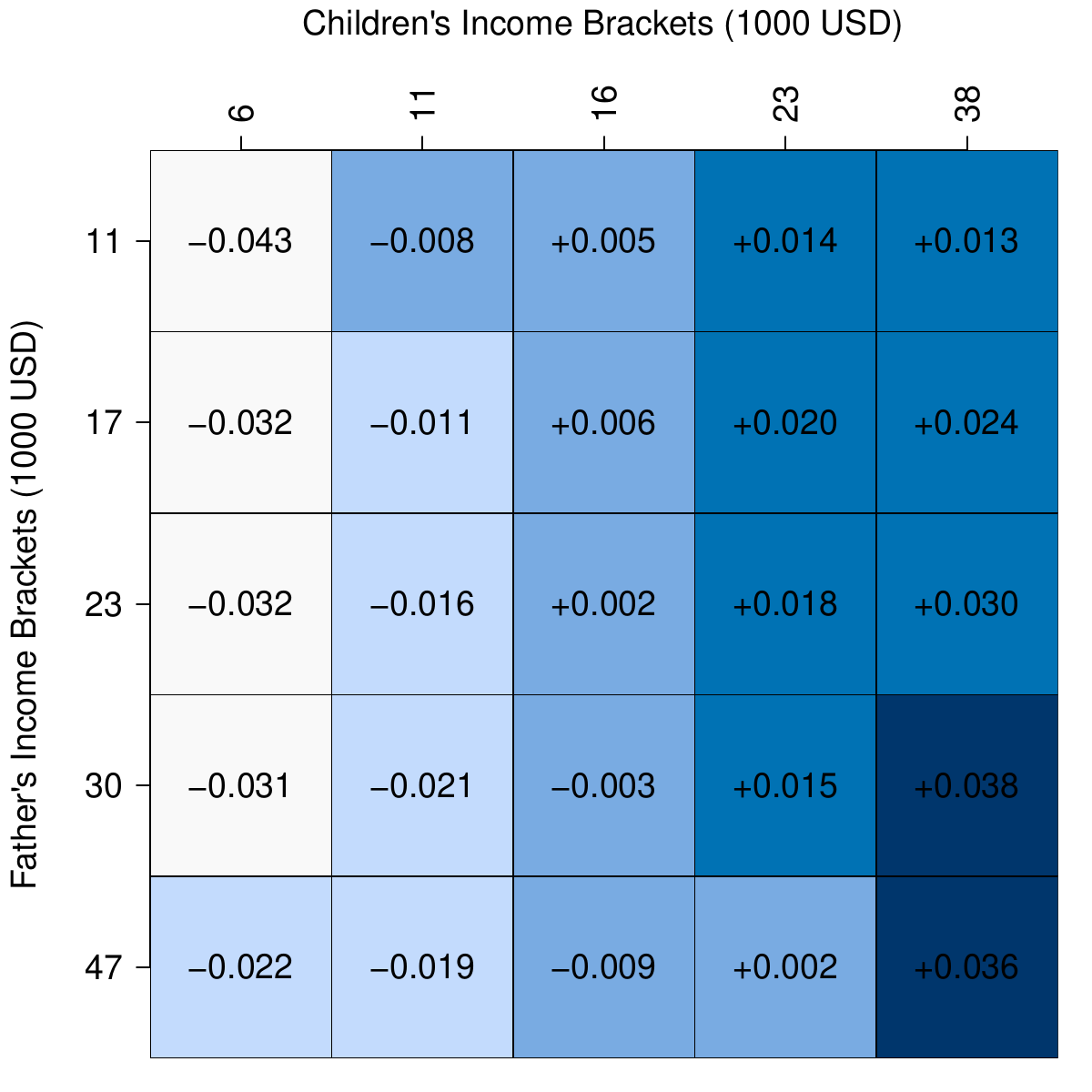}
		\caption{Total Difference}
	\end{subfigure} 
	\begin{subfigure}{.45\textwidth}
		\centering
		\includegraphics[scale=0.25]{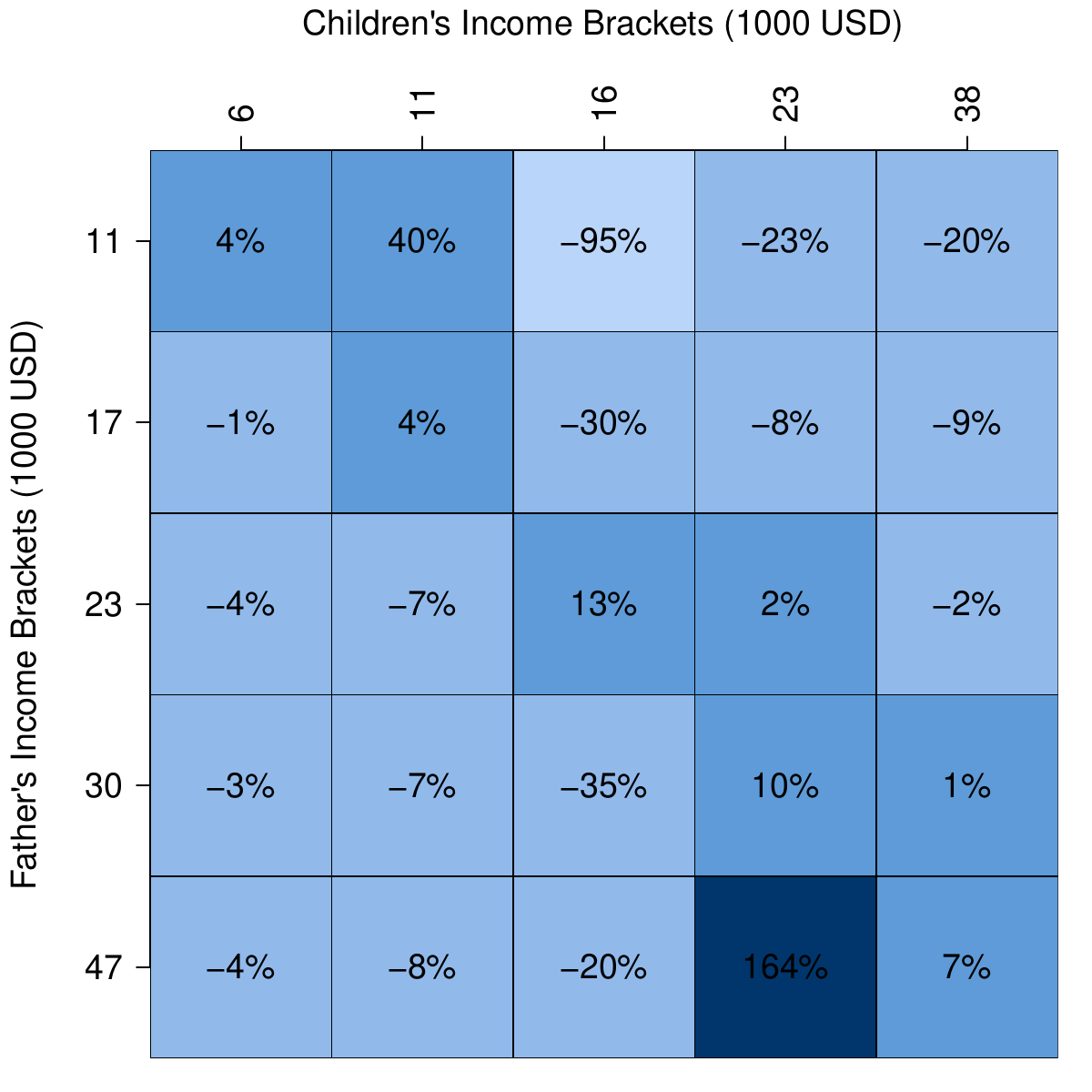}
		\caption{Composition}
	\end{subfigure} \\ 
        \vspace{.5cm}
	\begin{subfigure}{.45\textwidth}
		\centering
		\includegraphics[scale=0.25]{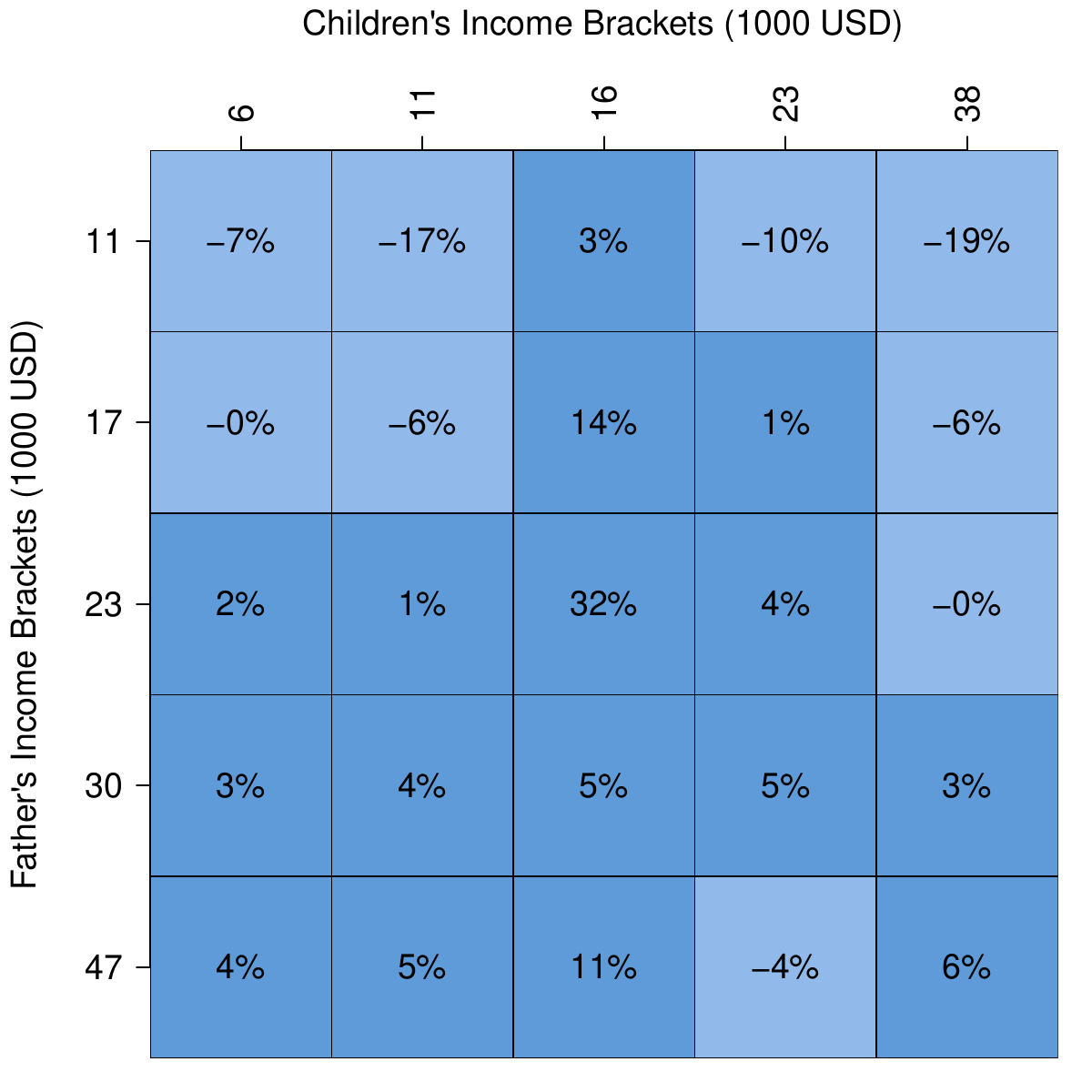}
		\caption{Sorting}
	\end{subfigure}%
        \begin{subfigure}{.45\textwidth}
		\centering
		\includegraphics[scale=0.25]{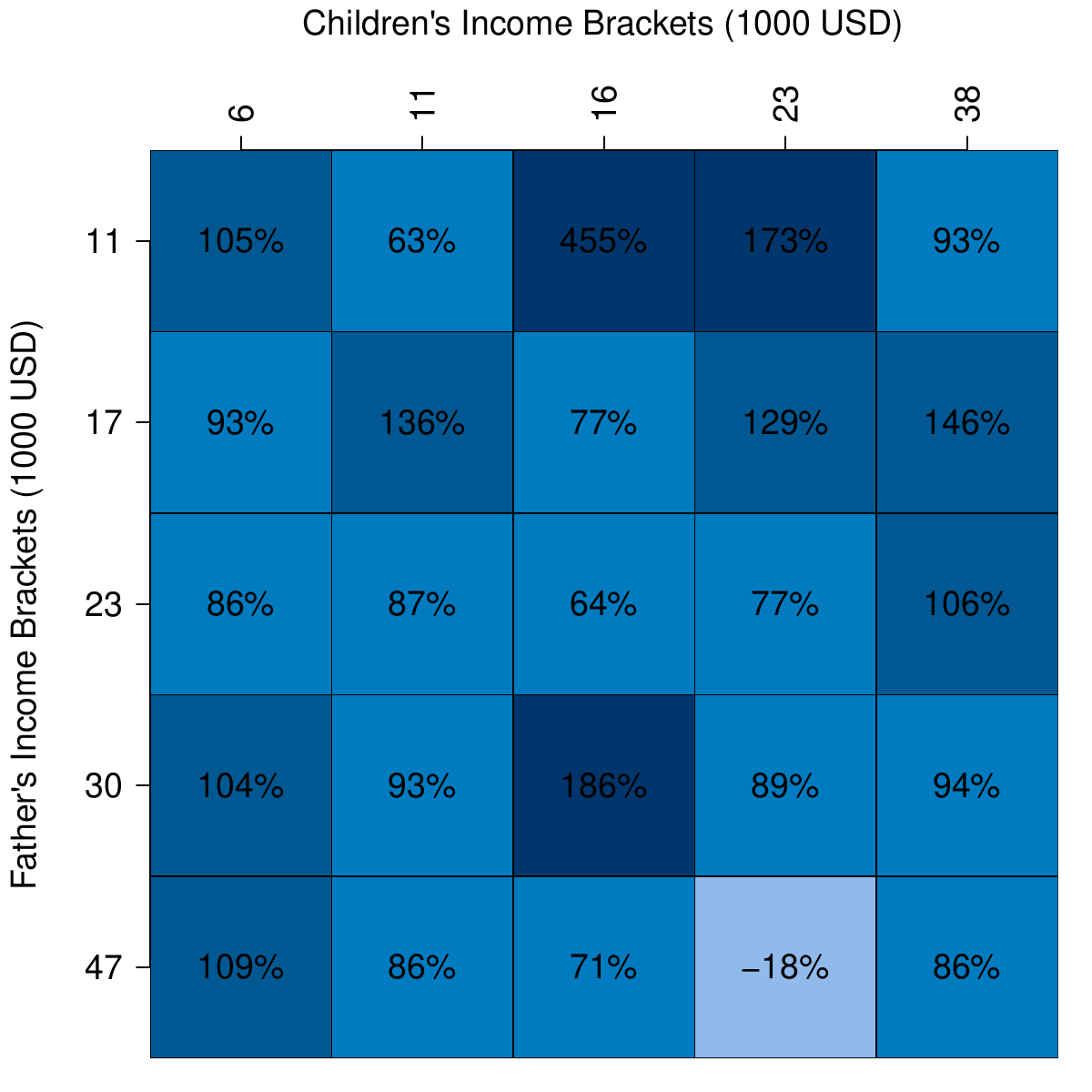}
		\caption{Marginals}
	\end{subfigure} \\ 
	\caption*{\tiny \textit{Notes: Figure \ref{f:tm_decom_1} shows the decomposition of the joint distribution between sons and daughters into three terms. 
    %The CDF was obtained using the BDR model that includes father covariates for the marginals and to estimate $\rho$. 
    %Note that there is not an equal number of observations in all rows as sons and daughters have different marginal distributions. For this figure, we evaluated the joint at the common labor income levels.
    }}
\end{figure}

\end{document}